\documentclass[11pt,a4paper]{article} 
\pdfoutput=1

\usepackage[no-natbib-sort]{my-jheppub}

\usepackage{aurical}
\usepackage[T1]{fontenc}

\usepackage{amsmath, amssymb}
\usepackage{environ}
\usepackage{mathrsfs}
\usepackage{array,arydshln}

\usepackage{graphicx,epsfig}
\usepackage{epic}
\usepackage{color}
\usepackage{youngtab}
\usepackage{float}

\usepackage{slashed}

\usepackage[nodayofweek]{date time}

\newcommand{\be}{\begin{equation}}
\newcommand{\ee}{\end{equation}}

\newcommand{\ads}{AdS$_5\times S^5$\ }

\newcommand{\mc}{\mathcal }
\newcommand{\Z}{\mathcal{Z}}
\newcommand{\N}{\mathcal{N}}
\newcommand{\E}{{\mathcal E}}
\newcommand{\A}{{\mathcal A}}
\newcommand{\M}{{\mathcal M}}

\newcommand{\s}{{\rm s}}
\newcommand{\ra}{{\rm a}}
\newcommand{\rc}{{\rm c}}

\newcommand{\la}{\longrightarrow}

\newcommand{\ds}{\displaystyle}
\newcommand{\vD}{\vec{\slashed{\DD}}}
\newcommand{\DD}{{\mathcal D}}

\def\XXint#1#2#3{{\setbox0=\hbox{$#1{#2#3}{\int}$}
     \vcenter{\hbox{$#2#3$}}\kern-.5\wd0}}

    \newcommand{\beq}{\begin{equation}}
    \newcommand{\eeq}{\end{equation}}
    \newcommand\bea{\begin{eqnarray}}
    \newcommand\eea{\end{eqnarray}}

\def \del{ \partial}
\def \la {\label}
\newcommand{\rf}[1]{(\ref{#1})}
\def\ov{\over}
\def\no{\nonumber} \def \aa {{\rm a}}

\def \ci {\cite}
\def \p {\phi}
\def \m {\mu}\def \n {\nu} 
\def \ed {\end{document}}
 \def \r {\rho} 

\def \foot {\footnote}
\def \b {\beta} 
\def \dd {{\rm d}} 
\def \om {\omega} 
\def \tr {{\rm tr}}
\def \D {\Delta} 
\def \vp {\varphi} 
 \def \ha {{{1 \ov 2}}}

\def \cc {{\rm c}}
\def \aa {{\rm a} }
\def \z {\zeta} 

\def \mZ  {{\rm Z}}
  \def \rk {{\rm k}}
\newcommand{\ts}{\textstyle}

\def \OO  {{\mc O}}   \def \tr  {{\rm tr }}   

\def \iffa  {\iffalse}

\title{Higher spins in AdS$_5$ at one loop:\\
   vacuum energy,  boundary  conformal anomalies
 \\
 and AdS/CFT}
\author[a]{Matteo Beccaria} 
\author[b]{, Arkady A. Tseytlin\footnote{Also at Lebedev Institute, Moscow}}

\affiliation[a]{Dipartimento di Matematica e Fisica Ennio De Giorgi,\\
Universit\`a del Salento \& INFN, Via Arnesano, 73100 Lecce, 
Italy} 

\affiliation[b]{Blackett Laboratory, Imperial College, London SW7 2AZ, U.K.}

\emailAdd{matteo.beccaria@le.infn.it}
\emailAdd{tseytlin@imperial.ac.uk}

\abstract{
We consider general-symmetry  higher spin  fields in  AdS$_5$   and derive  the  expressions 
for their   one-loop  corrections to vacuum energy $E_c$ and  the associated 
4d boundary   conformal anomaly  $\aa$-coefficient. 
We  propose   a  similar   expression for the  second   conformal anomaly $\cc$-coefficient. 
We  show that all the three quantities  $(E_c, \aa, \cc)$  computed for  $\N=8$   gauged 5d supergravity 
are equal to    $-\ha$  of  their   values  for  $\N=4$ conformal 4d supergravity  and also to  twice    the  values
  for $\N=4$ Maxwell  multiplet.
 This gives a 5d derivation of the   fact  that the system  of $\N=4$ conformal supergravity
   and  four  $\N=4$ Maxwell multiplets   is  anomaly free. 
   The values  of   $(E_c, \aa, \cc) $  for  the   states  at   level $p$  of  Kaluza-Klein tower of  10d type IIB supergravity  compactified 
   on $S^5$   turn out to be  equal to those   for   $p$ copies of   $\N=4$  Maxwell multiplets.
   This   may be   related to the fact that   these states  appear in the tensor product of $p$    superdoubletons. 
   Under a natural regularization of the sum over $p$, the full    10d supergravity 
    contribution is then  minus  that     of  one  Maxwell multiplet, in agreement with  the standard  adjoint AdS/CFT   duality  ($SU(N)$  SYM  contribution   is   $N^2-1$ times  that 
 of  one  Maxwell multiplet).  We  also verify the  matching of   $(E_c, \aa, \cc)$
  for spin 0 and $\ha$ boundary theory  cases of vectorial AdS/CFT duality. The consistency conditions for vectorial AdS/CFT 
  turn out to be equivalent  to the  cancellation of anomalies in the  closely related  4d  conformal higher spin theories. 
  In addition, we   study   novel example  of the vectorial AdS/CFT duality when the boundary theory is described  
 by   free spin 1  fields  and  is dual to a particular higher spin theory in AdS$_5$ containing fields in mixed-symmetry representations.
 We also discuss     its supersymmetric generalizations.
 }

\iffa 
we explain relation between   5d sugra and conf anomaly of N=4 CSG 
\begin{itemize}
\item papers of some potential use -- 1207.1079, 1201.0217
\item we should stress that we discuss  ads5 where
mixed reps  are general situation. so we go  through S1x S3 $\to$ $E_c$,
$S^4$ $\to$ a, $R_{mn}=0$ $\to$ c-a    cases ;   and we also relate to CHS in
4d.
possible application is also to string massive multiplets.
\item add a reference somewhere to the final Appendix with the more general Ansatz for $\rc$
In section 3 we will  find the general expression   for the $\aa(\De; j_1, j_2)$   conformal anomaly coefficient in \rf{1.12},\rf{1.11},
generalizing the computation of    \cite{Giombi:2013yva,Giombi:2014iua}
in the bosonic  totally symmetric $(2+s;\, \frac{s}{2},\frac{s}{2})$  case. 
Combined together, the results for $E_c$  and $\aa$ determine   also the  form of the coefficient $\gg(\De; j_1, j_2)$ 
in \rf{1.12},\rf{1.14}.  
In section 4   we will  discuss  similar  expression for the  coefficient $\cc$.  %(\De; j_1, j_2)$. 
While we are presently unable to give its systematic derivation  we shall make   a proposal   
for $\cc(\De; j_1, j_2)$  that  matches all known expressions  in  special cases 
 and  leads  to  very non-trivial  consistency checks  and predictions    in the context of AdS/CFT. 
\end{itemize}
\fi 

\allowdisplaybreaks

\begin{document}
%\date{\currenttime}
%\begin{flushleft} \boxed{\small{\tt \today \ \ - \ \  \currenttime }}\end{flushleft}

\begin{flushright}\small{Imperial-TP-AT-2014-05}\end{flushright}

 \maketitle

\flushbottom
%%%%%%%%%%%%%%%%%%%%%%%%%%%%%%%%%%%%
\def \De {\Delta} 
\def \ads {AdS$_{5}$\ }
\def \te {\textstyle} \def \iffa {\iffalse} 

\def \ha {{\te {1 \ov 2}}}

 \def  \ba { \begin{align} }
 \def  \ea { \end{align} }

\def \gg   {{\rm g}}
\def \cc    {{\rm c}} 
\def \aa  {{\rm a}}

\def \ep {\epsilon}
 \def \k {\kappa} \def \r {\rho} 

\def \RR {{\rm R}}
\def \OO {{\cal O}} 
\def \edd {\end{document}} 
\def \td {\tilde} 

\def \tO {{\td \OO}}\def \rZ {{\rm Z}}

 \def \Deltat  {{\bar \OO}}
 
\def \de {\delta} 
\def \KK {{\rm K}}

\iffa
We consider general-symmetry  higher spin  fields in  AdS_5 and derive  expressions for their one-loop corrections to vacuum energy E and the associated 4d boundary  conformal anomaly  a-coefficient.  We a propose  a  similar  expression for the  second   conformal anomaly  c-coefficient.  We  show that all the three quantities  (E, a, c) computed for  \N=8  gauged 5d supergravity  are  -1/2   of  the   values  for  \N=4 conformal 4d supergravity  and also  twice  the  values  for \N=4 Maxwell  multiplet. This gives  5d derivation of the   fact  that the system  of \N=4 conformal supergravity and  four  \N=4 Maxwell multiplets is anomaly free. The values of (E, a, c) for the states at level p of  Kaluza-Klein tower of  10d type IIB supergravity  compactified  on  S^5 turn out to be  equal to those for p copies of  \N=4  Maxwell multiplets. This may be related to the fact that these states  appear in the tensor product of p superdoubletons.  Under a natural regularization of the sum over p, the full 10d supergravity  contribution is then minus that of  the Maxwell multiplet, in agreement with  the standard  adjoint AdS/CFT   duality  (SU(N)  SYM  contribution  is N^2-1 of  one  Maxwell multiplet).  We also verify the matching of  (E, a, c)  for spin 0 and 1/2 boundary theory  cases of vectorial AdS/CFT duality. The consistency conditions for vectorial AdS/CFT turn out to be equivalent  to the cancellation of anomalies in the closely related  4d  conformal higher spin theories. In addition, we  study  novel example  of  vectorial AdS/CFT duality when the boundary theory is described  by  free spin 1 fields and  is dual to a particular higher spin theory in AdS_5 containing  fields in mixed-symmetry representations. We also discuss  its supersymmetric generalizations.
\fi
%%%%%%%%%%%%%%%%%%%%%%%%%%%%%%%
\section{Introduction} %  and summary}
%%%%%%%%%%%%%%%%%%%%%%%%%%%%%%

AdS$_{d+1}$/CFT$_{d}$    framework  leads to interesting connections  between  properties   of conformal fields  in  dimension $d$ 
 and their  counterparts in  $d+1$. In  particular, there are ``kinematic''   relations   based   on symmetries and 
 special  properties of  AdS  type spaces. One   set of such relations involves   
 singlet sector of    free   CFT$_{d}$,   dual  higher spin   theory in AdS$_{d+1}$
 and  ``shadow'' conformal higher spin theory in $d$  dimensions  (see, e.g., \ci{Giombi:2013yva,Giombi:2013fka,Tseytlin:2013jya,Tseytlin:2013fca, Giombi:2014iua,Giombi:2014yra,Beccaria:2014jxa}
 for some recent  discussions related to the topic of this paper). 
 Here we will be interested in the case of $d=4$.
 
  Starting, e.g.,  with a   free massless   complex scalar  theory   $ \int d^4 x \,   \Phi^*_r \del^2  \Phi_r $  % \ ($r=1, ..., N$)
 one  gets   a tower of    conserved  symmetric traceless  higher spin currents  
 $J_s \sim  \Phi^*_r  \del^{s}   \Phi_r$  which are 
 primary  conformal fields of dimension $\Delta=2+ s \equiv \Delta_+ $. 
 Adding these  currents to  the action with the  {source}  or shadow  fields $\vp_s(x)$  %(which may be interpreted as ``shadow''  fields) 
  one observes that   $\p_s$ has the   
 same   dimension   $\Delta_-= 4-\Delta_+= 2-s$ and  effectively  the same algebraic and gauge   %(due to   properties of $J_s$)
 symmetries   as   (in general, non-unitary)  conformal higher  spin (CHS)   fields.     % in \rf{3}.
  %  It also has  (due to the  properties of $J_s$)  the same algebraic and gauge  symmetries. 
   Integrating  out the  free fields $\Phi_r$ % in the path integral  then
    gives 
 an ``induced''  action  for $\vp_s$   with the kinetic term $\KK(x,x')  \sim \langle  J_s (x) J_s  (x')   \rangle $. The  
 leading (logarithmically divergent)  local    part of  this  action  is  the same as the  
 CHS  action   $\int d^4 x \ \vp_s  \del^{2s} \vp_s + ... $  (with $s=1$ being Maxwell vector, $s=2$ being Weyl graviton, etc.).
 From the  dual \ads   perspective  (implying matching between  the correlators of currents  and amplitudes 
 for dual AdS fields $\p_s$)  
  this induced action  can   be  found upon the  substitution 
 of the solution of the Dirichlet problem  with $\p_s\big|_{\del} =\vp_s$ % as the boundary data) 
 into   the  classical  5d action for  a massless  spin $s$ field  $\p_s$. 
% The free CHS   action  defines  (in general, non-unitary)  higher  spin  CFT   in $d=4$. 
 
 In addition to this ``tree-level''  relation between 5d fields $\p_s$ and  and 4d  conformal higher spin   fields  $\vp_s$ 
(or shadow  counterparts  of the conserved currents $J_s$)  
 there is  also a relation   between 
 the corresponding one-loop   partition  functions, i.e.  between the   determinant  of the 4d 
 kinetic operator $\KK\sim   \del^{2s} \delta(x,x')  $ 
 and  the ratio  of  determinants of     2nd-order   5d operators  for  the field $\p_s$ 
 with Neumann-type $(\De_-$)    and Dirichlet-type ($\De_+$)   boundary conditions. 
 This relation  has  essentially  a   ``kinematic'' origin  belonging to a  general class of bulk-boundary relations 
 discussed   in \ci{Barvinsky:2005ms};  for scalar operators   it was  also implicit in mathematics literature as   discussed in \ci{Diaz:2007an,Diaz:2008hy}. In the context of AdS/CFT  it  appeared in the  context of   the discussion of the bulk  counterpart 
 of  a ``double trace''  deformation  of the boundary CFT
  (see \ci{Witten:2001ua,Gubser:2002zh,Gubser:2002vv,Hartman:2006dy,Diaz:2007an,Diaz:2008hy,Giombi:2013yva}). 
  
  The generalization to higher symmetric  tensors was made explicit in \ci{Giombi:2013yva,Tseytlin:2013jya,Tseytlin:2013fca}).
  In the case when the 4d boundary is a sphere $S^4$  this   leads to  an   expression 
  for the   conformal anomaly  $\aa$-coefficient of the 4d  CHS  field      in terms of the properties
   of the   \ads   determinants \ci{Giombi:2013yva}.\foot{From the AdS/CFT  point of view this  is related, at the same
    time,   to the change of the $\aa$-coefficient    under  the RG flow  induced by  double-trace deformation.}
  In the  case of the ${\mathbb R} \times S^3$  boundary  one  gets   a relation  for the \ads vacuum energy or 
   the Casimir energy on $S^3$  for totally   symmetric CHS fields    \ci{Beccaria:2014jxa}.
 For  a  more general  curved 4d  boundary one should be able  to  obtain  also  a 
 5d expression  for the second  conformal anomaly coefficient $\cc$.

 The  point  which will be important   below 
  is that instead of a  4d CHS field   we may consider a generic primary 4d conformal  field 
 that will  be associated to a particular (in general, massive or massless higher spin)  field in \ads   which will effectively encode its 
 quantum characteristics. %\foot{Generic  $(\De; j_1, j_2)$ field in 5d will  correspond   to a non-unitary  4d  primary field.}
 Dimension 4   is the first  case when the   conformal fields  and  the dual   higher spin  fields in \ads 
 are not  only totally symmetric,   but   may   also appear in    mixed-symmetry   representations  (described 
  by $SO(4)$ Young tableau  with two rows). 
 We shall  use the  $SU(2) \times SU(2)$   weights  $(j_1,j_2)$ to label a representation of the Lorentz   group
 (with spin $s= j_1 + j_2$),  i.e. 
 a  conformal group $SO(2,4)$   representation   with scaling dimension $\De$  
 will be denoted   as $(\De; j_1, j_2)$.

 Our  aim   will be   to  determine the  expressions  for  the $S^3$ Casimir (or vacuum) energy $E_c$  and 4d 
  conformal anomaly coefficients $\aa$ and $\cc$    corresponding  to a generic \ads  field  for the 
   representation  $(\De; j_1, j_2)$.  Our results    will   generalize those for 
    $j_1=j_2= {s\ov 2}$  found for $\aa$  in  \ci{Giombi:2013yva,Tseytlin:2013jya,Giombi:2014iua}  and for $E_c$ 
    in \ci{Giombi:2014yra,Beccaria:2014jxa}. 
    We shall also   propose  a general 
     expression for the $\cc(\De; j_1, j_2)$ coefficient  which  matches   all known values  in special cases  and  
      provides   very non-trivial 
      consistency checks of  AdS/CFT.  
          %5d derivation of conformal anomaly  of conf sugra 
    We shall   then  discuss   applications of our general relations to the adjoint  and vectorial  AdS/CFT  dualities.
  %  with several  new results detailed  below. 
  
  %%%%%%%%%%%%%%%%%%%%%%%%%%%%%%%%%%%%%%%%
 \iffa  \item further checks  of vectorial ads/cft -- also for c-coeffs
\item understand quantum contributions of mixed reps --   they are
generic in ads5 and important also for  understanding zero tension
limit of ads5 x s5 string
\item check 
full consistency (a=0, c=0)   of CHS theories
\item possible generalizations of N=4 CSG to higher spins without
anomalies (though theories with finite number of fields probably do
not exist).But these will appear as building blocks of Fradkin-Linetsky theories
with      and that  is indication of their possible  consistency.
\fi 
%%%%%%%%%%%%%%%%%%%%%%%%%%%%%%%%%%%%%%%%%%%%%%%%
  %The two conformal anomaly coefficients $\rm a, c$ are functions
%of the quantum numbers $(\Delta;\, j_{1}, j_{2})$. 

\subsection{Structure of 4d conformal anomaly}

 Let us  first recall the general expression for the stress  tensor 
  trace  anomaly in a free 4d  CFT  defined on a  curved space 
  \cite{Duff:1977ay,Christensen:1978md}\footnote{Our choice of normalisation is such that  for a real conformal
scalar $\ra= {1 \ov 360}, \ \cc=  {1 \ov 120}, \ \gg= { 1 \ov   180   } $.} 
\be
\label{1.12}
\A = \sum b_4 = -{\rm a}\, \E + {\rm c}\, C^{2}+{\gg}\,D^{2}R \ . 
\ee
Here $b_4$  is the Seeley coefficient (often called also $a_2$)  for the  corresponding kinetic  operator.  
There  may  be several operators  in the case of gauge symmetries and they may be of higher order than 2 in general. 
$\E=R^*R^*$ is the Euler density and $ C$ the Weyl tensor  ($C^2 = \E + 2R^2_{\m\n}- {2\ov 3} R^2$). 
The coefficient {\rm g} of the 
total derivative term 
 is a priori ambiguous (regularization-dependent)  as it can be changed by adding a local $R^2$ counterterm.\footnote{
If one   uses  dimensional regularisation     \cite{Duff:1977ay}  and defines  Weyl tensor in $d$ dimensions then 
 $\rm g = \frac{3}{2}\rc$.
This implies $\A = (\rc-\ra) \E - 4\, \rc\, Q$
where $Q= {1 \ov 4 }[ \E - ( C^2 + \frac{2}{3}\, D^2 R)]$ is the   ``Q-curvature''. 
  % \cite{Eastwood:2008uf}.
This relation is not   true in the   standard heat kernel  (proper time cutoff)   \ci{Christensen:1978md}
 or $\zeta$-function regularization that we shall assume.   For example, for standard spin $\le 1$   fields  one then finds 
 $\aa=  { 31 \ov 180}   n_1  +  { 11 \ov 720 } n_{1\ov 2}  + {1 \ov 360} n_0, $\ \ \
$ \cc= { 1 \ov 10}   n_1  +  { 1 \ov 40 } n_{1\ov 2}  + {1 \ov 120} n_0, $\ \
$ \gg= - { 1 \ov 10}   n_1  +  { 1 \ov 60 } n_{1\ov 2}  + {1 \ov 180} n_0$, 
 where $n_i$ are  the numbers of gauge vectors, Majorana fermions and  real conformal scalars.
}
It  enters the expression for the Casimir energy on $ S^3$  that can  be found from the stress tensor 
\cite{Cappelli:1988vw} 
\be
\label{1.14}
E_{c} = \frac{3}{4}\,\big(   \ra+%\frac{3}{8}\,
{ 1 \ov 2}  \gg \big) \ .
\ee
Like $\gg$,   the vacuum  energy $E_c$ also depends on a choice of regularization.\foot{$E_c$  computed from the 
spectrum of the Hamiltonian   is given by a formally 
divergent sum   which  may be  defined using spectral $\z$-function regularization.} 
  Computed in the standard  heat kernel  or $\zeta$-function  scheme   the  coefficient $\gg$   happens 
  to vanish in     theories %with  a large amount  $\N \geq 3$ 
  with  large   amount   of supersymmetry\footnote{This 
 was found  \ci{Fradkin:1983tg}
in  $\N=4$ SYM
and also appears to be the case     in $\N=3,4$ conformal supergravity as we shall see below.}
so that $E_c$ and $\aa$-coefficient   become   directly proportional (see also    \cite{Herzog:2013ed,Huang:2013lhw}). 
  In   UV finite  theories   with (extended)   supersymmetry    one  also   finds that $\cc$ is equal to $\aa$, 
  and the  two  conditions appear to hold  at the same time  if the number of global  supersymmetries is  $\N \geq 3$, i.e. 
\be
\label{1.13}
  \N \geq 3\ \  {\rm susy}: \ \ \ \ \ \ \ \ \ \ \ \
E_c = { 3 \ov 4} \rm a \ , \ \ \ \ \qquad       \ra= \rc\ , \ \ \ \ \ \ \ \ \    \gg=0 \ . 
\ee
%$\gg$ depends on scheme; so does $E_c$ as  it generally requires regularization. Natural scheme 
%is  heat kernel cutoff   or zeta-function regularization in which $E_c$ is usually computed. In this scheme 
%$\gg$    is not  0 and not proportional to $\cc$ -- this is the scheme we assume below. 
  %aim $a, c , E$; $E$ is also related to conf anomaly  5d perspective 
%%%%%%%%%%%%%%%%%%%%%%%%%%%%%%
\iffa
we  know that d=5, N=8 sugra  has  a-anomaly of - 1/2  N=4 CSG;
adding KK tower   for ads5 x s5 we get a-anomaly as that of -1 SYM .
So  one might think that there is another way of cancelling anomaly
of N=4 CSG -- add  infinite tower of states that in 4d correspond to
KK tower in ads5 x s5. But this is misleading in general as these
massive fields
in 5d   will  lead to  strange  fields in 4d -- more precisely,
states in KK tower are like   $tr F_{mn}^2 Z^p$  --   BPS  operators
generalizing d=5 modes -- $tr F^2$ is dilaton Delta=4 operator, etc.
these are fields  with $\Delta_- < 0$ as I mentioned earlier -- so they
are very non-unitary, and using them to change anomaly directly in 4d
is not a good idea -- just adding SYM  itself is enough in 4d.
\end{enumerate}
\fi
%%%%%%%%%%%%%%%%%%%%%%%%%%%%%%%%%%
We would like to find the general  expressions for  the conformal anomaly   coefficients $ \aa, \cc$   and also $E_c$ (or $\gg$) 
as functions of  the representation labels $\De, j_1, j_2$  by starting with a dual 5d  description of 
a given conformal 4d field. 

Consider a  2nd-order operator $\OO= - D^2 + X $   defined  on a  5d field $\p$  which
corresponds to a representation $(\De; j_1, j_2)$. In the case   when 5d space is \ads  \ $X$ 
is  a constant ``mass'' term 
 (we shall  make the definition of $\OO$ precise below). More generally, we may consider a generalization of 
 \ads to an Einstein space 
 $ds^{2} = \frac{1}{z^{2}}\,[dz^{2}+g_{\m\n}(x,z)\,dx^{\m}\,dx^{\n}] $   which  asymptotes to a curved 
  boundary metric $g_{\m\n}(x)\equiv g_{\m\n}(x,0)$.\foot{In  general, 
   for a higher spin field $\p$ in a 5d  Einstein background the corresponding  kinetic operator may 
  contain non-minimal curvature couplings and  its   consistency    may require 
 an existence of a  proper embedding into an   interacting higher spin theory.}
  The corresponding one-loop partition function (with Dirichlet-type  ``$+$'' or Neumann-type  ``$-$'' boundary conditions) 
\be
\label{1.2}
Z^{\pm} = (\det\, \OO)_{_{\pm}}^{-1/2} \ ,
\ee
 will then   be a  functional of the boundary metric $g_{\m\n}$.
 One may define the associated  boundary conformal  anomaly $\A^{\pm }$ 
 as the  variation of $Z^{\pm}$   under the variation of the conformal factor of the boundary metric:
 $\delta\log Z^{\pm } = - { 1 \ov (4 \pi)^2} \int d^{4}x\,\sqrt g\, \delta\sigma\,\A^{\pm }$,\  $\delta g_{\m\n} = 2\,\delta\sigma\,g_{\m\n}$
 (generalizing the ``tree-level''  5d  derivation of 4d conformal anomaly \ci{Henningson:1998gx}). 
 It was  argued in  \cite{Mansfield:2000zw,Mansfield:2003gs}  that  one should   find 
 \be
\label{1.3}
\A^{+} = (\Delta-2)\,  \bar \A  \ , %  \ \ \ \ \ \ \ \ \ \    \bar \A  = - \ha    b_{4}(\bar \OO) \ , 
\ee
where  according  to   \cite{Mansfield:2003gs} \   $ \bar \A = - \ha    b_{4}(\bar \OO)$ and 
 $\bar \OO$   is a 4d  operator   corresponding to  a ``restriction'' of $\OO$   to the boundary. 
 In this case  
 $\bar \A$ (which should  have  the  same structure as  \rf{1.14})
    can  not depend on $\De$. 
    
    As we shall see below, while   \rf{1.3}  is indeed true, 
    i.e.   both $\aa$ and $\cc$   are proportional to $\De-2$, 
    %$\A^{+} $ is indeed  proportional to $\Delta-2$, 
 the coefficient  $\bar \A$     should have a  non-trivial dependence on $\D$  (in addition to its dependence on 
  $j_1,j_2$).\foot{To find  the $\aa$-coefficient   it is enough to consider the case of \ads   with conformally flat boundary, 
  while to  determine $\cc$ one may  specialize to the case of Ricci flat boundary metric. 
  That   $\aa$ coefficient in 
  $ \bar \A $   should   have   4-th order  polynomial dependence on $\De$  follows  already from 
  the results for a  general massive 5d scalar in euclidean 
   \ads  with  boundary $S^4$ \ci{Diaz:2007an,Diaz:2008hy,Giombi:2013yva}.}
   Our expressions for $\aa$ and $\cc$   will  thus be  different from the ones   proposed in \ci{Mansfield:2003gs} %Nolland:2003kc}
      for spins $ j_1 + j_2 \leq 2 $. 
      The  individual field   contributions  to  $\cc-\aa$ will   also disagree with the general ansatz  in 
       \ci{Ardehali:2013xya,Ardehali:2013gra,Ardehali:2014zba} based on the prescription of \ci{Mansfield:2003gs},  
       though  the agreement  (for $\cc-\aa$ but not for $\aa$ in \ci{Mansfield:2003gs})   will be restored 
       when fields are combined into  for $\N=1$ superconformal multiplets. 
  
%where $b_{4}$ is the Seeley coefficient of a certain 4d restriction $\mc O_{4}$ of $\mc O$.~\footnote{
%Notice that $b_4$ in general is a sum of terms. For instance, it can include ghost contributions.}  
%in the literature about scalar field case \cite{Giombi:2013yva,Diaz:2007an}. As we shall discuss below, 
%it may be true only for $\N=1$ multiplets.

\subsection{Relation between 5d and 4d  partition functions }% and  5d representation for 4d  conformal anomaly} 

To understand  the precise relation   between  the 5d determinants \rf{1.2} and the  conformal anomaly of the associated 4d
  operator  let us start with a 5d action $S_5=\int d^5 x \  \p \OO \p + ...$ and evaluate it on a solution of 
  the Dirichlet problem $\p\big|_{\del} = \vp$, i.e. symbolically 
   \be 
   S_5=  \int d^5 x\  \p \OO \p + ...\ \   \to \ \  S_4 = \int  d^4x\ \vp \KK \vp \  \sim\  \log  \varepsilon  \int  d^4x\  \vp \td \OO \vp + ...  \ . \la{133}\ee
   Here $\varepsilon= {\rm R}^{-1} \to 0  $ is an IR cutoff in 5d. In the case of $\De =2 +s$   when $\p$ is  a
    massless    higher spin field
     the  4d  field $\vp$  is the conformal  higher spin field  and $\td \OO \sim D^{2s} + ...$ is
      the corresponding Weyl-invariant 4d operator depending on $g_{\m\n}$.\footnote{The boundary operator   becomes local
     only  for special values of $\Delta$ (see, e.g., a discussion of the scalar case in  \cite{Diaz:2008hy}).  
In general, we shall   assume analytic  continuation  in $\Delta$.} % Physical applications will all correspond to a local $\mc O_{\rm CHS,4}$. }

%%%%%%%%%%%%%%%%%%%%%%%%%%%%%%%%%%%%%%So, we can evaluate the path-integral
Let us now  consider the following path integral\foot{In the AdS/CFT context 
   %[ give also boundary theory perspective
   this  should be equal to the  generating  functional  for  correlators of bilinear currents $J \sim \Phi^* \del^s \Phi$   in the   boundary CFT, 
   $\rZ(\vp) = \int  d \Phi \  \exp [ - S_4 (\Phi)   + J\cdot \vp ] $. Integrating over $N$   fields $\Phi$  gives induced action for $\vp$ 
   starting   with  $ N  \int \vp \KK \vp\sim   N \log  \varepsilon  \int   \vp \td \OO \vp + ... $   where $ \varepsilon$ is playing the role of a 
   UV 4d cutoff.}
   \be
\label{1.6}
\rZ(\varphi) =   \int_{\phi\big|_{\del} = \varphi}  d\phi\,\ e^{-S_{5}(\phi)} =
\ Z^{+}\,  e^{-S_4 (\varphi)} \big( 1  + ...\big)\  , 
\ee
where in the r.h.s. we  considered semiclassical expansion near the solution of the Dirichlet problem.
Here $Z^+$  is the ``free''  one-loop 5d partition function in \rf{1.2}. 
Next, let us integrate \rf{1.6} over the  4d field $\vp$.  As  was argued in  a similar   context   in 
 \cite{Barvinsky:2005ms}, this results  in path  integral over $\phi$ with ``free'' Neumann  boundary conditions, 
 with the leading 1-loop term then being $Z^-$  in \rf{1.2}
\be\la{1.7} 
\int  d\varphi\,\,  \rZ(\varphi) =    \int_{-}  d\phi\,\ e^{-S_{5}(\phi)}   =      Z^{-} \big(1 + ...\big) \ . 
\ee
Combining this   with  \rf{1.6}    we  find  at the  one-loop order 
\be
Z^{-} =  Z^{+} \,\,   Z   \ , \ \ \ \ \ \ \ \   Z = ( \det \KK )^{-1/2} \to ( \det\tO )^{-1/2}  \ .  \la{1.8}
\ee
Here we assume that $\Delta$ is such that $\KK$  has   leading  local term $\td \OO$ 
as in \rf{133} and the subleading terms  can be ignored in the  limit. The overall singular constant will not contribute to 
observables like conformal anomaly.  
The  case of an   arbitrary  $\Delta$ will  be defined by an analytic continuation, 
which should  give consistent   results  at least for the boundary 
conformal anomaly parts of the corresponding determinants.

%v2
We thus get 
%In the cases when we are allowed  to replace the  determinant of 
%(in general, non-local) operator   $\KK $  by  the  determinant  of   its leading singular  part $ \tO$  we thus get 
 a   relation   between the 5d and 4d determinants of local operators. 
In general,  for  a    5d field    corresponding to a massive or massless  representation 
 $(\Delta;\, j_{1}, j_{2})$  of $SO(2,4)$ 
the associated boundary conformal field   will have  canonical dimension equal to $\De_-= 4- \De$. Thus $\De \ge 4$  cases 
will correspond to 4d  fields with  higher $2(\De-2) \ge 2( j_1 +j_2) $ derivative  kinetic  operators
%v2
$\sim D^{2 (\De-2)} + ... $ 
which   should  give a  Weyl-invariant  action in    curved 4d background.
This implies, in particular, that the corresponding anomaly should   vanish at $\De=2$ 
as in \rf{1.3} as then the operator becomes  algebraic. 
 One simple  case  (cf. \ci{Diaz:2008hy})   is    when $\OO=- D^2 + X $ is the   5d scalar operator  with 
$X= \De (\De-4)=0$, i.e.  corresponding to the  representation $(4; 0,0)$. Then 
%for $\De=3$  the operator   $\tO$  is the standard 2nd order   4d  massless  conformal scalar  operator  and 
%for $\De=4$  the operator   
$\tO$  is the  4-derivative Weyl invariant scalar  operator of \ci{Fradkin:1981jc,Paneitz:1983}.\foot{Weyl-invariant operators  are not unique in 
general: for example, one can add a Weyl-invariant  $ C^2$ term to the $D^4 + ...$  Weyl-invariant 
 operator   with an arbitrary coefficient \ci{Fradkin:1981jc,Fradkin:1983tg}   and the same is true for the 2nd-derivative 
 Weyl-invariant operator defined on symmetric traceless tensor  \ci{Deser:1983tm,%Leonovich:1984cf,Gusynin:1986kz,
 Erdmenger:1997wy,Achour:2013afa}
  corresponding to representation $(3; 1,1)$  and on 4th rank tensor  with symmetries of Weyl tensor  \ci{Erdmenger:1997gy}
  corresponding to representation $(3; 2,0) + (3; 0,2)$.
The relation to  a consistent  5d  operator should  fix  this ambiguity.
 This ambiguity is absent in the case of $D^4$ operator defined  on dimension zero  
  tensor  or $(4; 1,1)$ coming out of the expansion of the $C^2$  Weyl action  related   \ci{Liu:1998bu} to  the   Einstein gravity  action in 5d.}

 As another example, we  may   consider   $\OO$   being  a massless   higher spin gauge  field  operator in (a generalization of) \ads
 space. Then  
 $\tO$   will be the  kinetic operator of the corresponding 
 4d CHS   field   and we will  get the following 5d representation 
for its  1-loop partition 
\be 
Z%_{\rm CHS} 
 = { Z^- \ov Z^+}    \ . \la{1.9}
\ee
This relation was   verified    for the leading (log divergent)  part of    $Z_{\rm CHS} $    on $S^4$ 
and the  corresponding IR divergent parts  of $Z^\pm$  in the euclidean \ads space, i.e. 
 for the  conformal  anomaly $\aa$-coefficient     \ci{Giombi:2013yva,Tseytlin:2013jya}.
In the case  of the  ``thermal'' cover of \ads    with $S^1 \times S^3$   boundary  
eq. \rf{1.9}  was    demonstrated  explicitly (for any value of the length   $\b = -\ln q$ of $S^1$) 
in \ci{Beccaria:2014jxa}.
In particular, it  then  relates the  Casimir energy  $E_c$  of a CHS  field on $S^3$ to the vacuum energy of the  corresponding 
  massless  higher spin field in \ads space.   

The above heuristic  argument   makes clear    the  simple  kinematic origin   of  the relation 
 \rf{1.8} or \rf{1.9}  and suggests 
that  it  should  also extend   to the case  when \ads is  deformed  to an Einstein space asymptotic to 
a generic  curved 4d boundary.  Then  the variation over   the  boundary metric  should   provide a 5d 
representation  for the   4d conformal anomaly %relation between  the corresponding conformal anomalies
\be \A=  \A^{-}-\A^{+} \ ,   \la{1.99} 
\ee
which should apply to all ($\aa$, $\cc$ and $\gg$)  coefficients in \rf{1.12}.
%where $\A$ refer to the 4d CHS field.
It was noticed  in the special case of the symmetric tensor representation $(2+s; {s\ov 2}, {s\ov 2})$   that the $\aa$-coefficients 
corresponding to  $\A^\pm$ 
obey \ci{Giombi:2014iua}  $\aa^+=-\aa^-$. 
Then \rf{1.99} implies that   $\aa= -2 \aa^+$. 
  Similar relation is true   \ci{Beccaria:2014jxa}   for  the Casimir energy and thus 
for the $\gg$ coefficient in   \rf{1.12},\rf{1.14}.  

We shall   see  below  that  the same  applies also  for the general representations $(\De; j_1, j_2)$. 
This is a consequence of the change  of  sign of the expressions for $\aa^+$ and $E_c^+$ under $\De_-\to \De_+$, i.e. 
under $\De-2 \to - (\De-2)$. 
It is  then natural to assume that the same should be true  also   for the $\cc$-coefficient,\foot{A (not directly related)  indication that local properties   of  variations of  5d determinants 
may have opposite signs for the Dirichlet and Neumann boundary conditions is that this is 
what happens for  the coefficients  of $\E$ and $C^2$  in the expression for the boundary $b_5$ Seeley coefficient  in 
\cite{Branson:1999jz}.}
i.e. that 
in general\foot{Equivalently, in   the notation of \rf{1.3}   that means $\bar \A(\De)=\bar \A(4-\De)$.}
\be
\label{1.11}
\A^{-} = -\A^{+} \ , \ \ \ \ \ \ \  {\rm i.e. }  \  \ \ \ \ \ \  \A= - 2 \A^+ \ . 
\ee
%%%%%%%%%%%%%%%%%%%%%%%%
\subsection{Higher spin operators in \ads}

%v2
Let  us now describe  the structure of the 5d operators $\OO$   we will be considering  below. 
Let $\p$  be a    massive ($ \De > 2 + j_1+j_2$  for $j_1 j_2 \not=0$ or $ \De > 1 + j_1+j_2$  for $j_1 j_2=0$) or massless 
($ \De = \De_0\equiv  2 + j_1+j_2$, \  $j_1 j_2 \not=0$)  field in AdS$_{5}$  corresponding  the $SO(2,4)$ representation 
$(\Delta;\, j_{1}, j_{2})$ (see  \rf{a1}).
%\footnote{%v2
%Here we   assume  that   $j_{1}j_{2}>0$. For a  general  discussion see Appendix \ref{A:FF} (cf. \rf{a1}).}
%\foot{The associated boundary conformal field   will have  canonical dimension equal to $\De_-= 4- \De$. Thus $\De \ge 4$  cases 
%will correspond to 4d  fields with  higher $2(\De-2) \ge 2( j_1 +j_2) $ derivative kinetic  operators.}
%will correspond to non-local  boundary operators
One may also 
define the weights 
\be  h_{1}= j_{1} + j_{2}\equiv s\ , \qquad \qquad h_{2}= j_{1}-j_{2}\ , \qquad h_1 \ge h_2 \ , \la{112} \ee
 which  are  integer for bosonic fields and  half-integer for fermionic fields. 
In the bosonic case, $h_{1}$ and $|h_{2}|$ are the lengths of a two-row Young tableau.
According to \cite{Metsaev:1994ys,Metsaev:1995re,Metsaev:2003cu}, the  covariant equation of motion for such 
bosonic transverse   field $\p$   is (for  $j_{1}\ge j_{2}$)
\be
\label{1.16}
 \OO \phi =0 \ ,\quad  \OO =-D^{2}+X\,, \quad \qquad  X =  \Delta\, (\Delta-4)  - h_{1} - |h_{2}|  = (\Delta-2)^2 -  2\,j_{1}  \ ,
\ee
where $D^{2}$ is the standard Laplacian in AdS$_{5}$. This equation is also valid not only for the bosonic, but also for 
the fermionic fields after squaring the 5d Dirac  operator. For a generic fermion  spinor-tensor field 
 $\Psi $ %= \Psi^\alpha_{m_1, \dots, m_{\rm s}}$,  ${\rm s}\equiv = s -\ha =  j_{1}+j_{2}-\ha $
one has     $(\slashed{D}+\Delta-2)\,\Psi=0$ \ \ci{Metsaev:1998xg}. 
 After squaring, this turns out to be 
$\big[ -D^{2}+ {1\ov 4} R -h_{1}-|h_{2}|+1 +(\Delta-2)^{2}\big] \,\Psi=0$  (see  \cite{Metsaev:2013wza}  for details), 
 where $R=-20$ is the scalar curvature of  AdS$_{5}$  assumed to  have  unit  scale.    This  gives  the 
  same $X$ as in (\ref{1.16}). 
%Alternatively, in fermion case another definition of mass is Dirac one:
%$(\slashed{D} + m_D)\, \Psi=0$, so $m_D = \Delta -2$.} %We thank R.R. Metsaev for related explanations.}
A  natural definition of mass of a bosonic field in  \ads  is such that it vanishes  for the  massless representation
  with   $\De=\De_0=2 + s$, i.e.\foot{In the fermionic case  there is a  possible  alternative  definition of mass as  the  parameter  in the Dirac equation: 
$(\slashed{D} + m_{_{\rm D}})\, \Psi=0$,\ \  $m_{_{\rm D} }= \Delta -2$.}
  % \be \la{1.18}
$m^{2} % \stackrel{\rm def}{=}  
\equiv  \Delta\,(\Delta-4) - \Delta_{0}\, ( \Delta_{0}-4) =  (\Delta-2)^{2}-  s^{2} \ ,  $
so that $X= m^2 + (j_1 + j_2)^2 - 2 j_1$.
%\ee
%and, accordingly,  $X$ can be written \be
%X = m^{2}+\Delta_{0}\, (\Delta_{0}-4) - h_{1} - |h_{2}|.
%\ee

The partition function of a massive higher spin field  with   standard (Dirichlet)   boundary  conditions corresponding to $\De=\De_+$ 
is then given by \rf{1.2}  with $\OO$  defined on transverse  fields  in representation $(j_1,j_2)$. We shall denote the massive case 
quantities with \ $\widehat {}\ $\  in what follows, i.e. 
\be 
 Z^+_{\rm massive} \equiv \widehat Z^+(\De; j_1, j_2)  = \big[ \det ( - D^2 + X)_\perp \big]^{-1/2} \ . \la{1.19} \ee
In the massless   case of $\De=\De_0= 2 +s$  we need to take into  account 
  the contribution  of the corresponding ghosts 
that belong to representation $(\De_0+1; j_1-\ha, j_2-\ha) $ (the gauge transformation parameters  $\xi$  in $\delta \phi \sim \del \xi$ 
have  one unit of spin and canonical  dimension $4-\De$  less):
\be 
 Z^+_{\rm massless} \equiv Z^+(\De; j_1, j_2)  =  { \widehat Z^+(\De + 1 ; j_1-\ha , j_2-\ha )  \ov \widehat Z^+(\De; j_1, j_2) }   \ , \ \ \ \ \ \quad 
 \De=2 + j_1 + j_2 \ .  \la{1.20} \ee
% To find  $Z^-$  one needs to 
For example, in the case of  the  totally symmetric massless  higher spin field   representation  one finds 
 \ci{David:2009xg,Gupta:2012he} %\foot{Here   $D^2$ 
 %  operators are defined on symmetric traceless  transverse  tensors in \ads  (with scale set to 1). 
% The energies (dimensions) of the corresponding representations in the  standard boundary conditions 
% case  are $\Delta_+= s + 2$    for the  physical field  \ci{Metsaev:1994ys} 
% and $\Delta_+= s + 3 $ for the ghost one.
%In the case of alternative    boundary conditions $\Delta_- = 4-\Delta_+ = 2-s$ and 
%$\Delta'_- = 4-\Delta'_+ = 1-s$. The ``mass terms'' in the operators  $X= \De (\De -4) -s $  are the same for $\De_+$ and $\De_-$. }
\ba 
 &\qquad \qquad \qquad\qquad  \Z^+(2+s; {\te {s\ov 2}, {s\ov 2}} )\equiv Z^+_{s}  =
 \Big[{\det \, \big(-D^2  +   X'  \big)_{s-1\, \perp} \ov
 \det \, \big(-D^2    +  X \big)_{s\, \perp}}\Big]^{1/2}\  ,  \  \ \   
       \la{111} \\ 
 &  X(\De,s) = \De (\De-2) -s = s^2 - s -4  \ ,  
  \ \ \ \ \qquad  X' = X(\De +1, s-1) = s^2 + s-2  \ . \no\end{align} 
 Below we will use \rf{1.16},\rf{1.19},\rf{1.20}   to compute the corresponding 
 $E_c$ and $\aa$ coefficients. 
 A  direct  5d computation  of $\cc$   or $\cc-\aa$ 
 would require a generalization of $\OO $ in \rf{1.16} to  an Einstein space  which is  asymptotically \ads  with Ricci flat boundary
 which  is not known in general for $s >2$ (cf. \cite{Zinoviev:2008ck,Boulanger:2008tg}). 
 However,  the expresions  for $E_c$ and $\aa$  and known  results in  special cases will allow us 
 to suggest a unique   expression for the $\cc$-coefficient  which will then pass  AdS/CFT consistency checks.

 %%%%%%%%%%%%%%%%%%%%%%%
 \subsection{Summary}
 
 Let us  summarize  the content of  the rest of this paper. 
In section 2  we   consider 
 the  $S^1 \times S^3$  partition function   $Z$ in \rf{1.9}   and   also 
 find  the     corresponding  $S^3$ Casimir  energy 
 for the case of generic  representation  $(\De;{ j_1}, {j_2} )$.  
 The resulting expression for $E_c$ will  follow the pattern in \rf{1.11}. 
 The  one-particle partition functions 
 corresponding to $Z^+$   will be  given directly by the  $SO(2,4) $  characters  but the  case of $Z^-$  will be  more subtle, 
 and  we will  determine it  in few special cases. 
 
In section 3 we   find the general expression   for the $\aa(\De; j_1, j_2)$   conformal anomaly coefficient in \rf{1.12},\rf{1.11},
generalizing the computation of    \cite{Giombi:2013yva,Giombi:2014iua}  done 
in the   totally symmetric $(2+s;\, \frac{s}{2},\frac{s}{2})$ bosonic case. 
Combined together, the results for $E_c$  and $\aa$ determine   also the  form of the coefficient $\gg(\De; j_1, j_2)$ 
in \rf{1.12},\rf{1.14}.  

In section 4   we   determine   a  similar  expression for the second conformal anomaly 
 coefficient $\cc$.  %(\De; j_1, j_2)$. 
While we are presently unable to give its systematic derivation,   we shall make   a proposal   
for $\cc(\De; j_1, j_2)$  that  reproduces  all known special cases 
 and  leads  to   non-trivial  consistency checks  and predictions    in the context of AdS/CFT. 
% We will  also comment on earlier  discussions in 
 %\ci{Mansfield:2003gs,Ardehali:2013xya,Ardehali:2013gra} and in \ci{Tseytlin:2013jya}.
  
% (the expressions in \ci{Mansfield:2003gs,Ardehali:2013xya,Ardehali:2013gra} do not lead to consistent results when applied to individual fields) we shall 
In section 5 we   apply  our  general expressions for $E_c$ \rf{2.32},\rf{2.33}, $\aa$ \rf{3.3},\rf{3.4} 
and $\cc$ \rf{4.16},\rf{4.3} to compute the  corresponding  quantities for sets of
  fields  forming  long or short  $SU(2,2|\N)$ superconformal multiplets. 
We shall  find that the total $\aa$ and $\cc$ vanish  for long ``massive''  $\N=1$  supermultiplets  and observe
that  $\cc-\aa$  for  short  $\N=1$  supermultiplets   agrees with the expressions in \ci{Ardehali:2013xya,Ardehali:2013gra}
formally extended to all  values of spins $j_1,j_2 \ge 1$. 
We will also     rederive  from the 5d   approach 
the values of $K=(E_c, \aa, \cc)$ for $\N \le  4$   Maxwell  and conformal supergravity supermultiplets, verifying the 
relation \rf{1.13}  for $\N=3,4$ cases. We  will    demonstrate 
 that  all the three quantities vanish when $\N=4$ conformal supergravity is combined with exactly   four  
$\N=4$  Maxwell multiplets as in  \ci{Fradkin:1981jc, Fradkin:1985am}. 
The 5d approach provides   a  direct  relation   between  the  conformal anomaly  of $\N=4$  conformal supergravity 
 and the  one-loop contribution   of fields  of $\N=8, \ d=5$ gauged supergravity
   as the  two theories  are  described by the equivalent  
 short  $PSU(2,2|4)$ supermultiplet  (this generalizes to the one-loop level the known tree-level relation  \ci{Liu:1998bu}). 
We will also show  that $K=0$ for a general long  massless  supermultiplet of $PSU(2,2|4)$.

%%%%%%%%%%%%%%%%%%%%%%%%%%%
In section 6 we   turn to applications of our expressions for $K=(E_c, \aa, \cc)$   to AdS/CFT dualities.  
We first   consider  in section 6.1  the ``adjoint'' duality between $\N=4$  $SU(N) $ SYM  and string theory in AdS$_{5}\times $S$^{5}$ .
%We  show   that  the one-loop contribution of 10d type IIB supergravity compactified on $S^5$  
% is exactly the opposite of  one  $\N=4$ Maxwell  multiplet, in agreement with ``$N^2-1$'' rule. 
 We find that   the values of  $K$  for  the   states  at   level $p$  of  Kaluza-Klein tower of  10d type IIB supergravity  compactified 
   on $S^5$   are equal to  the values of  $p$ copies of   $\N=4$  Maxwell multiplets, in line  with  the fact that 
   these states  appear in the tensor product of $p$    superdoubletons \ci{Gunaydin:1998jc}. 
  Under a  particular    regularization of the sum over $p$, this is 
    consistent  with the   adjoint AdS/CFT   duality  with   $SU(N)$  SYM  contribution to $K$ being  $N^2-1$ 
times  that of one  $\N=4$ Maxwell multiplet. 
 As we explain on the example 
 %Our derivation is different 
 %from  the one     in \ci{Mansfield:2003gs} as our expressions for $\aa$ and $\cc$ in general differ. 
  of the   vacuum energy $E_c$,  the  required regularization of the sum over the KK states 
 is,  in fact,  a    spectral $\zeta$-function one   applied to 10d instead of 5d energy states. 
% We then   turn to vectorial AdS/CFT where   we present several new results. 
 
 In section 6.2   we compute $(E_c, \aa, \cc)$  on both sides of the vectorial AdS/CFT   examples. 
 We consider the earlier studied cases  of type A and type B  higher spin theories in \ads 
 corresponding to  the scalar  and  spin $\ha$   fermion 4d  boundary  theories and also a novel example 
  of  ``type C''  theory   dual to a singlet sector of $N$   Maxwell  fields at the  boundary. 
 We also discuss   supersymmetric generalizations of vectorial AdS/CFT. 
In section 6.3 we  point out that consistency  conditions  of vectorial AdS/CFT in  non-minimal  scalar and fermion theory cases 
implying cancellation of total $\aa$ and $\cc$ coefficients  are  equivalent to the 
  consistency (cancellation of conformal anomalies or UV finiteness) 
of  the corresponding 4d conformal higher spin theories.  
Some concluding remarks are made in section 7. 

In Appendix \ref{A:FF}   we   summarize basic representations of $SO(2,4)$, decompositions 
of products of two  doubleton and superdoubleton  representations and  present 
 useful  relations for 
 their  characters that play important  role in the discussion of one-particle partition functions in vectorial AdS/CFT examples.  
   Appendix \ref{A:CSG}   contains  the  computation of  $S^1 \times S^3$ partition functions   of  low-spin  conformal 4d fields 
 that  appear in  extended conformal supergravities  and   provide useful examples for the discussion in section 2. 
  In Appendix \ref{A:heatkernel}  we give details of   the  derivation of  the spectral $\zeta$-function for 
  massive higher spin \ads operator $\OO$ in \rf{1.16}   which  is used in 
  section 3. 
   In Appendix \ref{A:cc}   we complement the discussion in section 4 by presenting a more general ansatz  for the 
  $\cc$-coefficient   that  contains one free parameter.
  Appendix \ref{A:KK}  summarizes the spectrum of  5d fields  appearing in 10d type IIB  supergravity compactified on  $S^5$
  which we use in sections 5 and 6. 
  \iffa 
  In Appendix \ref{A:zz}   we present the total ``twisted'' partition function   on $S^1 \times S^3$  (with fermions being periodic on $S^1$)
  for the field content of 10d type IIB  supergravity compactified on  $S^5$  
   and demonstrate,  that like the vacuum energy and 
   conformal anomalies discussed in section 6.1, 
    it is also  equal  minus the $\N=4$  Maxwell theory   contribution, in agreement with adjoint AdS/CFT. 
  \fi

%%%%%%%%%%%%%%%%%%%%%%%%%%%%%%%%%%%%%%%%%%%%%%%%%%%%%
\section{Partition function on $S^{1}\times S^{3}$  and Casimir energy}%   for general conformal fields}
%%%%%%%%%%%%%%%%%%%%%%%%%%%%%%%%%

In this section we shall  consider  the expressions for one-particle partition function  and  $S^3$ Casimir  energy. 
% for the case of generic  representation $(\De;{ j_1}, {j_2} )$.  
We shall start with the previously  discussed  case of  totally symmetric $(2+s;{ s\ov 2}, {s\ov 2} )$ 
 representation  and then turn to  the  case  of mixed representation  $(\De;{ j_1}, {j_2} )$.

\subsection{Totally symmetric bosonic  spin $s$ conformal fields}

The canonical partition function of a free CFT in $S^{1}\times S^{3}$   % (with  length $\b of  $S^1$   being    $\beta$)   
can be computed by direct evaluation of the free QFT path-integral, { i.e.} by finding the eigenmodes of the
 quadratic kinetic operator. The same  expression can be obtained   by the 
 %An alternative approach is the so-called 
 operator counting method \cite{Cardy:1991kr,Kutasov:2000td,Beccaria:2014jxa}.
  In radial quantisation,  
conformal operators in $\mathbb R^{4}$  with dimensions $\Delta_{n}$ are related to 
eigenstates  of the Hamiltonian  %(namely, the dilatation operator) 
on $\mathbb R_t \times S^{3}$.
From the spectrum  of    eigenvalues $\omega_{n} = \Delta_{n}$ and their
degeneracies ${\rm d}_n$   one  gets  the {\em one-particle}, or canonical, 
 partition function 
\be
\label{2.1}
\Z(q) = {\rm Tr}\, e^{-\beta H} =  \sum_n  {\rm d}_n\, e^{-\beta\,\omega_n} =   \sum_n  {\rm d}_n\,   q^{\Delta_n},
\qquad q\equiv e^{-\beta}\ .
\ee
The    multi-particle, or grand canonical,  partition function  is then given, in the bosonic  and fermionic cases,  by 
\ba
&B: \ \ \   \log Z(q)  = - \sum_n {\rm d}_n \log (  1 - e^{-\beta \omega_n})  = \sum_{m=1}^\infty  \frac{1}{m}\,  \Z(q^m)\ , \la{22}\\
&F: \ \ \    \log Z(q)  = - \sum_n {\rm d}_n \log (  1 + e^{-\beta \omega_n})  = -\sum_{m=1}^\infty  \frac{(-1)^m}{m}\,  \Z(q^m)\ . \la{2222}
\end{align}
The  analysis  of the counting of states  implies 
 the following   structure  of $\Z(q)$  \ci{Beccaria:2014jxa}
\be
\label{2.3}
\Z(q) = \Z_{-}(q)-\Z_{+}(q),\qquad\qquad  \Z_{-}= \Z^{\rm off-shell},\qquad \Z_{+}  = \Z^{\rm e.o.m.} \ . 
\ee
Here % the  ``{shadow} field''  partition function 
$\Z_{-} $    counts  the 
 off-shell components (and  their derivative descendants)  of a suitable gauge invariant 
  field strength    modulo non-trivial gauge identities  while  
  %The  ``conserved current''  partition function 
  $\Z_{+} $  counts the components  of  the  equations of motion  for the field strength (and their derivatives). 
  
  In the case of  totally symmetric  conformal higher spin  gauge field with spin $s$, canonical dimension $2-s$ 
   and   generalized $s$-derivative  field   strength 
 of  dimension $\Delta= 2$ 
   (with $s=1$ being Maxwell   vector, $s=2$ being Weyl graviton, etc.) 
  one finds  \ci{Beccaria:2014jxa}
\ba
   \Z_{+,s} =  { (s+1)^2    q^{s+2}  - s^2   q^{s+3}      \over (1-q)^4}  \ , \ \ \ \quad  \ \ \ \ \ \ \ \  \label{2.4} 
 \Z_{-,s} =  { 2  ( 2 s +1) q^2  %- \big[ (s+1)^2    q^{s+2}  - s^2   q^{s+3}  \big]    
  \over (1-q)^4}  -    \Z_{+,s}   \ .  %\la{2.5}
   % \qquad \qquad
 % \Z_{+,s} =  { (s+1)^2    q^{s+2}  - s^2   q^{s+3}      \over (1-q)^4}  \ .  \ \ 
  %  { 2 q^2 \big[  (s+1)^{2} ( 1 - q^{s})-  s^{2} (1- q ^{s+1} )    \big]     \over (1-q)^4}  \ , \
\end{align}
  The  form of  $\Z_{-,s} $   reflects the fact  that the  counts of gauge identities and  of equations of motion are  isomorphic. 
  
  These expressions   can be   interpreted  also from the  \ads   perspective. In general,   
   \ci{Beccaria:2014jxa}
%In \cite{Beccaria:2014jxa}, we considered totally symmetric bosonic CHS fields with spin $s$ and showed that 
%\be \label{2.4} \Z_s(q) =   { 2 q^2 \big[ (s+1)^{2} ( 1 - q^{s})-  s^{2} (1- q ^{s+1} )    \big]     \over (1-q)^4}. \ee
%This includes the Maxwell vector $A_{\mu}$ for $s=1$ and the conformal graviton $g_{\mu\nu}$ for $s=2$.
%\subsection{AdS/CFT perspective}
%According to the discussion in \cite{Beccaria:2014jxa}, it is also possible to prove that 
\be
\Z(q) = \Z^{-}(q)-\Z^{+}(q)\ , \ \qquad   \Z_{+}(q) = \Z^{+}(q)\ ,\qquad \qquad  \Z_{-}(q) = \Z^{-}(q) \ , \la{2.6} 
\ee
where $\Z^{\pm}(q)$ are the  one-particle  partition functions  \rf{22} 
for  the one-loop  partition  function $Z^\pm$  of  the corresponding  massless  higher spin gauge fields 
in thermal quotient of \ads %(with  boundary $S^{1}\times S^{3}$)
   computed with teh 
standard  (``Dirichlet'')  or alternative (``Neumann'')    boundary conditions.  
This  is the special   case of the general relation \rf{1.9}  with \rf{111}. 
%%%%%%%%%%%%%%%%%%%%%%%%%%%%%%%%%%%%%%%%%%%%%%%%
\iffa 
In the  present case of 
 totally symmetric 
 massless   higher spin   gauge fields  one finds  \ci{Gupta:2012he,Giombi:2013yva,Giombi:2014yra}\foot{The  $D ^2$ 
   operators are defined on symmetric traceless  transverse  tensors in \ads  (with scale set to 1). 
 The energies (dimensions) of the corresponding representations in the  standard boundary conditions 
 case  are
$\Delta_+= s + 2$    for the  physical field  \ci{Metsaev:1994ys} 
 and $\Delta'_+= s + 3 $ for the ghost one.
In the case of alternative    boundary conditions $\Delta_- = 4-\Delta_+ = 2-s$ and 
$\Delta'_- = 4-\Delta'_+ = 1-s$. The ``mass terms'' in the operators  $X= \De (\De -4) -s $ 
 are the same for $\De_+$ and $\De_-$. }
\be  
 Z^\pm_{s}  =
 \Big[{\det_\pm \, \big(-D ^2  +   X'   \big)_{s-1\, \perp} \ov
 \det_\pm \, \big(-D ^2    +  X \big)_{s\perp}}\Big]^{1/2}\  ,  \  \ \   
   \ , \ \ \      \la{11}
 \ee
where $ X = \De (\De-2) -s
 = s^2 - s-4 $ and $ X' =   \De' (\De'-2) -(s-1)=   s^2 + s-2 $ with $\De'= \De+1$.
 \fi 
 %%%%%%%%%%%%%%%%%%%%%%%%%%%%%%%%%%%%%%%%%%%%%%%%
 Explicitly, one finds from the \ads  heat kernel  expression  \ci{Gupta:2012he,Gopakumar:2011qs} 
 that $\Z^+_s$  is given by the same expression as $ \Z_{+,s}$ in \rf{2.4}. 
 The full singlet-sector  partition function of the boundary CFT is then given by the sum of $ \Z_{+,s}= \Z^+_s$ 
 contributions  over all spins.

Massless  higher spin $s$   field  in \ads with standard boundary  condition 
   is dual to the  conserved   spin 
$s$  current  operator of dimension $\De_+=2+s$  in the  free  complex   scalar  CFT$_4$.  
 $\Z^+_s= \Z_{+,s} $ has  the interpretation  of  counting the  bilinear   current field $J_s$   components 
(and its derivative descendants) modulo the on-shell   conservation condition. 
%The reason for $\Z^+_s= \Z_{+,s} $  is that 
This counting problem is isomorphic   to that of 
counting the equations of motion  for the 4d  conformal  higher spin field. 
Similarly, $\Z^-_s= \Z_{-,s} $  is counting the     components  of CHS fields  $\p_s$ 
modulo gauge identities  and  also counting the  components of 
  the shadow   spin $s$   conformal field  of dimension $\De_- = 2-s$  (conjugate to $J_s$)
in the  4d  scalar  CFT modulo gauge identities.
%\foot{Explicitly, starting with   scalar CFT path integral, 
%one may    introduce the  source  field $\p_s$   for the current, i.e. 
% consider  $\int d^4 x \p_s  J_s$,    and then  the current conservation condition will imply gauge redundancy 
% for $\p_s$.} 
 The negative term   in  $ \Z_{+,s} $  in  \rf{2.4}  corresponds  \cite{Beccaria:2014jxa}    to  
the subtraction of the contribution of  identities   among     equations of motion  from the  4d  CHS theory point of view,  of the 
current conservation condition  from the  4d  scalar  CFT   point of view and
of the   ghost  spin $s-1$  field contribution from the \ads  bulk point of view. 
%{\bf comment on conformal killing tensor  here ?} 

%Eq.~(\ref{2.7}) is the boundary CFT partition function counting
%conserved currents or same as corresponding to
%massless  HS in AdS$_{5}$ \cite{Gopakumar:2011qs,Gupta:2012he,Giombi:2014yra}. 

%%%%%%%%%%%%%%%%%%%%%%%%%%
\subsection{Mixed-symmetry  conformal fields}

Let us now consider  the case of a conformal  primary  field in $SO(2,4)$ representation $(\Delta;\, j_{1}, j_{2})$.
%where $(j_1,j_2)$   are $SO(3) \times SO(3)$  spin labels  (
%with total spin $s= j_1 + j_2 $. 
For generic $\De$   the     character of this long representation of $SO(2,4)$
 should be  equal  to the  one-particle partition function for 
the   massive \ads  higher spin  field 
 partition function \rf{1.19}   which 
 % with  operator  $-D ^2 + X$, $X= \De (\De-2) - 2 j_1$   and 
%without the ghost numerator in \rf{11}. It 
should just count 
 all the components of  (derivative descendants of)  such field   weighted  with
   its dimension $\De$ (see  \rf{a8}   in  Appendix \ref{A:FF})
  \be
\label{2.10}
\widehat{\Z}^+{(\Delta;\, j_{1}, j_{2})} =   d(j_{1},j_{2})\,  \frac{q^{\Delta}}{(1-q)^{4}}\ , \ \ \ \ \ \ \ \ \ \qquad 
d(j_{1},j_{2})\equiv (2j_{1}+1)(2j_{2}+1) \ . 
\ee
%Here  $d(j_{1},j_{2})$ is  the dimension of the $(j_{1},j_{2})$ representation of $SO(4)$. 
In the special case of   $\De= 2 + j_1 + j_2$   such primary field 
should   correspond  to a conserved  current in the boundary CFT  or to  its dual 
mixed-symmetry  massless \ads   higher spin gauge field. 
In this case $\Z_+ = \Z^+$ should be given  by the   character of the associated short representation  of 
 $SO(2,4)$ \rf{a9}, i.e.   should correspond  to \rf{1.20} 
%or  by the  direct analog of  the \ads partition function \rf{11}  with 
where the  ghost   contribution is  included. 
Taking into account  the  current conservation condition  or, equivalently, 
subtracting the 5d  ghost contribution gives the massless  partition function \cite{Dolan:2005wy,Gibbons:2006ij}
\ba
\label{2.9}
&\Z^{+} {(\Delta;\, j_{1},j_{2})}=
\widehat{\Z}^+ {(\Delta;\, j_{1}, j_{2})}-\widehat{\Z}^+ {(\Delta+1;\, j_{1}-\ha , j_{2}-\ha )}\ , \\
&
\label{2.8}
\Z^{+}{(\Delta;\, j_{1},j_{2})}=\Z_+{ (\Delta;\, j_{1},j_{2})} = \frac{q^{\Delta}}{(1-q)^{4}}\Big[
(2j_{1}+1)(2j_{2}+1)-4\,q\,j_{1}\,j_{2}
\Big]\ .\end{align}
%with the two dual interpretations being discussed in \cite{Gibbons:2006ij} and \cite{Dolan:2005wy}
%respectively. 
%This expression applies for $\De= \De_+ = 2 + j_1 + j_2$   but we may formally consider it for generic  values of $\De$. 
Note that eq. \rf{2.8}   reduces  to \rf{2.4}  for $j_1=j_2 = { s \ov 2}$,\ $\De= 2 + s$.

%The meaning of $\widehat{\Z}_{(\Delta;\, j_{1}, j_{2})}$  is clear:
%it counts  states of dimension $\Delta$ in the $(j_{1},j_{2})$ representation and 
%subtraction takes into account current conservation condition.
To find  the    partition  function $\Z$ in \rf{2.6}  corresponding to \rf{1.9} 
for a 4d  conformal  spin $(j_1,j_2)$  field  of canonical dimension $\De_- = 4-\De$ 
it  remains   to  determine  the expression for the shadow partition function $\Z_{-} { (\Delta;\, j_{1},j_{2})}$. 
Let   us   start   with  the special case  of   ``matter''  conformal  fields  in  $SO(4)$ representation $(j,0) + (0, j)$
corresponding to   massive  5d fields 
(here the subtraction  term  in \rf{2.8} is absent as  $j_1 j_2=0$ so formally $\widehat \Z^+=\Z^+$).
In this non-degenerate  case  it is natural   to  expect     $\Z_{-}=\Z^{-}$
to be  related to  $\Z_{+}=\Z^{+}$    by the   substitution 
\be   \De=\De_+\quad  \to \quad    \De= \De_- = 4- \De \ ,  \la{sub} \ee
which, according to \rf{2.10},   is equivalent to  %given by the simple rule 
\be
\label{2.11}
\Z^{-}(q) = \Z^{+}(q^{-1} ) \ .
\ee
%that amounts to replacing $\Delta\to 4-\Delta$ in (\ref{2.8}).
 Then   using \rf{2.10} we get for $\Z$ in (\ref{2.6})\footnote{We split the two cases in (\ref{2.12}) 
 because $(j,0)+(0,j)$ counts scalars as complex for $j=0$. Instead,  we shall always   assume that scalars are real.}
%convention is that $\phi$ is a real scalar.}
\be
\label{2.12}
\Z{(\Delta;\, 0,0)} = \frac{q^{4-\Delta}-q^{\Delta}}{(1-q)^{4}}\ , \ \ \ \ 
\qquad
\Z(\Delta;\, j,0) =\Z(\Delta;\, 0,j)=  (2j+1)\,\frac{q^{4-\Delta}-q^{\Delta}}{(1-q)^{4}} \ . 
\ee
Examples of such 4d conformal fields  are provided by matter fields  appearing in extended conformal supergravities \ci{Bergshoeff:1980is, 
%For example, we can consider the {\em matter} fields that appear in extended coFnformal supergravity \cite{
Fradkin:1985am} (see  Table \ref{T2} below)
\bea
%\begin{array}{l}
&& \phi\sim (3;\, 0,0), \ \ \ \ \ \ \ \ 
\Phi\sim (4;\, 0,0), \ \ \ \   \ \ \  T\sim (3;\, 1,0)+(3;\, 0,1), \no  \\
&&\te  \psi\sim (\frac{5}{2};\,\frac{1}{2},0)+(\frac{5}{2};\,0,\frac{1}{2}), \ \ \ \ \ \ \ \ \ \ \ 
%\end{array}
%\qquad
%\begin{array}{l}
\ \ \ \ \ \Psi\sim (\frac{7}{2};\,\frac{1}{2},0)+(\frac{7}{2};\,0,\frac{1}{2}) \ .  \la{212}
%\end{array}
\eea
Here  $\phi$ and $\psi$
are the  standard 4d massless scalar and spinor, $\Phi$ and $\Psi$  are conformal fields  with  $\partial^{4}$ and $\slashed{\partial}^{3}$
kinetic operators and  $T$ is conformal antisymmetric 2-tensor  with $\partial^{2}$
kinetic term and no gauge invariance. $\De$ in $(\De; j_1,j_2)$  stands  for $\De_+$  dimension  associated to 
the corresponding massive  5d field with standard boundary conditions  while the canonical  dimensions  of these
4d fields  are $\De_- = 4-\De$  (i.e. $\phi$ has  dimension 1, $\Phi$ has   dimension 0, etc.). 
The partition functions  $\Z$    for these fields are derived  in Appendix \ref{A:CSG}
by the explicit  path-integral computation on $S^1 \times S^3$ and also by the operator counting method 
and the results  are  consistent with \rf{2.12}. 

%\footnote{A  comment about the assignment of $\Delta$ is the following. For a massless  spin in AdS$_{5}$ we have $\Delta_{+}=2+s$  that is same as dimension  of  a conserved current. The associated shadow field has $\Delta_{-}= 4-\Delta_{+}=2-s$ that is the dimension of the corresponding CHS   field in 4d. In a similar way, for a physical scalar in 4d we have $\Delta_{-}=1$
%so $\Delta_{+}=3$. For the $(\partial^{2})^{2}$ scalar we have $\Delta_{-}=0$ and $\Delta_{+}=4$.}

Turning to the  massless  
gauge field  case  with $j_1 j_2 \not=0$  let us  recall the derivation \cite{Beccaria:2014jxa}  of the expression \rf{2.6} for 
$\Z^-=\Z_-$ in the  case of the bosonic totally symmetric field   with  $(\Delta;\, j_{1}, j_{2})= (2+s;\, \frac{s}{2},\frac{s}{2})$. 
The presence of gauge  degeneracy or ghost contribution implies that in this case 
%Another application is to CHS gauge fields. The case of bosonic totally symmetric fields has been discussed at length
%in \cite{Beccaria:2014jxa}. Eq. ~(\ref{2.8}) reduces to (\ref{2.7}) for $(\Delta;\, j_{1}, j_{2})\equiv(2+s;\, \frac{s}{2},\frac{s}{2})$ suitable for a totally symmetric (massless) gauge field of spin $s$. The non vanishing second term 
%in (\ref{2.8}) is the ghost  contribution. As shown in
%implies   that  \cite{Beccaria:2014jxa}, 
the simple  relation  (\ref{2.11})     between $\Z^+$ and $\Z^-$  is no longer  true. % in this case. 
The  shadow field with   dimension $\Delta_{-}=2-s$ corresponds to a  non-unitary   $SO(2,4) $ 
 representation  which   in general  contains singular states with their 
 associated submodules.  The AdS$_{5}$ counterpart  of this complication is  that
  in the case of the alternative  boundary condition
one has additional gauge transformations  allowed by non-normalizability \cite{Giombi:2013yva}.
 These can be put
in  one-to-one  correspondence with the conformal Killing tensors that  may be associated  to the finite dimensional 
$SO(6)$ representation  $(s-1,s-1,0)$  labelled  by  the  Young tableau with two rows with $s-1$ columns.
 Then \rf{2.11} is replaced by   \cite{Beccaria:2014jxa}    (same for lower $\pm$ labels)
\be
\label{2.15}
\Z_{s}^{-} (q) = \Z^{+}_{s} (q^{-1} ) +\sigma_{s}(q)\  ,
\ee
where $\sigma_{s}(q)$ is the character  of the representation
 for  the conformal Killing tensors. Computing $\sigma_{s}(q)$ one then arrives at the expression in \rf{2.4}. 
 
 A similar   derivation should be possible in the   mixed  representation case leading to 
 \be\la{230}
\begin{split}
\Z(q) &= \Z^{-}(q)-\Z^{+}(q)  = \left[ \Z^{+}(q^{-1} ) + \sigma(q)\right]  -\Z^{+}(q)= \bar \Z(q) -2\,\Z^{+}(q) \ ,  \\
&\qquad\qquad \qquad   \bar \Z (q) \equiv \Z^{+}(q^{-1} )+\Z^{+}(q) + \sigma(q) \ . %\right]-2\,\Z^{+}(q).
\end{split}
\ee
Below we  will  demonstrate this on  the example of the  fermionic  conformal higher spin gauge 
 fields described by  totally    symmetric spinor-tensor  with  one spinor index   and  $\s=0,1,2,...$  vector indices. 
 Its   total spin is  $s=\s+\frac{1}{2}$  and  it is represented by  the sum of two 
 mixed $SO(4)$ representations: 
%$\s=0,1,2,\dots$. The spin content of these fields is 
\be \te \la{216}
\left[\left(\frac{1}{2},0\right)+\left(0,\frac{1}{2}\right)\right]\times \left(\frac{\s}{2},\frac{\s}{2}\right) = 
\left(\frac{\s+1}{2},\frac{\s}{2}\right)+\left(\frac{\s}{2},\frac{\s+1}{2}\right) \ . 
\ee
%The   scaling dimension  of  associated  conformal group representation is 
Here  $\Delta =\De_+ = 2+j_{1}+j_{2} = 2 + s$.  %   which is $\De_+$ dimension 
%is    5d massless  higher spin  gauge field.
The  $\Z^+$ % AdS$_{5}$ 
partition function  is  given by \rf{2.8}, i.e.  
\be\la{217}
\begin{split}
 \Z_{\s+\frac{1}{2}}^{+}(q) &\equiv \te  \Z^{+}{\left(2+\s+\frac{1}{2};\, \frac{\s}{2},\frac{\s+1}{2}\right)}+ 
\Z^{+}{\left(2+\s+\frac{1}{2};\, \frac{\s+1}{2},\frac{\s}{2}\right)} \\
&=2\frac{\,(\s+1)(\s+2)\,q^{\frac{5}{2}+\s}-\,\s\,(\s+1)\,q^{\frac{7}{2}+\s}}{(1-q)^{4}}.
\end{split}
\ee 
%Following the above discussion, we can introduce the {\em naive}  partition function with alternate boundary conditions \be
%\widetilde\Z_{\s+\frac{1}{2}}^{-}(q) \equiv \Z_{\s+\frac{1}{2}}^{+}(q^{-1} ) = \frac{2\,(\s+1)(\s+2)\,q^{\frac{3}{2}-\s}-2\,\s\,(\s+1)\,q^{\frac{1}{2}-\s}}{(1-q)^{4}},
%\ee 
%and obtain the {\em correct}  partition function $\Z_{\s}^{-}$ as 
Then  by analogy  with  the bosonic CHS case \rf{2.15}   we should  find 
\be\la{219} 
\Z_{\s+\frac{1}{2}}^{-}(q) = %\widetilde
 \Z_{\s+\frac{1}{2}}^{+}(q^{-1} ) + \sigma_{\s+\frac{1}{2}}(q)\ ,
\ee
where $\sigma_{\s+\frac{1}{2}}(q)$ is the character for  the conformal algebra representation corresponding to the conformal Killing spinor-tensors.  The latter   may be   associated to  the $SO(6)$ representation 
$\left(\s-\frac{1}{2}, \s-\frac{1}{2}, \pm\frac{1}{2}\right)$ 
with  dimension\footnote{See also footnote 24 of  \cite{Giombi:2013yva}.}
\be\la{220}
\te \dim \left(\s-\frac{1}{2}, \s-\frac{1}{2}, \pm\frac{1}{2}\right) = \frac{1}{3}\,\s\,(\s+1)^{3}\,(\s+2)\ .
\ee
The relevant character   can be  found by  a specialization of the   discussion in   \cite{Beccaria:2014jxa} 
%is a specialisation already discussed in the bosonic case, i.e.
\ba
\la{sii}
\sigma_{\s+\frac{1}{2}}(q) &= \lim_{x\to 1}\chi_{\left(\s-\frac{1}{2},\s-\frac{1}{2},\pm\frac{1}{2}\right)}(q,x,1) = 2\, \lim_{x\to 1}
\frac{\det M(\s-\frac{1}{2}; x, q)}{\det N(x, q)}
\ , \\
M(\s-{\te \frac{1}{2}}; x, q) &= \left(
\begin{array}{ccc}
 2 & 2 & 2 \\
 x^{-\s-\frac{3}{2}}+x^{\s+\frac{3}{2}} &
   x^{-\s-\frac{1}{2}}+x^{\s+\frac{1}{2}} &
   \sqrt{x}+\frac{1}{\sqrt{x}} \\
 q^{-\s-\frac{3}{2}}+q^{\s+\frac{3}{2}} &
   q^{-\s-\frac{1}{2}}+q^{\s+\frac{1}{2}} &
   \sqrt{q}+\frac{1}{\sqrt{q}} \\
\end{array}
\right), \ \\
& N(x, q) = \left(
\begin{array}{ccc}
 2 & 2 & 2 \\
 x^2+\frac{1}{x^2} & x+\frac{1}{x} & 2 \\
 q^2+\frac{1}{q^2} & q+\frac{1}{q} & 2 \\
\end{array}
\right)\ . 
\end{align}
This gives $\sigma_{\s+\frac{1}{2}}(q)$   as a  finite sum\footnote{Some explicit values are 
$
\sigma_{\frac{1}{2}}(q) = 0, \  \sigma_{\frac{3}{2}}(q) = 4 \sqrt{q}+\frac{4}{\sqrt{q}}, \ \  $ 
$
\sigma_{\frac{5}{2}}(q) = 12 q^{3/2}+\frac{12}{q^{3/2}}+24 \sqrt{q}+\frac{24}{\sqrt{q}}  .$
% \ \ \  $
%\sigma_{\frac{7}{2}}(q) =24 q^{5/2}+56 q^{3/2}+\frac{56}{q^{3/2}}+\frac{24}{q^{5/2}}+80
  % \sqrt{q}+\frac{80}{\sqrt{q}} $. %\nonumber
%\end{align}
}
\be
\label{2.23}
\sigma_{\s+\frac{1}{2}}(q) = \frac{\s+1}{3}\,\sum_{p=1}^{\s}(p-\s-2)(p-\s-1)(2p+\s)\,
(q^{p-\frac{1}{2}}+q^{\frac{1}{2}-p}),
\ee
obeying the important property
\be\la{224} 
\sigma_{\s+\frac{1}{2}}(q) = \sigma_{\s+\frac{1}{2}}(q^{-1} ) \ , 
\ee
which   was  also true for the bosonic $\sigma_s$ in \rf{2.15}. 
Doing the sum over $p$ in (\ref{2.23})   gives  %the  we get the following closed expression
\be\la{225}
\sigma_{\s+\frac{1}{2}}(q) = \frac{2 (\s+1)
   q^{\frac{1}{2}-\s}
   \left(q^{\s+1}-1\right)
   \left[\s\,
   q^{\s+2} -(\s+2) q^{\s+1}+
   (\s+2)q-\s\right]}{(1-q)^4} \ . 
\ee
Then using this in \rf{2.15},\rf{230}   leads to 
the final  result for $ \Z_{\s+\frac{1}{2}}= \Z_{\s+\frac{1}{2}}^{-}(q)-\Z_{\s+\frac{1}{2}}^{+}(q)$
%\be\Z_{\s+\frac{1}{2}}^{-}(q)-\Z_{\s+\frac{1}{2}}^{+}(q) = \Z_{\s+\frac{1}{2}}(q),\ee
%where $\Z_{\s+\frac{1}{2}}(q)$ is 
\be\la{227}
\Z_{\s+\frac{1}{2}} = 4\,\frac{(\s+1)\,q^{\frac{3}{2}}+(\s+1)\,q^{\frac{5}{2}}-(\s+1)(\s+2)\,q^{\frac{5}{2}+\s}
+\s(\s+1)\,q^{\frac{7}{2}+\s}}{(1-q)^{4}} \ . 
\ee
As  a check, for  the standard massless spin $\ha$  fermion ($\s=0$)   this  agrees with  \rf{2.12} with $j= \ha $ and $\De={5 \ov 2} $.
Also, for the  conformal gravitino  ($\s=1$) this   leads to  
\be
\Z_{\frac{3}{2}} = \displaystyle 8\,  \frac{\,q^{\frac{3}{2}}+\,q^{\frac{5}{2}}
- 3 \,q^{\frac{7}{2}}+\,q^{\frac{9}{2}}}{(1-q)^{4}} \ ,  \la{228}
\ee
 which is  the same   expression \rf{A15} as 
derived  in Appendix ~(\ref{A:CSG})   by directly 
computing the conformal gravitino partition function on 
 $S^{1}\times S^{3}$.

%%%%%%%%%%%%%%%%%%%%%%%%%%%%%%%%%%
\subsection{General expression  for the Casimir energy on $S^3$}
%%%%%%%%%%%%

The Casimir energy on $S^{3}$ can be extracted from the one-particle partition function $\Z(q)$ in (\ref{2.1})
using  the standard relations (see, {\em e.g.},   \cite{Gibbons:2006ij})\footnote{Given 
the  data $({\rm d}_{n}, \omega_{n})$ the formal sum over $n$  is usually divergent and requires a regularization. 
A natural regularization is a spectral $\zeta$-function one  as above  which is also equivalent 
 to computing 
%are available, the above expression can be computed 
%using a spectral  cutoff  or spectral $\zeta$-function  regularization  
$E_c$   
as the finite part of the $\epsilon\to 0$ expansion of the following  regularized expression (see, e.g.,  \ci{Giombi:2014yra})
\be
\nonumber
 E_c= \ha (-1)^{F} \, \sum_n  \dd_n\, {\om_n}\,e^{-\epsilon\,\omega_{n}} \Big|_{\epsilon\to 0, \ {\rm finite}} \ . 
% = {\rm poles}+E_{c}+\mc O(\epsilon).
\ee}
\ba
\label{2.29}
 &E_c= \ha \, (-1)^{F} \sum_n  \dd_n\, {\om_n}  = \ha (-1)^{F}\,\zeta_E (-1) \ , \ \ \ \ \ \ \  \\
 & \zeta_E (z) =\sum_n  {\dd_n \ov {\om^z_n}}   =  {1\ov \Gamma(z) } \int^\infty _0 d \beta \, \beta^{z-1} \, \Z(e^{-\b}) \ . \la{2281}  
\end{align}
The  representation  in terms of  $\zeta_E (-1)$     has   the advantage   that it 
allows one to show  that the Casimir energy {\it  vanishes}  if the partition function  obeys
 $\Z(q) = \Z(q^{-1} )$  \cite{Giombi:2014yra} (see  also \ci{Basar:2014hda}). 

If   we start with $\Z_{+}$   corresponding  to a   primary  field  $(\Delta;\, j_{1},j_{2})$, 
the associated Casimir energy $E_{c}^{+}(\Delta;\, j_{1}, j_{2})$ is then 
the  same as  the vacuum energy 
of a single  massless  higher spin  field in AdS$_{5}$  with standard boundary conditions. 
If we  consider a 4d conformal higher   spin  field, its  Casimir energy on $S^3$ can be found 
from the corresponding one-particle partition function in \rf{230}. % (where $\sigma$ term is absent for fields without gauge symmetries). 
The Killing tensor character    should  in general  obey the property \rf{224}, implying that 
the same should be true  for $\bar \Z(q)$  in \rf{230}, and if $\bar \Z(q)=\bar \Z(q^{-1} )$   then it does not contribute to $E_c$. 
As a result, we conclude  that 
%The quantity in square brackets is even under $q\to q^{-1} $ and therefore does not contribute the Casimir energy. Thus
 the  Casimir energy of a 4d conformal field in representation $(\Delta;\, j_{1}, j_{2})$   is given by -2   of 
   the AdS$_{5}$ vacuum  energy of the  corresponding 5d  field  with the standard  boundary condition  
\be
\label{2.31}
E_{c}(\Delta;\, j_{1}, j_{2}) = \,E_{c}^{-}(\Delta;\, j_{1}, j_{2}) - \,E_{c}^{+}(\Delta;\, j_{1}, j_{2}) =   -2\,E_{c}^{+}(\Delta;\, j_{1}, j_{2})\ .
\ee
In the   non-gauge  4d field  case  (corresponding to  a massive 5d field)   we 
thus get  from $ \widehat \Z^+$ in  \rf{2.10} 
that $E_c = \widehat{E}_{c} $,  where 
%A straightforward evaluation, using (\ref{2.8}), gives the following general formula for the Casimir energy of the 4d CHS field in the 
%{\bf massive}
%$(\Delta;\,j_{1},j_{2})$ representation (see (\ref{2.9}))
\ba
&\widehat{E}_{c}(\Delta; \, j_{1}, j_{2})  = -2 \widehat{E}^+_{c}(\Delta; \, j_{1}, j_{2}) \no \\
&\ \ =-\frac{1}{720} (-1)^{2 j_1+2 j_2}\,
(2j_{1}+1)(2j_{2}+1)(\Delta-2) \Big[6\,(\Delta-2)^{4}-20\,(\Delta-2)^{2}+11\Big]\ .\label{2.32}
\end{align}
For  a  gauge conformal   field  (or a {massless} 5d field)  with $\Delta=2+j_{1}+j_{2}$  %and $j_{1}j_{2}>0$, 
we  get  according to \rf{1.20} 
\be
\label{2.33}
\te E_{c}(\Delta; \, j_{1}, j_{2}) = \widehat{E}_{c}(\Delta; \, j_{1}, j_{2}) 
-\widehat{E}_{c}\left(\Delta+1; \, j_{1}-\frac{1}{2}, j_{2}-\frac{1}{2}\right).
\ee
 As in \rf{2.8}, the  second term here vanishes  if  $j_1j_2=0$.
%Notice that for massless matter fields with $j_{1}j_{2}=0$, the second term in (\ref{2.33}) vanishes so that 
%(\ref{2.33}) covers all massless cases.
%\be
%\begin{split}
%E_{c}(\Delta; \, j_{1}, j_{2}) &= -\frac{1}{720} (-1)^{2 j_1+2 j_2}\,\Big[
%(2j_{1}+1)(2j_{2}+1)(\Delta-2)\times \\
%& \times (6\,\Delta^{4}-48\,\Delta^{3}+124\,\Delta^{2}-112\,\Delta+27)\\
%&-4\,j_{1}\,j_{2}\,(\Delta-1)\,(6\,\Delta^{4}-24\,\Delta^{3}+16\,\Delta^{2}+16\,\Delta-3)
%\Big].
%\end{split}
%\ee
%\be
%\begin{split}
%E_{c}(\Delta; \, j_{1}, j_{2}) &= -\frac{1}{720} (-1)^{2 j_1+2 j_2}\,\Big[\\
%&
%(\Delta-2)\,(6\,\Delta^{4}-48\,\Delta^{3}+124\,\Delta^{2}-112\,\Delta+27)\,(2\,j_{1}+2\,j_{2}+1)\\
%&-12\,j_{1}\,j_{2}\,(10\,\Delta^{4}-60\,\Delta^{3}+120\,\Delta^{2}-90\,\Delta+19)
%\Big].
%\end{split}
%\ee

Special cases  include  fields   of extended conformal supergravity   with values of $E_c$ listed in Table \ref{T2}. 
For  the  general spin $s$ totally symmetric bosonic
$(2+s; { s \ov 2}, {s\ov 2} )$   \cite{Beccaria:2014jxa}    and 
 fermionic  $(2+s; { s + { 1 \ov 2}  \ov 2}, { s - { 1 \ov 2}\ov 2} )   +     (2+s; { s - { 1 \ov 2}   \ov 2}, {s + { 1 \ov 2} \ov 2} ) $    conformal 4d fields
 we obtain  from  \rf{2.32},\rf{2.33} (or directly from \rf{2.4} and \rf{217}) 
 the following expressions for the Casimir energies
%(the bosonic case has been previously obtained in \cite{Beccaria:2014jxa})
\ba
\label{2.34}
%\begin{split}
&E_{c,s}= E_c {\te (2+s; { s \ov 2}, {s\ov 2} )}  = \frac{1}{720}\,\nu_b\,(18\,\nu_b^{2}-14\,\nu_b-11)\ , \qquad \qquad 
\ \ \ \ s=1,2, ...  \\
&
E_{c,s} = 2 E_c {\te (2+s; { s + { 1 \ov 2}  \ov 2}, { s - { 1 \ov 2}\ov 2} ) } 
=   \frac{1}{5760}\,\nu_{f}\,(36\,\nu_{f}^{2}+140\,\nu_{f}+85), \ \ \ \   \ \  {\te s= {1\ov 2}, {3\ov 2} , ...}\ \ \  \la{234}\\
& \nu_b\equiv s(s+1),\qquad \qquad  \nu_{f}\equiv -2(s+\ha)^{2 } = -2 \nu_b - \ha \ . \la{235}
\end{align}
Here  $\nu_b$ and $\nu_f$ are the numbers of  dynamical degrees of freedom of the  bosonic and fermionic CHS fields \ci{Tseytlin:2013jya}. 
The coefficient 2  in the fermionic case  accounts for the  equal contributions of the two $ j_1 \leftrightarrow  j_2$ representations. 
%sum of the two 
%Here the $s=1/2, 1, 3/2, 2$ 

%%%%%%%%%%%%%%%%%%%%%%%%%%
\section{Conformal anomaly  $\rm a$-coefficient } % conformal anomaly}
%%%%%%%%%%%%%%%%%%%%%%%%%%%%

Next, let  us   turn to the computation  of the conformal   anomaly $\aa$-coefficient
of 4d conformal  field  with canonical dimension $4-\De$ and $SO(4)$  spins $(j_1,j_2)$ 
 corresponding to a generic representation $(\Delta;\, j_{1}, j_{2})$. 
 
As follows from (\ref{1.12}), to find the $\aa$-coefficient    it is sufficient to 
consider the case of conformally flat  $S^{4}$  background
(for unit-radius sphere $\A^{+}  =-24\aa^+$). We shall use \rf{1.9},(\ref{1.11})  to give the \ads derivation of the $\aa$-anomaly 
generalizing the computation of  \cite{Giombi:2013yva,Giombi:2014iua}
  in the  totally symmetric $(2+s;\, \frac{s}{2},\frac{s}{2})$  
  bosonic  case. The  expressions 
  for $\aa$  for  both bosonic and fermionic totally symmetric  conformal higher spin   fields were   found 
   directly in 4d in \cite{Tseytlin:2013jya}. 

%The general expression (\ref{1.12}) shows that $\A$ is proportional to the 
%$\rm a$-anomaly when the boundary of AdS$_{5}$ is the conformally flat sphere $S^{4}$. 
%In such case, the actual calculation of ${\rm a}$ has been performed for totally symmetric 
%CHS fields of any spin in \cite{Tseytlin:2013jya}, working directly in 4d.

%The relation (\ref{1.11}) shows that  $\A$ can also be obtained from 
$\A^{+}$  in \rf{1.11} is  associated with the variation of the 
one-loop  partition  function of  5d  field   corresponding to the  representation $(\Delta;\, j_{1}, j_{2})$ 
under a local  conformal  variation of the boundary  metric. 
In the case of the Euclidean AdS$_{5}$   with boundary $S^4$  (i.e.   hyperboloid  $\mathbb H^{5}$)
the conformal    anomaly is proportional to  the  logarithmic IR   singular part of the 
 one-loop partition function (see, e.g., \ci{Diaz:2007an,Giombi:2013yva})
%The advantage of considering 
%$\A^{+}$ is that one can easily cover the general case of a HS field in a $(\Delta;\, j_{1}, j_{2})$ representation by 
%spectral methods (or, in principle, by means of (\ref{1.3})).
 %This approach has been exploited in \cite{Giombi:2013yva} for HS fields in the  representation 
%$(2+s;\, \frac{s}{2},\frac{s}{2})$, i.e. for totally symmetric bosonic CHS fields. 
%To this aim, one considers
 %a free  spin $s$ field propagating in Euclidean AdS$_{5}$, that is the hyperbolic space $\mathbb H^{5}$. 
% The partition function can be evaluated by spectral methods as 
 \be
 \log Z^{+} = -\ha\log{\det}_{+}\, \mc O = \ha\,\zeta'(0) = - 4 \aa^+ \log \RR  + ... \ . \la{31}
 \ee
 Here  $\zeta(z)$ is the spectral zeta function defined by evaluating the trace of the  $\mathbb H^{5}$ heat kernel 
 associated with the ``massive''  5d operator $\mc O$ in \rf{1.16} 
    (see  \cite{Camporesi:1994ga,Giombi:2013yva}). 
 %\footnote{The structure of the  operator $\mc O$ has been discussed in the Introduction, see (\ref{1.15},\ref{1.16}).}
  The trace  is proportional to the regularised volume of $\mathbb H^{5}$ that has a factor $\log \RR$  depending on  IR cutoff. % (see Appendix~(\ref{A:heatkernel} for details). 
 % The  anomaly  $\A^{+}$  is   the coefficient of $\log R$ in $\log Z^{+}$. 
 
  The explicit derivation of $\zeta(z)$  for the operator 
 $\OO$  acting on a transverse  field  in a general representation $(\Delta;\, j_{1}, j_{2})$ is given in 
 Appendix  \ref{A:heatkernel}. 
 Using  \rf{B.16} 
 the   ${\rm a}$-coefficient  for the 4d conformal   field  associated to   ``massive''   $(\Delta;\, j_{1}, j_{2})$  representation 
 can be represented as  
 (-2  factor  is as in  (\ref{1.11}))
 \be\la{32}
 \begin{split}
 \widehat{\rm a}(\Delta;\, j_{1}, j_{2})&= - 2\widehat \aa^+(\Delta;\, j_{1}, j_{2})
  = \frac{1}{4\,\log \RR}\,\zeta'(0) = { \frac{1}{48\,\pi} }(-1)^{2(j_{1}+j_{2})}(2j_{1}+1)(2j_{2}+1)
  \\
 &\times\lim_{z\to 0} \frac{\partial}{\partial z}
 \int_{0}^{\infty}d\lambda\,\frac{\left[\lambda ^2+(j_{1}-j_{2})^2\right] \left[\lambda ^2+(j_{1}+j_{2}+1)^{2}\right]}{\big[\lambda^{2}+(\Delta-2)^{2}\big]^{z}}\ . 
 \end{split}
 \ee 
   A straightforward computation   gives 
   \be
 \label{3.3}
\begin{split}
&  \widehat{\rm a}(\Delta;\,j_{1},j_{2}) = \frac{1}{720} (-1)^{2 (j_1+j_2)} (2 j_1+1)
   (2 j_2+1) (\Delta-2)\\ 
  \times  & \Big[- 3 (\Delta-2)^4
   +{10} \big(   j_1^2 +  j_2^2 +   j_1 +  j_2   + \ha   \big) (\Delta-2)^2-15  (j_1-j_2){}^2 (j_1+j_2+1){}^2\Big].
\end{split}
\ee
This  expression  is  odd   under $\De \to 4- \De$, i.e. under \rf{sub}. This 
implies   that  the anomaly 
 corresponding to $Z^-$ computed with the alternative boundary condition    has the opposite sign, 
 i.e.  
 we have   $\widehat \aa=\widehat  \aa^- - \widehat \aa^+ = - 2\widehat  \aa^+$.  
 This is also the same pattern that was found  for  the Casimir energy  \rf{2.31}.\foot{As was discussed in section 2,   in the  case of  $S^1 \times S^3$ boundary 
  the  $Z^-$  partition function  is not simply given by $Z^+$ 
with $\De_+ \to \De_-=4-\De_+$ (eq. \rf{2.15}  contains   non-trivial $\sigma$  term) but this  relation still 
 holds for $E_c$  in \rf{2.31}. 
Same   may be true   in the case of $S^4$   boundary:  while  the  IR  divergent  parts of $\log Z^-  $  and $\log Z^+$ 
 proportional to $\aa^-$ and $\aa^+$  are the same up to sign, 
 the relation between   the finite (non-universal) parts of the partition functions may be more
  involved.}
 
%As is clear from \rf{32}, the full  expression  for $\

In the  {massless}  field  case   %with $j_{1}j_{2}>0$
 one is to subtract the ghost contribution in \rf{1.20}, i.e. 
%that  % the ${\rm a}$-anomaly is given by 
\be
\label{3.4}\te
 {\rm a}(\Delta;\, j_{1}, j_{2}) =  \widehat{\rm a}(\Delta;\, j_{1}, j_{2})
 - \widehat{\rm a}\big(\Delta+1;\, j_{1}-\frac{1}{2}, j_{2}-\frac{1}{2}\big)\ . 
 \ee
 %where the second term subtracts the ghost contribution. 
 As in the case of $E_{c}$ in \rf{2.32},\rf{2.33}, the second term in (\ref{3.4}) vanishes for $j_{1}j_{2}=0$.
 %so that case covers
 % and we can conclude that (\ref{3.4}) covers all massless cases.
 
 It is easy to check that in the special cases of conformal fields appearing in extended conformal supergravity 
 the expressions  \rf{3.3},\rf{3.4}   reproduce  the known  values  \ci{Fradkin:1981jc,Fradkin:1985am}
 of the corresponding $\aa$-coefficients (see Table \ref{T2}). 
 Also,  for the  totally symmetric  bosonic and fermionic conformal higher spin  gauge   fields    we  find 
 as in \rf{2.34},\rf{234} 
 % In the bosonic and fermionic cases, one has 
%(see (\ref{2.34}) for the definition of $\nu_{b}$ and $\nu_{f}$)
%\footnote{The factor 2 in fermionic case is to take 
%into account the two equal   contributions in  $(j_{1},j_{2})+(j_{2},j_{1})$.}
\begin{align}
\aa_s={\te {\rm a}\big(s+2;\,\frac{s}{2},\frac{s}{2}\big) }  &= \frac{1}{720}\,\nu_{b}\,(14\,\nu_{b}^{2}  + 3\,\nu_{b} )\ , \ \ \ \qquad\qquad \     s=1,2, ...\la{35} \\
\aa_s={\te 2\, {\rm a}\big(s+2;\,\frac{s+\frac{1}{2}}{2},\frac{s-\frac{1}{2}}{2}\big)}&= \frac{1}{2880}\,\nu_{f}\,\big(
14\nu_{f}^{2} + 45\nu_{f}+12\big)\  , \ \ \ \quad   s={\te  {1 \ov 2}, {3\ov 2} ,} ... \la{36}
\end{align}
Eq.\rf{35} was first  found  in the  5d  approach in  \cite{Giombi:2013yva};
both  expressions were also obtained  by  direct computation in 4d   \cite{Tseytlin:2013jya}.
% found directly for CHS fields in 4d while  bosonic
%expression was found through this ads5 route in \cite{Giombi:2013yva}. 
\iffa 
Another useful specialisations of (\ref{3.4}) is
%\be
%\begin{split}
%{\rm a}_{\left(\Delta;\,0,0\right)} &= -\frac{1}{720}\,(\Delta-2)^{3}\,(3\,\Delta^{2}-12\,\Delta+7), \\
%{\rm a}_{\left(2+j;\,j,0\right)+\left(2+j;\,0,j\right)} &= -\frac{1}{180}\,(-1)^{2j}\,j^{3}\,(2j+1)\,(4j^{2}+10j+5), \\
%{\rm a}_{\left(3+j;\,j,0\right)+\left(3+j;\,0,j\right) }&= -\frac{1}{180}\,(-1)^{2j}\,(j+1)^{3}\,(2j+1)\,(4j^{2}-2j-1), 
%\end{split}
%\ee
\be
\label{3.7}
\begin{split}
{\rm a}(\Delta;\,j,0) &= \widehat{\rm a}(\Delta;\,j,0) = -\frac{1}{720}\,(-1)^{2j}\,(2j+1)\,(\Delta-2) \\
&\Big[
3\,(\Delta-2)^{4}-5\,(\Delta-2)^{2}\,(2j^{2}+2j+1)+15\,j^{2}(j+1)^{2}
\Big].
\end{split}
\ee
\fi 
%%%%%%%%%%%%%%%%%%%%%%%%%%%%%%%%%%%%%%%%%%%%%
%Let us finish with few comments. 

Eq.(\ref{3.3}) is  a generalisation of  the one in  \cite{Giombi:2013yva}  (see also \ci{Giombi:2014iua})
derived there for the  $(\Delta;\,{\textstyle\frac{s}{2}, \textstyle\frac{s}{2}})$ fields  %which reads (in our notation) 
%  Indeed, in our normalisation, their result can be written
\begin{align}
\widehat\ra(\Delta;\,{\textstyle\frac{s}{2}, \textstyle\frac{s}{2}}) &= 
-\frac{(s+1)^{2}}{48\,\pi}\int_{2}^{\Delta}dx\,(x-2)(x+s-1)(x-s-3)\,\Gamma(x-1)\Gamma(x-3)\sin(\pi x)
\nonumber \\
&= \frac{1}{720}\,(s+1)^{2} \, (\Delta-2)^{3}\,\big( - 3\,\Delta^{2}+ 12\,\Delta+5s^{2}+10s-7\big)\ .\label{3.8}
\end{align}
%Using (\ref{3.3}), we obtain precisely the same result. This is not surprising since our approach is a 
%generalisation of the calculation in \cite{Giombi:2013yva}. An alternative check of this expression for a 
Also, a  special   case of  a    massive  
scalar field    with   $m^2 = \De (\De-d) $  in  AdS$_{d+1}$     with even $d$  
corresponding to a    conformal field $(\Delta;\,0,0)$   at the boundary 
was  considered  in  \cite{Diaz:2007an,Diaz:2008hy}, where it was found
($(...)_n$    is  Pochhammer symbol)%(in eq. (4.3))
\be
\label{3.9}
%\frac{1}{2\,(\Delta-2)}
\frac{\partial}{\partial\Delta}\widehat\ra(\Delta;\,0,0) 
 = -\frac{1}{2} %\,(\Delta-2)}
 \,\frac{(-1)^{d/2}}{\Gamma(d+1)}\,
(\Delta-2)_{2}\,(2-\Delta)_{2} \ .
\ee
In  $d=4$  this gives 
\be\label{3.10}
\frac{\partial}{\partial\Delta}\widehat\ra(\Delta;\,0,0) 
 = -\frac{1}{48}\,(\Delta-3)(\Delta-2)^{2}(\Delta-1) \ , 
\ee
in agreement   with \rf{3.8}. 
%that is the same as the derivative of (\ref{3.8}) at $s=0$. 
For general $(\Delta;\,j_{1}, j_{2})$, it follows from \rf{3.3} that 
the $\De$ derivative  of $\widehat \aa$   has  a simple factorized structure % the same derivative of $\ra$ as in (\ref{3.10}) is 
\begin{align}
&\frac{\partial}{\partial\Delta}\widehat\ra(\Delta;\,j_{1}, j_{2}) 
= -\frac{1}{48}\,(-1)^{2\,(j_{1}+j_{2})}\,(2j_{1}+1)\,(2j_{1}+1) \no  \\
 &
\times (\Delta-j_{1}-j_{2}-3)\, (\Delta-j_{1}+j_{2}-2)\,(\Delta+j_{1}-j_{2}-2)\,(\Delta+j_{1}+j_{2}-1)\ .  \la{310}
 \end{align}
 Let us note in passing that since    this  expression is an obvious   generalization of \rf{3.9},\rf{3.10} 
  %is obtained from (\ref{3.9}) 
  (obtained by  $\Delta\to \Delta-j_{1}-j_{2}$ in the Pochhammer
 symbols, etc.),  % and by multiplying times the dimension $(2j_{1}+1)\,(2j_{1}+1)$ and statistics sign. 
 this suggests that the general field bulk-to-bulk propagator  can be obtained 
 from the scalar one  by a similar replacement (with the prefactor coming from the trace over spin).
  This is indeed  consistent  
 with the  known  expressions in the case   of  totally  symmetric tensors  considered in \cite{Costa:2014kfa}. 
% is modified in the general case by the same replacement  (with the prefactor coming from the trace over spin).

%Finally, let us note that the  above  expressions for $\aa$   which are non-linear   in 
%in $\De$ disagree   in general   with the proposal   (\ref{1.3})  of  \cite{Mansfield:2003gs}  for the  structure  of  the 
%boundary conformal anomaly   corresponding to a  massive  \ads field. 

%As we shall see  below, the  linearity in $\De$ is restored  in $c-a$

%   has different 
%  structure  than  (\ref{1.3})  suggested (in low spin cases) in  .}
  %because the factor $(\Delta-2)$ multiplies a non trivial function of $\Delta$, while $b_{4}$ in (\ref{1.3}) is 
 % supposed to be independent on $\Delta$ in the approach of \cite{Mansfield:2003gs}.}
%  \fi 

%%%%%%%%%%%%%%%%%%%%%%%%%%%%%%%%%%%%%%%%%%%%

\section{Conformal anomaly  $\cc$-coefficient}
%%%%%%%%%%%%%%%%%

In this section  we shall propose   the general expression  for the $\cc(\De;j_1,j_2)$  coefficient 
in the 4d  conformal anomaly \rf{1.12}  which will be the counterpart of the expression for $\aa(\De;j_1,j_2)$   
in \rf{3.3},\rf{3.4}. We shall motivate it by imposing various consistency conditions and agreement with known
special   cases. 

%%%%%%%%%%%%%%%%%%%%%%%%%%%%%%%
\subsection{Expression  for $\cc$ in low spin  cases}   %$\leq 2$ case } % and its naive  generalization}
%%%%%%%%%%%%%%%%%%%%%%%%%%%%%%%%%%%%%%%%%

Once  the value of $\aa$ is known, to find  $\cc$   it is  sufficient to compute $\cc-\aa$  by 
considering the case of Ricci flat 
4d  space   when the   conformal anomaly \rf{1.12} becomes 
$\A= (\cc-\aa) \E$. 

In the case of   ``massive''   low spin  5d  fields  appearing in supergravity  (e.g., in the  KK spectrum of 10d  type IIB 
supergravity compactified on $S^5$)   
ref. \cite{Ardehali:2013xya,Ardehali:2014zba}  suggested,   following  the proposal  in \cite{Mansfield:2000zw}, 
 a   general parametrization 
of $\cc-\aa$ coefficient  in  the  boundary conformal anomaly\foot{To recall,  we use $\ { \widehat {}}\ $ to indicate massive 
representation and  $^+$ indicates the one-loop  5d  field contribution  computed  with standard (Dirichlet) boundary conditions. 
The normalization of $\cc-\aa$ in  \cite{Ardehali:2013xya,Ardehali:2014zba}   is such that 
that it corresponds to 1-loop contributions of 5d fields   dual to composite 4d  operators in the AdS/CFT picture; 
summing over all such contributions   should reproduce  the  conformal anomaly of the boundary CFT. 
Thus $\widehat \cc^+ -\widehat\aa^+$ for, e.g., a   scalar  field  corresponding  the  $(3;0,0)$ 
representation  is  $ - \ha$  of the standard value ${1 \ov 180}$.}
%If we now consider a Ricci-flat 4d manifold $\M_{4}$, the anomaly $\A^{+}$ measures the difference 
%$\rm c-a$ of the dual 4d fields. This quantity can be discussed in  massless 10d IIB supergravity compactified
%on S$^{5}$. This leads to AdS$_{5}$ supergravity together with massive Kaluza-Klein (KK) towers with spin $\le 2$.
%For massive fields, the authors of \cite{Ardehali:2014zba}, based on \cite{Mansfield:2000zw,Ardehali:2013xya},
%proposed the following expression for the difference ${\rm c}^{+}-{\rm a}^{+}$ of the conformal anomalies
%of the 4d fields dual to the 5d fields in $(\Delta;\,j_{1},j_{2})$ representations
\ba
\no &\widehat {\rm c}^{+}-\widehat{\rm a}^{+} = - {1\ov 2} (\De-2) b_4(\Deltat_{j_1,j_2}) =  -\frac{1}{360}\, (-1)^{2\,(j_{1}+j_{2})}(\Delta-2)\,d(j_{1},j_{2})\,\big[1+f(j_{1})+f(j_{2})\big], \\
& d(j_{1},j_{2}) = (2j_{1}+1)(2j_{2}+1), \qquad\qquad  f(j) \equiv  j\,(j+1)\,\left[6j\,(j+1)-7\right] \ .  \label{4.1}
\end{align}
This  expression follows from the ansatz  (\ref{1.3})  with $\bar \A= -\ha  b_4 (\bar \OO)$ 
 assuming that $ \bar \OO$, i.e. the   4d boundary restriction of the 5d  massive   kinetic operator 
 defined on an Einstein space which is a 
generalization of \ads space asymptotic to the Ricci-flat boundary, is 
 the standard  $\Deltat= - D^2 + U$   operator defined on  4d 
 field  in  Lorentz representation $(j_1,j_2)$  with    ``minimal''   curvature coupling. Then  
 applying  the standard  algorithm to compute its Seeley coefficient $b_{4}$   \cite{Christensen:1978md} gives \rf{4.1}.\foot{Here the 
4d operator obtained  by restricting  the  5d operator defined on transverse fields  to the boundary  acts on unconstrained  4d  fields.}  
%Note also   that   for generic  Ricci flat space the discussion in   \cite{Christensen:1978md}    applied only in the case of $j_1 + j_2\leq 2$:
%for higher spins  the  consistency of  ``minimal''  curvature coupling  operators  considered   in   \cite{Christensen:1978md}  require additional constraints on the curvature  formally invalidating the derivation of $\cc-\aa$.}

%where the sum  is over all fields (including KK towers).  The expression (\ref{4.1})
%is consistent with the Ansatz (\ref{1.3}) with the Seeley coefficient $b_{4}$  computed according to the  results of \cite{Christensen:1978md}. 

%It is tempting to assume that (\ref{4.1}) holds also for individual fields. So, we can apply it 
% to determine the $\rm c$-anomaly of the CHS fields, according to (\ref{1.11}), i.e. using 
% ${\rm c} = -2\,{\rm c}^{+}$. From (\ref{3.3}) and (\ref{4.1}), we then obtain
 
 Applying \rf{4.1}  together  with our result  \rf{3.3}  for  the  value  of $\aa$-coefficient 
  to compute the corresponding 4d conformal field  anomaly $\cc$-coefficient
  according to \rf{1.11}, %i.e. ${\rm c} = -2\,{\rm c}^{+}$,  
   we find 
  in  the non-gauge 5d  massive $\Delta>2+j_{1}+j_{2}$  case 
\begin{align}
&\widehat{\rm c}(\Delta;\,j_{1},j_{2}) = -2\,\widehat {\rm c}^{+} (\Delta;\,j_{1},j_{2})= \frac{1}{720}\,(-1)^{2(j_{1}+j_{2})}
(2j_{1}+1)(2j_{2}+1)\,(\Delta-2)\, \nonumber \\
& \ \ \times  \Big[-3\,(\Delta-2)^{4}+10\,(j_{1}^2 + j_2^2 +  j_1 + j_2   + \ha)(\Delta-2)^{2}\,  +9(j_{1}^{4}+j_{2}^{4})+30 j_{1}^{2}j_{2}^{2}
\nonumber \\
\label{4.2} &\qquad 
+18\,(j_{1}^{3}+j_{2}^{3})+30\,j_{1}j_{2}(j_{1}+j_{2}+1)-19(j_{1}^{2}+j_{2}^{2})-28\,(j_{1}+j_{2})+4
\Big].
\end{align}
%Consistently with our notation, this is the anomaly for a massive field with . 
To  get  $\cc$  for    CHS gauge   fields  corresponding to massless  5d fields with $\Delta=2+j_{1}+j_{2}$
we are to subtract   the 5d ghost  contribution  as in  \rf{1.20},\rf{3.4}: 
\be
\label{4.3}\te 
{\rm c}(\Delta;\,j_{1},j_{2}) = \widehat{\rm c}(\Delta;\,j_{1},j_{2}) - 
\widehat{\rm c}\left(\Delta+1;\,j_{1}-\frac{1}{2},j_{2}-\frac{1}{2}\right) \ .
\ee
This expression  reproduces   the  known  values of  $\cc$    for  spin $\le 2$   $\N=4$ conformal 
 supergravity fields  \ci{Fradkin:1981jc}  in Table \ref{T2}  (which are dual to fields  of 5d  $\N=8$ gauged supergravity).  
   % turns out to agree with the known values. 

If we formally  assume  \rf{4.2},\rf{4.3}   to be valid also for  all totally symmetric higher spin   fields with $j_{1}=j_{2}=\frac{s}{2}$ 
 then  we  find as in  \rf{35},\rf{36}\foot{This parametrization of $\cc_s$ in terms  of two a priori arbitrary   constants 
 $r_b,\, r_f$  was  introduced in  \cite{Tseytlin:2013jya}  to ensure the agreement  
 with known values for  low spins $s=\ha, 1, {3 \ov 2}, 2$.}
%one finds, using (\ref{4.3}), 
\begin{align}
\label{4.4}
\cc_s={\te {\cc}\big(s+2;\,\frac{s}{2},\frac{s}{2}\big) }&= \frac{1}{1080}\,\nu_{b}\,\big[ \nu_{b}\,(43\,\nu_{b}-59)+r_{b}(\nu_{b}-2)(\nu_{b}-6)\big], \\
\label{4.5}
\cc_s={\te 2\, {\cc}\big(s+2;\,\frac{s+\frac{1}{2}}{2},\frac{s-\frac{1}{2}}{2}\big)}&= \frac{1}{23040}\,\nu_{f}\,\big[\nu_{f}\,(173\,\nu_{f} + 490)+r_{f}(\nu_{f}+2)(\nu_{f}+8)\big],
\end{align}
with $\nu_b, \nu_f$ defined in \rf{235}  and 
\be
r_{ b}=\te \frac{1}{2}\ ,\qquad \qquad  \qquad r_{ f}=59\ .  \la{46}
\ee
These are the same expressions  as  obtained   in \cite{Tseytlin:2013jya}   by the direct computation in 4 dimensions. The 
 key assumption there was that  the factorization of  the  higher-derivative CHS  kinetic operator  on Ricci-flat background 
 into a product of standard  2nd derivative operators known to apply for $ s\le 2$ continues to be valid 
  also for $s>2$.

It is useful to understand   the reason for this agreement. 
%the derivation in  \cite{Tseytlin:2013jya} is based on a total factorisation of the wave operator in terms of 
%second order operators whose contribution to $\rm c-a$ is again evaluated by the algorithm in \cite{Christensen:1978md}.
Let us consider, for example, the bosonic  CHS field  on a curved Ricci-flat background. 
Assuming factorization of the  conformal $D^{2s}+...$ kinetic  operator  into a product of $s$ 
2nd-derivative massless spin $s$   operators  with  minimal coupling to curvature %as in  \cite{Christensen:1978md}  
%For illustration, let us discuss in more details the analysis of \cite{Tseytlin:2013jya} for totally symmetric bosonic 
%fields. The main assumption is that for any conformal higher spin $s$ field in a Ricci-flat background the kinetic operator 
%factorizes into $s$ factors of the massless spin $s$ 2nd-order operators. 
%Then the partition function should be given by
the corresponding CHS partition function can be written as \cite{Tseytlin:2013jya}
\be
\label{4.7}
Z_{s} = \left[\frac{(\det \Deltat_{s-1})^{s+1}}{(\det \Deltat_{s})^{s}}\right]^{1/2},
\ee
where %\foot{We put ``hat'' on the differential operators   $\Delta_s$  not to confuse it with scaling dimensions.}  
 $ \Deltat_{k}= (- D^2 +  U)_k, \ U= -  R^{ab}_{\ \ mn} \Sigma ^{mn} \Sigma_{ab} $ are 
 covariant 2nd-order differential operators defined on traceless rank $k$ 
tensors  and having the   standard massless  higher spin form that   was  assumed  also  in \cite{Christensen:1978md}.
  Then the conformal anomaly $\b_1 \equiv \cc-\aa$  coefficient for spin $s$ CHS field  can be expressed in terms of 
% corresponding to standard massless  fields with spin $k$. Let us follows the notation of \cite{Tseytlin:2013jya}
%and express the difference between the conformal anomalies in term of the $\beta$-function coefficient
%${\rm c}_{s}-{\rm a}_{s} = \beta_{1,s}$. From (\ref{4.7}), we can express the value of the $\beta_{1}$ 
% for a CHS field as a combination of
  $\beta_{1}$ coefficients for the operators $\Deltat_{s}$ 
 \be
 \beta_{1,s} = s\,\beta_{1}( \Deltat_{s})-(s+1)\,\beta_{1}( \Deltat_{s-1}) \ , \ \ \qquad  \ \qquad   \beta_1 \equiv \cc- \aa   \ . \la{48}
 \ee
 Here  the scaling dimension  is $\De=2 + s$   so that \rf{48}   has exactly the same  structure 
 $(\De-2) \beta_{1}( \Deltat_{s})-  (\De-1) \,\beta_{1}( \Deltat_{s-1})   $  as  required for a  massless  5d 
  field   anomaly (cf. \rf{4.2},\rf{4.3}). Since 
  $\beta_{1}(\Deltat_{s})$    was computed  in \cite{Tseytlin:2013jya}     from   the same expression
  for $ b_4 (\bar \OO_s)$  in  \cite{Christensen:1978md} 
 as used in \rf{4.1}  we   conclude that the expressions   for 
 $\cc-\aa$ should indeed match. 
 As the $\aa$ coefficients are already known to  agree, this implies   
  the agreement of the $\cc$ coefficients found from the 5d approach based on \rf{4.1}  and 
  from the 4d approach based on \rf{4.7}.

However, there are    good reasons to believe  that  both \rf{4.1}   and \rf{4.7}    are to be modified  for spins  
$j_1, j_2 > 1$.  First, the  expression   for the Seeley coefficient  of 4d operator  on $(j_1,j_2)$   field
used in \rf{4.1} was  taken from \cite{Christensen:1978md}   which  formally applies only for  spins $\leq 2$:
 %  for generic  Ricci flat space the discussion in   \cite{Christensen:1978md}   
for higher spins  the  consistency of  ``minimal   coupling''   operators  considered   in   \cite{Christensen:1978md}  requires 
extra  constraints on the curvature  (in addition to Ricci flatness)  invalidating the derivation of $\cc-\aa$.
Indeed,  kinetic   operators  of  higher spin    4d fields  should in general contain 
terms with   non-minimal (e.g., $R_{....}D_.D_.$)  coupling  to  the curvature \cite{Zinoviev:2008ck,Boulanger:2008tg}
which   does not allow  the  application  of the standard algorithm for computing the $b_4$ Seeley coefficient   used in 
\cite{Christensen:1978md}.

Second, the assumption  of factorization of the CHS operator on Ricci flat background  made in 
\cite{Tseytlin:2013jya}
 was questioned in  \cite{Nutma:2014pua}. 
It is likely  that $\cc-\aa$  for  CHS fields 
 may still   be  computed by  assuming that factorization formally 
 applies (extra terms obstructing factorization  appear to  involve derivatives of the curvature
  that  can  not produce   non-trivial contribution to  conformal anomaly in 4d)  
  but the  corresponding 2nd-derivative factor-operators   should  then also have   non-minimal 
  structure   rather than being minimal  operators  as  assumed in  \cite{Tseytlin:2013jya}.
  
While the  form  of  such 2nd-derivative  higher spin operators  that may appear in factorization 
of CHS   operator  on  a Ricci-flat    background  remains to be  understood,  
 below we shall present a   conjecture for  what  should be the  correct 
  generalization of   $\cc$ in \rf{4.2} to  higher spins $j_1,  j_2 >1$.
  Our expression  will   lead to    unique  consistency properties    when applied in the context of AdS/CFT. 
%  A more general ansatz   for $\cc$  dependending on one parameter
 %  (which is fixed once agreement with AdS/CFT  is required) is discussed in Appendix \ref{A:cc}. 

%,  implying, in particular, that $\cc$   satisfies  the same relations  as 
%$E_c$ and $\aa$   do in the context of AdS/CFT.  

%%%%%%%%%%%%%%%%%%%%%%%%%%%%%%%%%%%%%%%%%%%%%%%
\subsection{Proposal for general expression for $\rc(\Delta;\, j_{1},j_{2})$}
%%%%%%%%%%%%%%%%%%%%%%%%%%%%%%%%%%%%%%%%%%%%

%The problems discussed in the previous section are amended by the following expression that 
%replaces (\ref{4.2}) and is 

Our proposal for   $\cc$ that replaces \rf{4.2}  in the    massive  representation case  $(\Delta;\,j_{1},j_{2})$
 is 
\begin{align}
%{\rm c}(\Delta;\,j_{1},j_{2}) & =
&\widehat{\rm c}(\Delta;\,j_{1},j_{2}) =  -2\,\widehat {\rm c}^{+} (\Delta;\,j_{1},j_{2}) = 
 \frac{1}{720}\,(-1)^{2(j_{1}+j_{2})}
(2j_{1}+1)(2j_{2}+1)\,(\Delta-2) \nonumber\\
&\qquad \times \Big[   -6\,(\Delta-2)^4+20\,(\Delta-2)^2+
    6\,(j_{1}^{4}+j_{2}^{4})
  +20\,j_{1}^{2}j_{2}^{2}
   +12\,(j_{1}^{3}+j_{2}^{3}) \nonumber \\
\label{4.11}
& \qquad   \ \  \quad   +20\,(j_{1}^{2}j_{2}+j_{1}j_{2}^{2}) -6\,(j_{1}^{2}+j_{2}^{2})+20\,j_{1}j_{2}-12\,(j_{1}+j_{2})-8\Big].
\end{align}
The corresponding expression for  $\rc-\ra$   following from \rf{3.3}    and 
(\ref{4.11})  is then 
\begin{align}
\widehat{\rm c}(\Delta;\,j_{1},j_{2})-\widehat{\rm a}(\Delta;\,j_{1},j_{2})& =\frac{1}{720}\,(-1)^{2(j_{1}+j_{2})}
(2j_{1}+1)(2j_{2}+1)\,(\Delta-2)\, \nonumber\\
&\times \Big[   -3\,(\Delta-2)^4-  5(  2 j_{1}^2  + 2 j_2^2 + 2 j_1 + 2 j_2 - 3)   (\Delta-2)^2 \nonumber \\ &+
    21\,(j_{1}^{4}+j_{2}^{4})
  -10\,j_{1}^{2}j_{2}^{2}
   +42\,(j_{1}^{3}+j_{2}^{3})-10\,(j_{1}^{2}j_{2}+j_{1}j_{2}^{2}) \nonumber \\
\label{4.16}
&  \quad  +9\,(j_{1}^{2}+j_{2}^{2})-10\,j_{1}j_{2}-12\,(j_{1}+j_{2})-8\Big].
\end{align}
This  is different from   (\ref{4.1}) as  the dependence on $\Delta$ is not just  via  the 
    overall   $\Delta-2$ factor. 

Eq. (\ref{4.11})  and its  massless representation  counterpart \rf{4.3} 
 is  consistent with  all low-spin data,   giving, e.g.,  
 the correct values for all the fields of extended conformal supergravity (see Table \ref{T2}): 
scalars with $\Delta=3,4$,  spin $\ha $ fermions  with $\Delta=\frac{5}{2},\frac{7}{2}$,
non-gauge antisymmetric tensor,  conformal gravitino and  conformal graviton.
Applying  \rf{4.3},\rf{4.11}   to the cases of totally symmetric bosonic and fermionic CHS  fields
we find again the expressions in  \rf{4.4},\rf{4.5} but now   with  % different  values of the parameters 
\be
\label{4.12}
r_{b}=-1\ ,\qquad \qquad  \qquad r_{f}=51\ , 
\ee
instead  of  \rf{46}. 
These  values of the parameters 
 are  precisely the ones that lead to  the vanishing of  the  sum  $\sum_s \cc_s$   over all totally symmetric  CHS  fields 
\cite{Tseytlin:2013jya,Giombi:2014iua}, assuming the 
same regularization  that implies the vanishing of  $ \sum_s\aa_s$  \ci{Giombi:2013yva,Tseytlin:2013jya}
   and  $\sum_s E_{c,s}$ \ci{Beccaria:2014jxa}.

The crucial  feature of (\ref{4.11})  is that it  leads to   important consistency checks  of   vectorial AdS/CFT duality
which are  direct analogs of  the  earlier   checks  based on the expressions for $\aa$-coefficient  and $E_c$.\foot{We present a more 
general ansatz for $\cc$ that reduces to \rf{4.11} after imposing this  consistency constraint  in Appendix \ref{A:cc}.} 
These checks  will be discussed in detail  in section 6. Here we just mention
 two non-trivial relations in the case of a particular mixed representation
 satisfied  by 
$\cc$ \rf{4.3}  defined  by  \rf{4.11}  but  not by \rf{4.2}: 
\ba
\label{4.14}
\sum_{s=1,2,...}^{\infty}{\rm c}{\te \big(2+s;\,\frac{s+1}{2},\frac{s-1}{2}\big) }= -\frac{1}{120}\ , \qquad \qquad 
%\label{4.15}
\sum_{s=2,4,...}^{\infty}{\rm c}{\te \big(2+s;\,\frac{s+1}{2},\frac{s-1}{2}\big) }= -\frac{1}{30}\ .
\end{align}
%Here the sums are understood in a regularized sense (see section 6.2). 
%These  relations % Eqs.~(\ref{4.14}, \ref{4.15})
%are  direct counterparts   of the relations  for $\ra$ and $E_c$. % to be discussed in Sec.~(\ref{sec:applications}).

%%%%%%%%%%%%%%%%%%%%%%%%%%%%%%%%%%%%%%%%%%%%%%%%%%%%%%%
\section{$E_{c}, \ra, \rc$ for  superconformal  $SU(2,2|\N)$ multiplets}
\label{sec:multiplets}
%%%%%%%%%%%%%%%%%%%%%%%%%%%%%%

In  this section we  shall   compute  $E_{c}, \ra, \rc$   for  collections of  primary fields of  
$SO(2,4)$ representations $(\De; j_1,j_2)$     forming  superconformal multiplets. 
It turns  out that  the difference between $ \cc-\aa$ in \rf{4.1}  and  our proposal \rf{4.16}  disappears 
once one sums over  all fields in the supermultiplet, implying that  the  resulting 
$\cc- \aa$  is linear in $\De$ as in \rf{4.1}  (but separate values of  the coefficients 
$\aa$ and $\cc$ are still different from the ones  implied by the prescription of \ci{Mansfield:2003gs}). 
% i.e.  that for  both massive and massless cases  one has   
%\be\la{5.1}
%  (\cc-\aa)_{\rm supermultiplet} = - 2 ({\rm c}^{+}-{\rm a}^{+})_{\rm supermultiplet} 
 % =   \sum_{{\rm supermultiplet}}  (\De-2)\,  F(j_1, j_2)  \ . 
%  = \sum_{_{\rm supermultiplet}}  \frac{1}{180}\, (-1)^{2\,(j_{1}+j_{2})}(\Delta-2)\,d(j_{1},j_{2})\,\big[1+f(j_{1})+f(j_{2})\big]
%  \ee
%The explicit  expressions for $\cc-\aa$ then match those 

\subsection{Summary of   contributions of a  single conformal $(\Delta;\,j_{1},j_{2})$   field}

It  is useful  first  to summarize  the expressions for $E_{c}, \ra, \rc$  and $\cc-\aa$ in  \rf{2.32}), \rf{3.3}, \rf{4.11},\rf{4.16}
for a non-gauge (massive  5d)  field 
in a compact  form using  the variables $d_{1} = 2j_{1}+1$, $d_{2}=2j_{2}+1$:
\begin{align}
&\widehat{E}_{c}(\Delta;\,j_{1},j_{2}) = -\frac{1}{720}(-1)^{d_{1}+d_{2}}\,d_{1}\,d_{2}\,(\Delta-2)\,
\Big[6\,(\Delta-2)^{4}-20\,(\Delta-2)^{2}+11\Big], \la{51}
\\
&\widehat{\ra}(\Delta;\,j_{1},j_{2}) = \frac{1}{11520}(-1)^{d_{1}+d_{2}}\,d_{1}\,d_{2}\,(\Delta-2)
\nonumber\\
& \qquad\qquad \times \Big[
-48\,(\Delta-2)^{4}+  40\,(d_{1}^{2}+d_{2}^{2})\,(\Delta-2)^{2}-  15\,(d_{1}^{2}-d_{2}^{2})^2 %+30\,d_{1}^{2}\,d_{2}^{2}
\Big], \la{52}\\
&\widehat{\rc}(\Delta;\,j_{1},j_{2}) = \frac{1}{5760}(-1)^{d_{1}+d_{2}}\,d_{1}\,d_{2}\,(\Delta-2)
\nonumber\\
& \qquad\qquad\times  \Big[
-48\,(\Delta-2)^{4}+160\,(\Delta-2)^{2}+3\,(d_{1}^{4}+d_{2}^{4})+10\,d_{1}^{2}\,d_{2}^{2}
-40\,(d_{1}^{2}+d_{2}^{2})
\Big], \la{53}\\
&\widehat{\rm c}(\Delta;\,j_{1},j_{2})-\widehat{\rm a}(\Delta;\,j_{1},j_{2}) =\frac{1}{11520}\,(-1)^{d_{1}+d_{2}}
d_{1}\,d_{2}\,(\Delta-2)\,\la{54}
\\
&   \times  \Big[   -48\,(\Delta-2)^4  -40\,(d_{1}^{2}+d_{2}^{2}-8)\,(\Delta-2)^2
+  21\,(d_{1}^{4}+d_{2}^{4})-10\,d_{1}^{2}\,d_{2}^{2}-80\,(d_{1}^{2}+d_{2}^{2}) 
\Big]\no .
\end{align}
Note that these  expressions 
 are odd under $\De \to 4-\De$, cf. \rf{sub}.
%These expressions change sign under $\Delta\to 4-\Delta$
%so  going to conjugate or shadow field is like  changing statistics.
The values  in the   gauge (massless 5d)  field case with $\De= 2 + j_1 + j_2 $   follow  from \rf{2.33}, \rf{3.4},\rf{4.3}. 
Written   in terms of the  variables  
\be  s= h_1 = j_1 + j_2\ , \qquad    h_2  = j_1 - j_2 \ , \qquad  \nu = s (s+1)   \ , \ \ \ \ \   \De=2 + s  \ ,  \la{55} \ee
they  read 
\begin{align}
&{E}_{c} ( j_1,j_2) = \frac{1}{720}(-1)^{2s } 
\Big[\n  ( 18 \n^2 - 14 \n -11)   - 3 h_2^2  ( 10 \n^2 - 10 \n -1) \Big], \la{5.6}
\\
&{\ra} ( j_1, j_2) =\frac{1}{720}(-1)^{2s } 
\Big[\n  ( 14 \n^2   + 3 \n )   - 3 h_2^2  ( 20  \n^2 + 10 \n +1)  + 5 h_2^4 ( 6 \n +1)  \Big], \la{5.7}\\
&{\rc} ( j_1,j_2) =\frac{1}{360}(-1)^{2s } 
\Big[  \n  ( 14 \n^2  -17 \n -4)      -  h_2^2  ( 15  \n^2 -15  \n -7)  -  5 h_2^4  + h_2^6   \Big] \ ,   \la{5.8}
\end{align}
generalizing   \rf{2.34},\rf{234},\rf{35},\rf{36},\rf{4.4},\rf{4.5}.\foot{Here   we wrote the fermionic  contribution in terms of
 $\nu= s(s+1)$ rather than $\nu_f$.}
These expressions are symmetric under $j_1 \leftrightarrow j_2$  so  that  in the case of $j_1 \not= j_2$ 
when the physical combination is 
$(j_1, j_2)_c= (j_1, j_2) + (j_2, j_1)$  
an extra  factor of 2 is to be added  (in our notation  bosonic $j_1=j_2$ fields are real).

%%%%%%%%%%%%%%%%%%%%%%%%%%%%%%%%
\subsection{$\N=1$ superconformal multiplets}

%Above we   gave expressions for  $E_{c}, \ra, \rc$  for primary  conformal $SO(2,4)$  fields
%$(\Delta;\, j_{1}, j_{2})$.  
Let us   now  find  the total contributions
of  $\N=1$ superconformal multiplets  containing $(\Delta;\, j_{1}, j_{2})$  field  as the 
lowest dimension  member.  The structure of relevant multiplets was  given,  e.g.,  in \cite{Dobrev:1985qv}. 
In addition to  long massive multiplets there are  shortened ones:
  chiral and    right-handed semi-long (SLII), as well as  their  CP conjugates --  anti-chiral and 
left-handed semi-long (SLI). There are also CP self-conjugate (``conserved'') 
multiplets that  are the sums of one SLI and one SLII multiplet  (thus  they need not 
 be considered separately).
% If we denote by $D(\Delta;\, j_{1}, j_{2})$ the contribution from a conformal field with quantum 
% numbers $(\Delta;\, j_{1}, j_{2})$, then 

 $SO(2,4)$ representation content 
 of   massive long $\N=1$ superconformal
 multiplet is\footnote{The term $2\,(\Delta+1;\, j_{1}, j_{2})$
comes from two representations with the same $SO(2,4)$  labels but different R-charge. 
For the  computation  of $E_{c}, \ra, \rc$  we do not need to keep track 
 of the $R$ charge (in general, it is constrained by the shortening conditions).}
 \be\la{5.9}
\begin{split}
 [\Delta;\, j_{1}, j_{2}]_{\rm long} &= (\Delta;\, j_{1}, j_{2}) + \left(\Delta+\textstyle\frac{1}{2};\, j_{1}+\textstyle\frac{1}{2}, j_{2}\right) 
+ \left(\Delta+\textstyle\frac{1}{2};\, j_{1}-\textstyle\frac{1}{2}, j_{2}\right) \\
&
+ \left(\Delta+\textstyle\frac{1}{2};\, j_{1}, j_{2}+\textstyle\frac{1}{2}\right) 
+ \left(\Delta+\textstyle\frac{1}{2};\, j_{1}, j_{2}-\textstyle\frac{1}{2}\right)  
+ 2\,\left(\Delta+1;\, j_{1}, j_{2}\right)\\
&+ \left(\Delta+1;\, j_{1}+\textstyle\frac{1}{2}, j_{2}+\textstyle\frac{1}{2}\right)
+ \left(\Delta+1;\, j_{1}+\textstyle\frac{1}{2}, j_{2}-\textstyle\frac{1}{2}\right) \\
&+ \left(\Delta+1;\, j_{1}-\textstyle\frac{1}{2}, j_{2}+\textstyle\frac{1}{2}\right)
+ \left(\Delta+1;\, j_{1}-\textstyle\frac{1}{2}, j_{2}-\textstyle\frac{1}{2}\right) \\
%+ D\left(\Delta+1;\, j_{1}, j_{2}\right) 
& 
+\left(\Delta+\textstyle\frac{3}{2};\, j_{1}, j_{2}+\textstyle\frac{1}{2}\right)
+\left(\Delta+\textstyle\frac{3}{2};\, j_{1}, j_{2}-\textstyle\frac{1}{2}\right)\\
&
+\left(\Delta+\textstyle\frac{3}{2};\, j_{1}-\textstyle\frac{1}{2}, j_{2}\right)
+\left(\Delta+\textstyle\frac{3}{2};\, j_{1}+\textstyle\frac{1}{2}, j_{2}+\textstyle\frac{1}{2}\right)
+(\Delta+2;\, j_{1}, j_{2}).
\end{split}
\ee
Using the above  expressions we  find that the total  $\ra$ and $\rc$ anomalies of a long  massive  multiplet  vanish 
  but    the Casimir energy does not: 
\ba\la{56}
\ra_{\rm long} =\rc_{\rm long} = 0  , \qquad \qquad 
E_{c\ }{}_{\rm long} = - { 1 \ov 16} 
% \frac{1}{32}
\,(-1)^{2\,(j_{1}+j_{2})}\,(2j_{1}+1)(2j_{2}+1)\,(\Delta-1)\ .
\end{align}
The vanishing of $\cc-\aa$ 
for long multiplets  follows also from  \rf{4.1}   \cite{Ardehali:2013xya,Ardehali:2014zba}.
The  fact that $E_c$  is not proportional to $\aa$-coefficient  as in \rf{1.13} 
 means that the coefficient $\gg$  of the $D^2R$ term in 
the trace anomaly \rf{1.12} does not vanish  (in  the heat kernel  scheme we are using to define $E_c$);
indeed, $\gg$  is  expected to cancel  only  in $\N>2$  extended supersymmetric  cases (cf. \rf{1.13}).

The content of the  chiral short multiplet  is 
\be\la{57}
\begin{split}
&[\Delta;\, j,0] _{\rm chiral}= 
(\Delta;\, j,0) 
+(\Delta+\textstyle\frac{1}{2};\,j+\textstyle\frac{1}{2},0)
+(\Delta+\textstyle\frac{1}{2};\,j-\textstyle\frac{1}{2},0)
+(\Delta+1;\,j,0) \ ,  
\end{split}
\ee
and  thus we find 
\ba
\ra_{\rm chiral} &= \frac{1}{96}\,(-1)^{2j}\,(2j+1)\,(2\,\Delta-3)\,(-2\,\Delta^{2}+6\,\Delta+6j^{2}+6j-3),\no  \\
\rc_{\rm chiral} &=- \frac{1}{48}\,(-1)^{2j}\,(2j+1)\,(2\,\Delta-3)\,(\Delta^{2}-3\,\Delta+j^{2}+j+1), \no \\
E_{c\ }{}_{\rm chiral} &=- \frac{1}{384}\,(-1)^{2j}\,(2j+1)\,(16\Delta^{3}-72\Delta^{2}+94\Delta-33) \ , \no \\
(\rc-\ra)_{\rm chiral}& = -  \frac{1}{96}
% \frac{1}{192}
\,(-1)^{2j}\,(2j+1)\, (8j^{2}+8j-1)\,(2\,\Delta-3)\ .\la{59}\end{align}
The  SLII  short multiplet has  the content 
\be\la{510}
\begin{split}
 &[\Delta;\, j_{1},j_{2}]_{\rm SLII}= 
(\Delta;\, j_{1},j_{2})+ (\Delta+\textstyle\frac{1}{2};\, j_{1},j_{2}+\textstyle\frac{1}{2})+
(\Delta+\textstyle\frac{1}{2};\, j_{1}+\textstyle\frac{1}{2},j_{2})\\ & \qquad +
(\Delta+\textstyle\frac{1}{2};\, j_{1}-\textstyle\frac{1}{2},j_{2})  +  (\Delta+1;\, j_{1}+\textstyle\frac{1}{2},j_{2}+\textstyle\frac{1}{2})+
(\Delta+1;\, j_{1}-\textstyle\frac{1}{2},j_{2}+\textstyle\frac{1}{2}) \qquad\\   &\qquad    +
(\Delta+1;\, j_{1},j_{2}) (\Delta+\textstyle\frac{3}{2};\, j_{1},j_{2}+\textstyle\frac{1}{2}),
\end{split}
\ee
and we obtain
\ba
&\ra_{\rm SLII} = \frac{(-1)^{2\,(j_{1}+j_{2})  }} {96}\,(2j_{1}+1)\,(2\,\Delta+2\,j_{2}-1)
\big[2\,(\Delta+j_{2}-1)(\Delta+j_{2})-6j_{1}(j_{1}+1)- 1\big], \qquad \no \\
&\rc_{\rm SLII} =\frac{  (-1)^{2\,(j_{1}+j_{2})}   \,}{48}\, (2j_{1}+1)\,(2\,\Delta+2\,j_{2}-1)
\big[(\Delta+j_{2}-1)(\Delta+j_{2})+j_{1}(j_{1}+1)-1\big], \qquad \no \\
&E_{c\ } {}_{\rm SLII}=\frac{  (-1)^{2\,(j_{1}+j_{2})}   \,}{384}\,(2j_{1}+1)
\big[16\,\Delta^{3}-24\,\Delta^{2}-26\,\Delta+29+2\,(24\,\Delta^{2}-60\,\Delta+31)\,j_{2}\big], \qquad \no\\
  \la{511}
&(\rc-\ra)_{\rm SLII} = \frac{1}{96}
%-\frac{1}{192}
\,(-1)^{2\,(j_{1}+j_{2})}\,(2j_{1}+1)\,(8j_{1}^{2}+8j_{1}-1)\,(2\Delta+2j_{2}-1) \ .
\end{align}
The same expressions \rf{59}  and \rf{511}  for $\cc-\aa$  %(up to -2  normalization factor) %  $\cc-\aa= -  2  (\cc^+ - \aa^+)$)  
follow \cite{Ardehali:2014zba}   if we use  (\ref{4.1}) instead of our \rf{4.16},
i.e.  the  chiral  and SLII multiplet   expressions for $\cc$   are  not sensitive   to   the difference between \rf{4.2} and \rf{4.11}.\footnote{This  equality of $\rc-\ra$ for $\N=1$ multiplets
 computed  using  $\rc$  from  (\ref{4.2}) or from   (\ref{4.11}) is non-trivial. 
Consider the difference between (\ref{4.2}) and (\ref{4.11}) 
for the  basic combination  of representations  $\langle \Delta; j\rangle \equiv  (\Delta; \, j,0)+(\Delta+\textstyle\frac{1}{2};\,j+\textstyle\frac{1}{2},0)$.
This turns out to be a function of $\Delta+j$ multiplied by $(-1)^{2j}$  and  the contribution of 
%This can be used to show some cancellations. For instance,
 a chiral multiplet  happens to   be  the same as of 
$\langle \Delta; j\rangle  +\langle \Delta+\textstyle\frac{1}{2}; j-\textstyle\frac{1}{2}\rangle $. It is then possible to see  that  the contribution of this sum vanishes.
%The sum if the arguments is the same, and we get a relative minus sign from $(-1)^{2j}$. 
%So the sum is zero showing that (\ref{4.2}) or  (\ref{4.11})  give the same result.
}

%%%%%%%%%%%%%%%%%%%%%%%%%%%%%%%%%%%%%%%%%%%%%%%%%%%
\begin{table}[H]
\be
\def\arraystretch{1.3}
\begin{array}{|l|ccc|c|c|c|c|}
\hline
\mc N & \phi & \psi & V_{\mu} &  E_{c} & {\rm a} & \rc  \\
\hline 
1 & \mbox{--} & 1 & 1 &  \frac{7}{64} & \frac{3}{16} &  \frac{1}{8}\\
2 & 2 & 2 & 1 &  \frac{13}{96}  & \frac{5}{24} & \frac{1}{6}\\
3,4 & 6 & 4 & 1 & \frac{3}{16}  & \frac{1}{4} & \frac{1}{4} \\
\hline
\end{array}
\nonumber
\ee
\caption{Values of $E_{c}, \ra, \rc$ for   $ \N\le 4$ supersymmetric Maxwell multiplets.}
\label{T1}
\end{table}
\noindent

\subsection{$\N>1$ superconformal multiplets}

Next, let us  present  the  expressions  for  $E_{c}, \ra, \rc$   in the case of  some $\N > 1$ superconformal    multiplets.
 
  % with various amounts of supersymmetry, up to $\N=4$.

\subsubsection{Maxwell supermultiplets}

Considering massless 4d multiplets   with  the highest spin  1 we get the values in Table \ref{T1}.
%We can consider 4d extended supersymmetric multiplets with a vector as the top spin component.
%Application of our general expressions  leads to the following table:

We notice that for $\mc N=3, 4$ eq.\rf{1.13} is satisfied, i.e. 
\be  E_c = { 3 \ov 4 } \aa\ , \ \ \ \ \ \  \ \quad   \aa= \cc % \ , \ \ \  \ \ \ \ \   E_c= { 3 \ov 4} \aa
  \ ,  \la{517}
\ee 
and thus  the  coefficient $\gg$ of the  derivative term in  \rf{1.12} vanishes \ci{Fradkin:1983tg}. 
%in the natural  heat kernel   scheme  

Let us mention   that  $\N=4$ Maxwell multiplet is isomorphic  to the $\N=4$ 
%has the same field content as the
 superdoubleton multiplet  $\{\N=4\} = \{\ 1,0\}_c + 4 \{\ \ha, 0\}_c + 6  \{0,0\}$   of $PSU(2,2|4)$  
 %  from the   AdS$_{5}$    point of view 
 \ci{Gunaydin:1984fk} 
 and thus their  quantum characteristics  should be the same,  
  %(in particular, one-particle partition functions match, see \rf{a26}), i.e.
 \be 
 K(\{ \N =4\}) = K(\N=4 \ { \rm Maxwell}) \ , \ \ \ \ \ \ \ \ \   K\equiv  (E_c, \aa, \cc)  \ . \la{5177}
 \ee
  Also, the one-particle partition functions match, see \rf{a26}. 
 
\iffa
% Its representation content is 
 $(2; 1, 0) + (2; 0,1)$, $   % $(3\,; \textstyle\frac{1}{2}, \textstyle\frac{1}{2}) +
  4 \left[( \textstyle\frac{3}{2}\,; \textstyle\frac{1}{2} ,0)   + ( \textstyle\frac{3}{2}\,; 0, \textstyle\frac{1}{2})\right]$ + 6   (1\,;0,0)$,
 where, once again, $\Delta=\De_+$   corresponds dual 5d fields  and not to the  the canonical 
 dimension $\Delta_-$ of  the elementary 4d conformal field.
\fi 

%
%there is enough supersymmetry to make $\ra=\rc$. Also, we check the 
%relation between $E_{c}$ and $\ra$ suggested in \cite{Herzog:2013ed,Huang:2013lhw} 
%assuming absence of D-anomalies. In general even dimension, this is 
%\be
%E_{c} = \frac{1\cdot 3\cdot 5\cdots (d-1)}{(-2)^{d/2}}\,{\rm a},
%\ee
%and reduces to $E_{c} =\frac{3}{4}\,\ra$ for $d=4$.

%%%%%%%%%%%%%%%%%%%%%%%%%%%%%%%%%%%%%%%%%%
\subsubsection{Conformal supergravity multiplets}
%%%%%%%%%%%%%%%%%%
The case of short multiplets with   highest   spin  value is  2  is that of  4d extended conformal supergravity  (CSG)   multiplets. 
The relevant fields are listed in Table \ref{T2} 
together with their individual $E_{c}, \ra, \rc$ values. %  (canonical dimensions of the fields there are $\De_-=4-\De$).  
The  total values for  $\N\le 4 $ conformal  
 conformal supergravity  multiplets   are given  in Table \ref{T3}
(the numbers in the central square are   multiplicities of the fields, i.e.   dimensions of their $U(\N)$  or $SU(4)$ representations).

As in the case of  Maxwell supermultiplets, for $\mc N=3,4$ we find  the relation \rf{1.13} or \rf{517}  satisfied, 
implying  $\gg=0$  (cf. \rf{1.14}).   The values of   $E_c$  and $\gg$   %(related to $\aa$    by \rf{1.14})
 for conformal supergravities were not computed  previously.
\begin{table}[H]
\be
\def\arraystretch{1.6}
\begin{array}{|c|c|c|c|c|}
\hline
{\rm Field} & (\Delta;\,j_{1},j_{2}) & E_{c} & {\rm a} & {\rm c}\\
\hline
\phantom{\Big(}\phi\ (\Box) & (3; 0,0) & \ds\frac{1}{240} & \ds\frac{1}{360}&\ds \frac{1}{120}\\
\hline
\phantom{\Big(}\Phi\ (\Box^{2}) & (4; 0,0) &\ds-\frac{3}{40}&\ds-\frac{7}{90}& \ds-\frac{1}{15}\\
\hline
\phantom{\Big(}\psi\ (\slashed{\partial}) & (\frac{5}{2};\, \frac{1}{2},0)+(\frac{5}{2};\, 0,\frac{1}{2})&\ds\frac{17}{960}&\ds\frac{11}{720}&\ds\frac{1}{40} \\
\hline
\phantom{\Big(}\Psi\ (\slashed{\partial}^{3}) &  (\frac{7}{2};\, \frac{1}{2},0)+(\frac{7}{2};\, 0,\frac{1}{2}) 
&\ds-\frac{29}{960}&\ds-\frac{3}{80}&\ds-\frac{1}{120} \\
\hline
\phantom{\Big(}T_{\mu\nu}\ (\Box) & (3; 1,0)+(3; 0,1)&\ds\frac{1}{40}&\ds-\frac{19}{60}&\ds\frac{1}{20} \\
\hline
\phantom{\Big(}V_{\mu}\ (\Box) & (3; \frac{1}{2}, \frac{1}{2})&\ds\frac{11}{120}&\ds\frac{31}{180}&\ds\frac{1}{10} \\
\hline
\phantom{\Big(}\psi_{\mu}\ (\slashed{\partial}^{3}) & (\frac{7}{2};\,1,\frac{1}{2})+(\frac{7}{2};\,\frac{1}{2},1) 
&\ds-\frac{141}{80}&\ds-\frac{137}{90}&\ds-\frac{149}{60} \\
\hline
\phantom{\Big(}g_{\mu\nu}\ (\Box^{2}) &(4;\,1,1)&\ds\frac{553}{120}&\ds\frac{87}{20}&\ds\frac{199}{30} \\
\hline
\end{array}\nonumber
\ee
\caption{Values of $E_{c}, \ra, \rc$ for  fields of dimension  $4-\De$  of  conformal supergravities.} %(canonical dimensions are  $4-\De$).}
\label{T2}
\end{table}
%%%%%%%%%%%%%%%%%%%%%%%%%%%%%%%%%%%%%%%%%%%
\begin{table}[H]
\be
\begin{array}{|l|cccccccc|c|c|c|}
\hline
\mc N & \phi & \Phi & \psi & \Psi & T_{\mu\nu} & V_{\mu} & \psi_{\mu} & g_{\mu\nu}  & E_{c} & {\rm a} & \rc \\
\hline 
&&&&&&&&&&&\\ 
1 & \mbox{--} & \mbox{--} &  \mbox{--} &  \mbox{--} & \mbox{--} & 1 & 1 & 1 & \frac{47}{16} & 3 & \frac{17}{4}\\
&&&&&&&&&&&\\ 
2 & \mbox{--} & \mbox{--} &  2 & \mbox{--} & 1 & 4 & 2 & 1 &  \frac{145}{96} & \frac{41}{24} & \frac{13}{6}\\
&&&&&&&&&&&\\ 
3 & 6 & \mbox{--} &  9 & 1 & 3 & 9 & 3 & 1 &  \frac{3}{8}  & \frac{1}{2} & \frac{1}{2} \\
&&&&&&&&&&&\\ 
4 & 20 & 2 &  20 & 4 & 6 & 15 & 4 & 1 &  -\frac{3}{4} & -1 & -1\\
&&&&&&&&&&&\\ 
\hline
\end{array}\nonumber
\ee
\caption{Values of $E_{c}, \ra, \rc$ for $ \N\le 4$ extended conformal supergravity.
 }
\label{T3}
\end{table}
\noindent

 As  was  found in \cite{Fradkin:1981jc,Fradkin:1985am},  the   conformal  anomalies of the combined system of 
 $\N=4$   conformal supergravity and {\it four}  $\N=4$ Maxwell multiplets  cancel, i.e. this is a UV  finite theory. 
 This is readily seen from the values in Tables \ref{T1} and \ref{T2}: 
\be 
%both $\ra=\rc$ and $E_{c} = \frac{3}{4}\,\ra$.
%Notice also the well known cancellation between $\mc N=4$ conformal supergravity and four $\mc N=4$ Maxwell multiplets $
K(\N=4\ {\rm CSG}) + 4\,   K({\N=4 \ \rm Maxwell}) = 0\ , \ \ \ \ \ \ \ \ \qquad      K = (E_{c}, \ra, \rc) \ . \la{518}
\ee 
The vanishing  of the total $E_c$ is a  new result (implied  by \rf{517} which  is valid  for each  of the $\N=4$ multiplets). 

%For $E_c$ this is a new result indicating that $D^2R$ terms do not  cancel only in $\N=4$
%SYM as is known,  but also in pure $\N=4$ conformal supergravity.
The $\N=4$   conformal  supergravity 
multiplet\foot{In addition to fields listed in Table \ref{T2} this  $PSU(2,2|4)$  short   multiplet contains also 20 auxiliary scalars 
with $\De=2$ which do not contribute to physical quantities  (the total  number of  helicity $2 j_1 + 2 j_2 +1$  states is 256).} 
is isomorphic to the  supercurrent multiplet   of $\N=4$  Maxwell  theory  \ci{Howe:1981qj} and also to the  short 
massless multiplet of  fields of gauged $\N=8$ supergravity in 5 dimensions  whose \ads 
 vacuum isometry is   $PSU(2,2|4)$  \ci{Gunaydin:1984fk,Gunaydin:1984vz,Ferrara:1998ej,Liu:1998bu}.\foot{$\N=8$  supersymmetry of 5d supergravity  corresponds to   4  Poincare and 4 conformal    supersymmetries of   $\N=4 $ conformal supergravity in 4d.}
 The field content of the latter is given in  $p=2$ entry in  Table \ref{T5}  in Appendix \ref{A:KK}. 
 Indeed, the  5d  expression for  the  conformal anomaly and  the Casimir energy  for $\N=4$   CSG is  directly 
  given by  the  one-loop  contributions of fields  of $\N=8$  5d supergravity, i.e. 
 \be 
 K(\N=4\ {\rm CSG})  =\,  - 2\,  K^+ (\N=8\ {\rm 5d\ SG}) \ . \la{519} \ee 
 This one-loop relation between the two theories  generalizes the tree-level one in   \ci{Liu:1998bu}. 
 
 In view of \rf{518} this   also  implies that one-loop  contribution of $\N=8$  5d supergravity  is the same  as of  two  
 $\N=4$ Maxwell  multiplets, 
 \be 
  K^+ (\N=8\ {\rm 5d\ SG})  = 2\, K({\N=4 \ \rm Maxwell}) \ . \la{520} \ee 
  Remarkably, this  non-trivial  relation may be interpreted  as  expressing the fact that 
   the states of $\N=8$  5d supergravity  appear in the product of two $\N=4$ superdoubletons  \ci{Gunaydin:1998jc}.
  We shall return to this  observation  in section 6.1 below.

%Finally, we remark that in the table we do not list the auxiliary fields since they do not contribute $E_{c}, \ra, \rc$. 
%Of course, they are necessary to get the correct counting of states. For instance, the $\N=4$ fields
%are in a BPS multiplet of $SU(2,2|4)$ with length 256. This is obtained adding 20 auxiliary scalars and counting states
%by multiplying the $SU(4)$ multiplicity times the number of helicity states $2(j_{1}+j_{2})+1$. This multiplet
%has the same content as the massless $\N=8$ $d=5$ gauged supegravity multiplet.
\iffa 
We also    use our  expressions to  rederive  from 5d  perspective 
the values of $K=(E_c, \aa, \cc)$ for $\N \le  4$ super Maxwell  and conformal supergravity multiplets, verifying the 
relation \rf{1.13}  for $\N=3,4$ cases and also that  all three quantities vanish when $\N=4$ conformal supergravity is combined with  four  
$\N=4$  Maxwell multiplets \ci{Fradkin:1981jc, Fradkin:1985am}. 
The 5d approach provides direct  relation   between   anomalies of $\N=4$  conformal supergravity 
 and  one-loop contribution of fields  of $\N=8, \ d=5$ gauged supergravity  as the  two are  described by the equivalent  
 short  $PSU(2,2|4)$ supermultiplet  (this generalizes to one-loop level the known tree-level relation  \ci{Liu:1998bu}). 
We show also that $K=0$ for general long  massless  supermultiplet of $PSU(2,2|4)$.
\fi

%%%%%%%%%%%%%%%%%%%%%%%%%%%
%%%%%%%%%%%%%%%%%%%%%%%%%%%%%%%%%%%%%%%%%%%%%%%%%%%%%%%
\subsubsection{General long higher spin massless  $PSU(2,2|4)$  supermultiplet}
%%%%%%%%%%%%%%%%%%%%%%%%%%%%%%%%%%%%%%%%%

The general long massless multiplet of  $PSU(2,2|4)$  \cite{Gunaydin:1998sw,Gunaydin:1998jc}
% the $\mc N=8$ $d=5$
%AdS super algebra $SU(2,2|4)$ has been discussed in .
has spin range 4  (8 supercharges).  Its  conformal 
representation content is that of $[j_{1},j_{2}]\oplus [j_{2},j_{1}]$ 
where $[j_{1},j_{2}]$ is summarized in Table \ref{T4}.  
There  $j_{1},j_{2}\ge 1$  are the labels of the supermultiplet 
and  all    states 
have  $\Delta=2+j_{1}+j_{2}$. 
The  members of this multiplet may be viewed as  representing  massless  higher spin  \ads  fields  
 or the corresponding  4d conformal  higher spin gauge fields.   %5d states 

% as it should for a massless representation.
%%%%%%%%%%%%%%%%%%%%%%%%%%%%%%%%%%%%%%%%%%
\begin{table}[H]
{\small 
\be
\def\arraystretch{1.5}
\begin{array}{|l|l|}
\hline
\mbox{spin}\ (j_{L},j_{R})  & SU(4)  \\
\hline
\hline
(j_{1}+1,j_{2}+1) & 1 \\
\hline
(j_{1}+1,j_{2}+\frac{1}{2})+(j_{1}+\frac{1}{2}, j_{2}+1) & 4+4^{*} \\
\hline
(j_{1}+\frac{1}{2}, j_{2}+\frac{1}{2}) & 1+15 \\
(j_{1}+1,j_{2})+(j_{1},j_{2}+1) & 6+6 \\
\hline
(j_{1}+\frac{1}{2},j_{2})+(j_{1}, j_{2}+\frac{1}{2}) & 4+4^{*}+20+20^{*} \\
(j_{1}+1,j_{2}-\frac{1}{2})+(j_{1}-\frac{1}{2},j_{2}+1) & 4+4^{*}\\
\hline
(j_{1},j_{2}) & 1+15+20' \\
(j_{1}+\frac{1}{2},j_{2}-\frac{1}{2})+(j_{1}-\frac{1}{2},j_{2}+\frac{1}{2}) & 6+6+10+10^{*} \\
(j_{1}+1,j_{2}-1)+(j_{1}-1,j_{2}+1) & 1+1 \\
\hline 
\end{array}
\ 
\begin{array}{|l|l|}
\hline
\mbox{spin}\ (j_{L},j_{R})  & SU(4)  \\
\hline
\hline 
(j_{1},j_{2}-\frac{1}{2})+(j_{1}-\frac{1}{2},j_{2}) & 4+4^{*}+20+20^{*} \\
(j_{1}+\frac{1}{2},j_{2}-1)+(j_{1}-1,j_{2}+\frac{1}{2}) & 4+4^{*}\\
\hline
(j_{1}-\frac{1}{2},j_{2}-\frac{1}{2}) & 1+15 \\
(j_{1},j_{2}-1)+(j_{1}-1,j_{2}) & 6+6 \\
\hline
(j_{1}-\frac{1}{2},j_{2}-1)+(j_{1}-1,j_{2}-\frac{1}{2}) & 4+4^{*}\\
(j_{1}-1,j_{2}-1) & 1\\ 
& \\
& \\
& \\
\hline
\end{array}\nonumber
\ee
}
\caption{Spin and $SU(4)$ content of  general long massless supermultiplet $[j_1,j_2]$ of % the $\mc N=8$ $d=5$
%AdS super algebra
 $PSU(2,2|4)$. } %all states have $\Delta=2+j_{1}+j_{2}$. } %All fields are massless:  $\Delta=2+j_{1}+j_{2}$.}
\label{T4}
\end{table}
\noindent

Using  \rf{5.6},\rf{5.7},\rf{5.8}   we find that   for all choices of the $j_1,j_2$ labels of the supermultiplet %$[j_1,j_2]$ choices here 
\be   E_{c} = {\rm a} = {\rm c} = 0 \ . \la{5199}
\ee
Thus in contrast to the case the massive $\N=1$ long multiplet in \rf{56} 
here  the total  Casimir energy  vanishes along with $\aa$ and $\cc$. This    is   another   manifestation of 
the relation  \rf{1.13},\rf{517}  valid for $\N \geq 3$. 

%Using our expressions for the Casimir energy and the anomalies, we verify that 
%$E_{c} = {\rm a} = {\rm c} = 0$ for all $j_{1},j_{2}$.

%%%%%%%%%%%%%%%%%%%%%%%%%%%%%%%%%%%%%%

\section{Applications to AdS/CFT}
\label{sec:applications}

Let  us now apply the  general  expressions  for $(E_{c}, \ra, \rc)$   to specific examples of AdS/CFT duality.
This   will require summation 
of contributions of   infinite   collections 
 of 5d fields  (in the above discussion of supermultiplets the sets of fields were finite), and thus a choice of a regularization that 
 should be consistent with symmetries of   the underlying theory.  

  %%%%%%%%%%%%%%%%%%%%%%%%%%%%%%%%%%%%%%%%%%%%%%%%%%%%%%%%%%%%
\subsection{Adjoint AdS$_5$/CFT$_4$}

Let us start with the canonical example of  the duality between 
 type IIB superstring on AdS$_{5}\times $S$^{5}$   and    $\mc N=4$ $SU(N)$ SYM theory \ci{Maldacena:1997re,Gubser:1998bc,Witten:1998qj}. 
 The  partition function of SYM theory  defined on a curved   4d background  $M^4$ 
  should match the  one of the superstring defined on a generalization of \ads  asymptotic to  $M^4$.
  This implies, in particular,  the  matching of conformal anomalies  and  Casimir  energies    computed on the two sides of the duality. 
  The direct perturbative comparison is possible due to the expected non-renormalization of these   quantities, 
  with the SYM side  giving 
  \be
\label{61}
K(\N=4\ {\rm SU(N)\ SYM})   = (N^{2} -1)\,\rk \  , \ \ \ \ \ \ \qquad K\equiv (E_c, a, c) \ , 
\ee
where $\rk = ({ 3 \ov 16}, {1\ov 4}, {1\ov 4} )$  are the  single $\N=4$   Maxwell multiplet entries   in Table \ref{T1}. 
  
 At the  leading $N^2$  order  (string tree level  or classical type IIB supergravity)  
 this  matching  was demonstrated in  \ci{Henningson:1998gx} (for the conformal anomalies) 
 and  in \ci{Balasubramanian:1999re}  (for the vacuum  energy). 
 To   consider    the next -- string one-loop   order   it is natural to assume  
  that the contributions  of loops  of  all   massive string modes  should vanish. 
   %cancel out.  
   Indeed, string modes  form long  massive 
   $PSU(2,2|4)$ multiplets\foot{KK descendants of massive string excitations sit in long multiplets 
given by tensoring string primaries with the long Konishi multiplet  %on AdS$_{5}$ 
\cite{Bianchi:2003wx}.} 
  and  thus should give zero contribution 
 (cf. section 5).
%Thus, they give zero contribution
%to $E_{c}, \ra, \rc$ and do not contribute in this context.
Equivalently,      string mode masses  depend on 't Hooft coupling  ($m^2 \sim \alpha'{}^{-1}  \sim {\sqrt \lambda}$)
 and thus a non-trivial contribution from them would   contradict  the expectred non-renormalization of \rf{61}. 
 
 Assuming this,  the  subleading 
  $O(N^0)$  term in \rf{61} should be reproduced just  by the  loop of massless string modes, i.e.   by  the 
 one-loop  correction in 10d  type IIB supergravity 
   compactified on $S^5$. The latter  is   given 
     by the sum of the  contributions  of the massless  $\N=8$  $5d$ supergravity multiplet and an infinite tower of 
   massive KK  multiplets \ci{Kim:1985ez}. Thus  for consistency with \rf{61} 
     one should  find   that 
 \be
\label{62}
\mbox{one-loop 10d   IIB  supergravity on $S^5$:}\quad \quad  \ E^+_{c} = -\frac{3}{16}, \quad \ra^+ = -\frac{1}{4}, \quad \rc^+ = -\frac{1}{4}\ . 
\ee
Here we put superscript $+$ as we  are   interested in direct  contributions  
of 5d fields   with standard (``Dirichlet'')   boundary conditions
given by 
 \be   K^+=(E^+_{c}, \ra^+, \rc^+)= -\ha (E_{c}, \ra, \rc)  \ , \la{662} \ee
    in terms of the corresponding elementary 4d conformal field 
  values quoted in section 5.1.  Eq.\rf{62}   may  be written also as (cf. \rf{520})
  \be 
  K^+ ({\rm 10d  \ IIB  \ SG\ on \ S^5})\ \  = \ \  - K(\N=4 \ {\rm Maxwell}) \ .  \la{6622}\ee 
 %Comparing with  supergravity, we expect $-k_{I}$ to be equal to the 1-loop contribution. So, we should find 
%in agreement with (\ref{1.13}) due to cancellation of $D^2R$ terms in conformal anomaly of $\N=4$ SYM.
This  matching of both $\ra$ and $\rc$  coefficients at the one-loop supergravity level
was  earlier  claimed    in   \cite{Mansfield:2000zw,Mansfield:2003gs}.
In particular,  using \rf{4.1}  motivated by the prescription of  \ci{Mansfield:2003gs}, 
 the vanishing of the type IIB supergravity contribution to 
  $\rc-\ra$  implied by \rf{62} was  interpreted in  
\cite{Ardehali:2013xya,Ardehali:2013gra} as a consequence   of  the vanishing  of the  contributions of 
each of the long  KK  multiplet   and  the  separate cancellation of the  $\cc-\aa$ contributions  from states in 
 the massless multiplet.\foot{Similar pattern applies to matching of  axial anomalies  \ci{Bilal:1999ph}.}
Reproducing  the explicit value of $\aa$ in \rf{62} is  much  more  non-trivial, requiring a specific  choice  of a
regularization of  the sum over the  infinite number of  KK modes.
While our final conclusion is the same as in 
\cite{Mansfield:2003gs}  the intermediate steps of the derivation disagree. 

Starting with  our  general  expressions  for $E_{c}, \ra, \rc$    given  in section 5.1       we  shall  
explicitly  demonstrate the validity of  (\ref{62}) or \rf{6622}. 
 The  proportionality \rf{1.13}  of $E_c$ and $\aa$-coefficient  
is expected due to the maximal supersymmetry,   implying, in particular,  that 
 $E_c$ (i.e. the  \ads  vacuum energy)
  does not vanish  in the one-loop type IIB  supergravity compactified on $S^5$. 
This is different from  the vanishing of the vacuum energy 
in $\N>4 $  gauged   supergravities in 4  dimensions  \ci{Allen:1983an} and 
in also in 11d supergravity compactified on $S^{7}$ \cite{Duff:1982ev,Gibbons:1984dg,Inami:1984vp}.\foot{A possible  
way to reconcile these  different  conclusions 
   from  the AdS/CFT point of view  is to note that Casimir energy  
   should automatically vanish in the case of 3d boundary theory, but need not in the 4d case
(see also  below).}
The non-vanishing of the vacuum  energy in the pure $\N=8$  5d supergravity  was already noted in 
\cite{Gibbons:2006ij}  but the inclusion of the contribution of the KK  multiplets  leading to the value of $E_c$ in \rf{62} is a  new result.

%   For  notational   simplicity we shall omit the  $+$ superscripts  on the one-loop 
%   supergravity field contributions to $(E_c, \aa,\cc)$  in  the rest of this 
%     subsection. 

%This is no longer true for 10d supergravity on S$^{5}$ as expected by the above AdS/CFT argument.
%The explicit proof that in 5d we have $E_{c} = -\frac{3}{16}$ is thus a new test of Maldacena's 
%correspondence.
 % without  using any information about the special properties of $\rc-\ra$, and also to prove that the prediction for the Casimir 
%energy is indeed what expected. We remark that the Casimir energy for 
%Due to the  maximal supersymmetry  the Casimir energy is expected to be proportional to $\ra$
%as in  \rf{1.12}. 
%  due to the absence of total derivative  $D^2 R$ terms
% in the full expression for the trace anomaly \cite{Herzog:2013ed,Huang:2013lhw}. 
 %Thus, the prediction in (\ref{5.2}) for the Casimir energy is also  expected to be correct. 

The  $PSU(2,2|4)$  multiplet   content of 10d supergravity compactified on $S^5$ is recalled in Table \ref{T5} in Appendix \ref{A:KK} 
(where $p$ is  KK level). 
The degeneracies, i.e. the dimensions of the corresponding $SU(4)$  representations   can be found using \rf{C.1}. 
  Summing up the elementary 5d field  contributions  using   \rf{5.6}--\rf{5.8} and \rf{662} 
  we find for 
%Moving to the details of the calculation, we consider the fields of 5d supergravity compactified on S$^{5}$, 
%recalled in Table \ref{T5}. 
%We find the following expressions for the total contribution from level $p$
%multiplets, using (\ref{2.32}), (\ref{3.3}), and (\ref{4.10}), and the dimension formula
%(\ref{C.1}). First we have the 
 the  massless  $p=2$ supermultiplet   in Table \ref{T5}\foot{The value of $E_c$ is the same as  found    in \cite{Gibbons:2006ij}.}
\be
\label{63}
\begin{split}
p=2:  \qquad   E_{c}=\frac{3}{8}, \qquad\ \,   \ra =\frac{1}{2}, \qquad \rc = \frac{1}{2} \ .
\end{split}
\ee
The  $p=2$   multiplet   corresponding   to the states of    pure $\N=8$  5d  gauged supergravity  
 is isomorphic  to the $\N=4$  4d conformal supergravity multiplet.\foot{The full  set of states of  
 10d supergravity compactified on $S^5$ will then correspond in 4d  to $\N=4$  conformal supergravity
 coupled to  infinite collection of conformal fields   with canonical dimensions $\De_-=4-\De$ corresponding to massive $p \ge 3$ 
  states in 5d   spectrum in  Table \ref{T5}.}
 The   corresponding values  for  the conformal anomaly  and $E_c$ 
should  thus  be  related  as in \rf{662}:   indeed,  -2 times  the values in  \rf{63}   gives   the values in the  
 last line  of Table \ref{T3}, i.e.  we get  the  expression given   above  in  \rf{519}. 
The equivalent form of \rf{63} was given  in \rf{520}. 

% Multiplying  (\ref{5.3}) times $-2$ gives  last line  of Table \ref{T3}.
%This is because the 5d multiplet content  is same, 
%CSG  fields are boundary sources for  fields of $d=5$  supergravity
%with AdS$_{5}$ vacuum. The Casimir energy from the supergraviton multiplet $p=2$ is  in agreement with  and differs from (\ref{5.2}), the contribution %from KK states being non vanishing. 

For   both $p=3$ and $p\geq 4  $   massive KK multiplets in  Table \ref{T5}  we  obtain 
\be
\label{64}
\begin{split}
%p=2:  & \qquad E_{c}=\frac{3}{8}, \qquad\ \,   \ra =\frac{1}{2}, \qquad \rc = \frac{1}{2}, \\
%p=3: & \qquad E_{c}=\frac{9}{16}\ , \qquad  \ra =\frac{3}{4}\ , \qquad \rc = \frac{3}{4}\ ,\\
p\ge 3: & \qquad E_{c}=\frac{3p}{16} \ , \qquad \ra =\frac{p}{4}\, , \qquad \rc = \frac{p}{4}\ .
\end{split}
\ee
Remarkably,    despite  the   different  structure  of the $p=2,\,  p=3 $ and $p \geq 4$  multiplets  in Table \ref{T5}, 
their contributions to $K=(E_c,\aa,\cc)$  are thus universally described by\foot{Note, in particular, that the
 relation \rf{1.13} or \rf{517} 
applies level by level, i.e. for each $\N=4$ supermultiplet.} 
 \be 
  K^+ ({\rm  KK \  level} \ p \  {of \  10d  \ IIB  \ SG\ on \ S^5})\ \  = \  \  p\,  K(\N=4 \ {\rm Maxwell}) \  , \ \ \ \  \   p=2,3,4, ... \quad  \la{66}
  \ee 
 As the $p=1$ level  may  be interpreted as the $\N=4$ superdoubleton multiplet, this relation formally applies also  for $p=1$, becoming \rf{5177}.
  For $p=2$   eq.\rf{66}    is equivalent to \rf{520}, while for $p >2$ to \rf{64}. 
A  natural  interpretation of this  non-trivial identity  (which relies on the  particular values  of $E_c,\aa, \cc$  we used)\foot{For example, 
this  relation would not be true for the $\cc$-coefficient    had we used \rf{4.2} instead of \rf{4.11}. 
The expressions 
 for the  contributions  of each  $p$ level  to $\aa$ and $\cc$  coefficients 
  found    in  \cite{Mansfield:2003gs}  were  very different: 
 they were  not linear in $p$ but polynomials of order  5. 
 The reason  for this  was   that the expressions for the  individual  5d field  contributions to 
 $\aa$  and $\cc$ used  there (cf.  \rf{1.3})  were   linear in $\De-2$  and thus linear in $p$ (cf. Table \ref{T5}), 
 while  the  higher powers in $p$    were  coming from the  multiplicities given by the 
 dimensions \rf{C.1} of the corresponding $SU(4)$ 
 representations.  The   correct  expressions    for $E_c,  \aa$   and $\cc $   found here 
 are  instead  5th order polynomials in $\De -2$  (and thus in $p$, for the states in   Table \ref{T5}), 
 but, remarkably,  the non-linearity in $p$  cancels  out   after   multiplying by the  dimensions of $SU(4)$ 
 representations and summing over the members of each  supermultiplet.}
 is that it expresses the fact   that the 5d  states  at  the   KK level $p$   appear in the tensor 
 product of  $p$ copies of   $\N=4$ superdoubleton \ci{Gunaydin:1998jc}. 
%\be  \aa=\cc= {4\ov 3} E_c= {p\ov 4} \ , \ \ \   \qquad    p=2,3,4, ...\ ,  \la{66}  \ee
 %\rf{63}  may be formally viewed  as the  special  
 %i.e.  each  level $p$ contributes  exactly as  $p$  copies of  $\N=4$  Maxwell multiplets.\foot{This   may be related
 %to  the  fact   that the supergravity  states on level $p$   appear in the tensor 
% product of $p$   $\N=4$ superdoubletons \ci{Gunaydin:1998jc} (with $p=1$   being   Maxwell multiplet itself). 

%As  follows from \rf{63},\rf{64}  that for each  supermultiplet we already have the relation  \rf{1.13} or  $\aa=\cc= {4\ov 3} E_c$ satisfied. 
 
 It remains  to sum up the   supermultiplet contributions \rf{66} over the KK level $p$,
i.e. to  assign a consistent value to the  divergent sum $\sum_{p=2}^\infty p$.
The prescription that is  required to reproduce \rf{6622} is 
\be
\sum_{p=1}^\infty  p =0   \ , \qquad \ {\rm i.e.} \qquad
\sum_{p=2}^\infty  p =  -1 \ .   \la{67} \ee
This can be interpreted as follows. 
As was  noted  above,  the $p=1$   case of \rf{66}  is the same as the contribution   of  one  $\N=4$   Maxwell  multiplet
\rf{517} or  superdoubleton.  The contribution of  the  $p=1$  superdoubleton  should  not to be
included \ci{Gunaydin:1984fk}    in  the  list of physical   multiplets  in
Table \ref{T5} as it is gauged away \ci{Kim:1985ez}   but  if we would formally include it 
 then under \rf{67}  the total 10d  supergravity contribution 
would vanish.\foot{Adding the $p=1$ superdoubleton contribution
would   be equivalent to adding the decoupled $U(1)$  D3-brane
contribution, i.e.  the same as   replacing $SU(N)$ by 
$U(N)$  group on the dual SYM side and thus dropping -1
term in \rf{61}.  An alternative interpretation  might  
be  in terms of  an  effective bulk+boundary anomaly cancellation (conformal anomaly analog of ``anomaly inflow''). 
 That would also formally  imply  the  cancellation 
  of the total   AdS$_5 \times$ S$^5$ vacuum energy  as in %which is known to happen 
 in the   AdS$_4 \times$ S$^7$ case. %  as was already mentioned above.
 }
The  condition  \rf{67}  is  satisfied if 
 one defines the  sum over KK level $p$  with a sharp cutoff
and then drops all  cutoff-dependent terms.\foot{Explicitly, one   has $\sum_{p=1}^P
p = \ha P^2 + \ha P  \to 0 $.
The same  sharp cutoff  regularization of the sum over KK level was assumed  in \cite{Mansfield:2003gs}. 
In such a  regularization all sums $\sum_{p=1}^\infty  p^n$  with
positive integer $n$ are  just set to zero.
This formally   explains  why a different   expression for the
summand in \cite{Mansfield:2003gs} still 
led to the same  correct expression  for the result in \rf{62}.}

While the prescription \rf{67}  may look  artificial  (e.g., it is not the  ubiquitous  Riemann
 $\zeta$-function  rule)  %(e.g., not a standard $\zeta$-function one)
it is possible, in fact,    to justify it  by  starting with 
 the  standard  spectral    $\zeta$-function regularization.
The  key point  is that a    regularization  consistent with
symmetries of the theory 
should be  applied directly  at the 10d   rather than 5d level, i.e.  it should  be based on 
the spectrum of the original 10d  differential 
operators defined on  AdS$_5 \times$ S$^5$ or its generalization.

Let us demonstrate   this on the example  of the  sum of  $E_c$   contributions.
%Notice that, at each level,
%we already match (\ref{1.13}).  The prediction (\ref{5.2}) can be
%recovered after summing over all supermultiplets. This is divergent
%and require a regularisation. We adopt a spectral regularisation that
%we now explain in details. Notice that
%in this section we denote by $E_{c}$ the quantity $E_{c}^{+}$ in  (\ref{2.31})
%since we are not interested in the associated CHS theory.
% To perform the sum over  $p$  requires a choice of regularization.
The   expression for  the  contribution of a massive
$(\Delta;\, j_{1},j_{2})$  5d field to the vacuum energy $E_c$ 
can be  obtained  
 from the  partition function   (\ref{2.10})  which
may be written as
 %. Writing the partition function in the form
\be\la{651}
\widehat{\Z}{(\Delta;\, j_{1}, j_{2})} =  d(j_1,j_2)
\sum_{k=0}^{\infty}\te \binom{k+3}{3}\,q^{\Delta+k} \ .
\ee
Then (\ref{2.29}) implies that a  formal (divergent) expression for
$E_c$  is given by
%we obtain the Casimir energy as the divergent sum, see (\ref{2.29}),
\ba
\label{661}
&\widehat E_{c}(\Delta;\; j_{1}, j_{2}) = \sum_{k=0}^{\infty}  e_k
(\Delta;\; j_{1}, j_{2})   \ , \ \ \ \ \ \ \ \\
&e_k (\Delta;\; j_{1}, j_{2}) =
\frac{1}{2}\,(-1)^{2\,(j_{1}+j_{2})}\, d(j_1,j_2) \,
\te \binom{k+3}{3}\,(\Delta+k)\  .\la{677}
\end{align}
This  sum can be   computed   using  the  $\zeta$-function  prescription  \rf{2.29} applied   to
the full effective energy  eigenvalue $\Delta+k$,
or, equivalently,  by  introducing  an  exponential cutoff
%We regularise it by introducing the function
\be  e_k \ \  \to \ \  e_k \, e^{- \ep (\De +k)} \ , \la{6777} \ee
doing the sum, expanding in $\ep \to 0$ and finally dropping all singular terms.
Keeping $\ep$ finite we  may  find  the   contribution  to the sum \rf{661} from   all 
  states  of the $p\ge 4$    massive KK multiplet in Table \ref{T5}. This 
gives    the total  summand $e_k(p;\ep)$.
%To deal with the double summation over $k$ in (\ref{5.7}),
%and over the KK level $p$, we evaluate the full sum of $ \widehat
%e_{c}(\Delta;\; j_{1}, j_{2}; t)$ over the states in
%a $p\ge 4$ multiplet. This gives a (long expression for ) total
%$\widehat e_{c}(p,k; t)$.
Summing   over  both $k$ and $p$ we obtain 
%This can be summed over  $k\ge 0$ and $p\ge 4$ because of the
%exponential damping factor in (\ref{5.7}). The result is
\be\la{688}
\begin{split}
\sum_{k=0}^{\infty}\sum_{p=4}^{\infty} e_{k}(p; \ep) &=\te 
\frac{e^{-2 \ep} \left(95 e^\ep+120 e^{3 \ep/2}-220 e^{2 \ep}-420 e^{5 \ep/2}+50
  e^{3 \ep}+420 e^{7 \ep/2}+210 e^{4 \ep}-6\right)}{\left(e^{\ep/2}-1\right)^2
  \left(e^{\ep/2}+1\right)^{10}}\\
  & = \frac{249}{256\,\ep^{2}}-\frac{9}{8}+\mc O(\ep^{2})\ .
  \end{split}
\ee
Keeping only  the   finite part   and adding the   contributions of  the $p=2$ and $p=3$  multiplets
 in \rf{63},\rf{64} gives   finally for the total 10d supergravity contribution 
\be\la{68}
E^+_{c} = \frac{3}{8}+\frac{9}{16} -\frac{9}{8} = -\frac{3}{16}  \ .
\ee
This   is in agreement with \rf{62} and  thus confirms  the  prescription in \rf{67}.

\iffa 
In addition to matching   the values of $(E_c, \aa, \cc)$ in \rf{61},\rf{6622}
  one can also match (as we describe in Appendix \ref{A:zz}) the corresponding 
 ``twisted'' partition functions defined  on $S^1 \times S^3$ with fermions  being periodic  rather than anti-periodic 
 in euclidean time.
 Here the supersymmetry is preserved  even in presence of a non-trivial parameter $q= e^{-\beta}$;  
  that  should  be again  the reason for  non-renormalization of this quantity and thus  matching
  between   one-loop SYM and 10d supergravity. 
 At the same time, the standard  thermodynamic partition function   is known to  receive  non-trivial  string corrections 
 and cannot be matched  ignoring string modes.\foot{An interesting question for the future is 
  to try to  compare it on both sides   in the $g_{\rm YM} \to 0$  or  tensionless string limit,   accounting for an
   infinite number of contributions of  massless   higher spin fields.}
\fi 

\newpage
%%%%%%%%%%%%%%%%%%%%%%%%%%%%%%%%%%
\subsection{Vectorial AdS$_5$/CFT$_4$}
In the case of vectorial    AdS$_{d+1}$/CFT$_{d}$  correspondence  one considers $N$ free   fields % scalars  or fermions 
transforming in a   vector (fundamental) representation of $U(N)$  or $O(N)$. The  restriction to the singlet sector 
of  bilinear  conserved higher   spin   current  operators  implies  duality  to massless   higher spin 
fields  in AdS$_{d+1}$   described by  Vasiliev-type theories (see, e.g.,  \ci{Vasiliev:2001zy,Vasiliev:2003ev,Bekaert:2005vh,Didenko:2014dwa}).
%\foot{Higher-spin symmetries of  such 
%  boundary models   and  their higher-spin and  super-extensions where found in \ci{vasiliev:2001}.}
The coefficient in front of the classical action in    AdS$_{d+1}$ is proportional to 
 $ N$,   with  the cubic  and higher  amplitudes     supposed to match  free-theory correlators of conserved currents at the boundary
 in $1/N$ expansion. 
   %  at leading order in large $N$. 
   
The original examples were for $d=3$ \cite{Klebanov:2002ja,Sezgin:2003pt,Leigh:2003gk,Giombi:2009wh}  while  
  generalizations to  $d>3$  
%v3
   were   studied  in \ci{Didenko:2012vh,Bekaert:2013zya,
   Giombi:2013yva,Giombi:2013fka, Giombi:2014iua, Giombi:2014yra}
 (see also \ci{Konstein:2000bi,Mikhailov:2002bp,Schnitzer:2003zr,Bekaert:2012vt} for   related  work). 
In  $d=3$    one may build conserved higher spin  currents as bilinears  of  free scalars or  spin $\ha$  fermions  and then get the spectrum 
of dual massless  higher  spin  theories in  AdS$_4$
containing totally symmetric tensors 
 (these are the only options  to get a consistent 3d theory  with higher-spin symmetry 
   under natural assumptions \ci{Maldacena:2011jn}).  

%%%%%%%%%%%%%%%%%%%%%%%%%%%%%%%%%%%%%%%%%%%%%%
%Here we will   be interested in  the $d=4$  case, i.e.   AdS$_{5}$/CFT$_{4}$  vectorial duality. 

In $d=4$  case we will be interested in here
 in the  free  fermion case there is a  new feature: the corresponding conserved  currents  belong to 
  particular mixed-symmetry representations of $SO(4)$  \ci{Vasiliev:2004cm,Dolan:2005wy,Giombi:2014yra}.
Another novelty of the $d=4$ case 
  is that   here one  can also  use  spin 1   fields\foot{In $d=3$   Maxwell  vector is dual to a scalar.} 
   as building blocks  for  higher spin conserved currents 
 (free spin $0, \ha, 1 $  are    the only  options to get a 4d theory with a  higher spin symmetry
    if one assumes   unitarity  \ci{Boulanger:2013zza,Stanev:2012nq,
 Stanev:2013qra,Alba:2013yda}).\foot{In
 %v3
  priniciple, one can also  explore the  possibility of defining the boundary theory 
  in terms of higher spin   singletons  
 which are unitary and conformal  when described in terms of field strengths. 
  This possibility was noticed
in \ci{Bekaert:2009fg} where the corresponding  higher spin algebras were studied.}  
 The   corresponding conserved currents  appearing    in the  product of two   spin $1$   doubletons 
 %v2
   as in \ci{Gunaydin:1998km,Ferrara:1998jm,Gunaydin:1998sw,Sundborg:2000wp,
   Sezgin:2001zs,Vasiliev:2004cm,Dolan:2005wy, Boulanger:2011se}
     are also in specific mixed-symmetry representations of $SO(2,4)$. 
   The singlet sector of a theory of $N$  Maxwell  vectors   should then   be dual to a particular   version of 
    higher spin theory in \ads involving mixed-symmetry fields  which should exist  but  was not studied 
    detail so far (we shall call it ``type C'' theory).\foot{Interacting higher spin  theory for totally symmetric fields in \ads   was  considered
      in \ci{Vasiliev:2001wa,Alkalaev:2002rq}. 
     Mixed-symmetry fields  in \ads  and the associated currents  were  discussed   in 
    \ci{Metsaev:1995re,Metsaev:2003cu,Alkalaev:2003qv,
    Alkalaev:2006rw,Alkalaev:2012ic,Alkalaev:2012rg,Metsaev:2013wza}.
Cubic interactions  of mixed-symmetry  higher spin    fields  in flat  space  were studied  in  \ci{Metsaev:2005ar}  and in  \ads they 
    were  considered  in  \ci{Alkalaev:2010af,Boulanger:2011se,Boulanger:2011qt,Boulanger:2013zza,Lopez:2012pr,Joung:2012hz}.  
    The   question of  consistency   of an interacting \ads   theory  involving mixed-symmetry fields 
     goes   beyond the cubic order  and requires, in particular,     the  closure  of  the
       symmetry algebra  \ci{Boulanger:2013zza}.
     Unitarity imposes additional constraints, excluding, e.g., partially massless   fields.}

The   field content  of  the  corresponding dual  pairs     is summarized in Table \ref{T8}  where  we
 use the notation $(\Delta;\,j_{1},j_{2})_c \equiv (\Delta;\,j_{1},j_{2})+(\Delta;\,j_{2},j_{1})$. 
 %%%%%%%%%%%%%%%%%%%%%%%%%%%%%%%
\iffa
The  AdS/CFT correspondence  
can also be formulated in the so-called vectorial case. The gravitational bulk theory is then dual to a conformal 
theory whose dynamical fields 
transform in the fundamental (instead of adjoint) representation of the $U(N)$ or $O(N)$ symmetry group \cite{Klebanov:2002ja}. Supersymmetry is no more necessary, but it is required that there is an infinite tower of conserved higher spin currents that are $U(N)$ or $O(N)$ singlets. The dual gravitational theories in AdS  
contain the corresponding tower of massless higher-spin gauge fields \cite{Sundborg:2000wp}. Natural candidates
for these dualities are the theories extensively explored by Vasiliev and 
others \cite{Fradkin:1987ks,Vasiliev:1990en,Vasiliev:1992av,Vasiliev:1995dn,Prokushkin:1998bq,Vasiliev:1999ba,Vasiliev:2003ev,Bekaert:2005vh}.
 In particular, one can consider the  dualities summarised in the following
 Table %(see \cite{Didenko:2012vh,Giombi:2014iua} for results in general dimension) 
 \fi
 %%%%%%%%%%%%%%%%%%%%%%%%%%%%%%%%%%%%%
\begin{table}%[H]
\be
\begin{array}{|c|c|}
%%%%%%%%%%%%%%%%%%%%%%%%%%%%%%%%%%%%%%%%%%%%%%%%
\hline
\mbox{AdS$_{5}$} & \mbox{CFT$_{4}$}\, \mbox{(singlet sector)} \\
\hline
\mbox{non-minimal type A theory} & N \mbox{ complex scalars}:   U(N) %\ \mbox{vector }
  \\
(2; 0,0) +   \bigoplus_{s=1}^{\infty} (2+s; \frac{s}{2},\frac{s}{2}) & \\
%%%%%%%%%%%%%%%%%%%%%%%%%%%%%%%%%%%%%%%%%%%%%%
\hline
\mbox{minimal type A theory  } & N \ \mbox{real scalars}:   O(N) %\ \mbox{vector }  
  \\
(2; 0,0) +   \bigoplus_{s=2, 4,  \dots }^{\infty} (2+s; \frac{s}{2},\frac{s}{2}) & \\
%%%%%%%%%%%%%%%%%%%%%%%%%%%%%%%%%%%%%%%%
\hline
\mbox{non-minimal  type B theory  } &  \\
2\,(3;0,0) +    %\oplus
 & N  \mbox{ Dirac fermions}:    U(N)\  %\mbox{vector }   
   \\
2\bigoplus_{s=1}^{\infty} (2+s; \frac{s}{2},\frac{s}{2})+
\bigoplus_{s=1}^{\infty} (2+s; \frac{s+1}{2},\frac{s-1}{2})_c & \\
%%%%%%%%%%%%%%%%%%%%%%%%%%%%%%%%%%%%%%%%%%%%
\hline
\mbox{minimal  type B theory  } &  \\
2\,(3;0,0)+%\oplus
 &N  \mbox{ Majorana fermions}:    O(N)  %\ \mbox{vector }    
   \\
\bigoplus_{s=1}^{\infty} (2+s; \frac{s}{2},\frac{s}{2})+
\bigoplus_{s=2, 4,  \dots}^{\infty}  (2+s; \frac{s+1}{2},\frac{s-1}{2})_c
&\\
%_c 
%%%%%%%%%%%%%%%%%%%%%%%%%%%%%%%%%%%%%%%%%%%%%%
\hline
\mbox{non-minimal  type C theory  } &  \\
2\,(4;0,0)+ (4;1,0)_c  + %\oplus
 &N  \mbox{\ complex Maxwell  vectors}:    U(N) %\ \mbox{vector }    
   \\
2 \bigoplus_{s=2}^{\infty} (2+s; \frac{s}{2},\frac{s}{2})+
\bigoplus_{s=2}^{\infty}  (2+s; \frac{s+2}{2},\frac{s-2}{2})_c
 &\\
%_c 
%%%%%%%%%%%%%%%%%%%%%%%%%%%%%%%%%%%%%%%%%%%%%%
\hline
\mbox{minimal  type C theory  } &  \\
2\,(4;0,0)+%\oplus
 &N  \mbox{\ real Maxwell  vectors}:    O(N) %\ \mbox{vector }    
   \\
 \bigoplus_{s=2}^{\infty} (2+s; \frac{s}{2},\frac{s}{2})+
\bigoplus_{s=2, 4,  \dots}^{\infty}  (2+s; \frac{s+2}{2},\frac{s-2}{2})_c
%_c 
%\hline
& \\
\hline
\end{array}
\nonumber
\ee
\caption{Field content  of  vectorial  AdS$_{5}$/CFT$_{4}$  dualities.} \label{T8}
\end{table}
%%%%%%%%%%%%%%%%%%%%%%%%%%%%%%%%
 Higher spin theory content  matches  the  list of   bilinear     conserved currents  in the  boundary theory. 
 It   can be obtained   by taking the product of two doubleton   representations  corresponding to the boundary fields
 (see Appendix \ref{A:FF}).  In addition to conserved currents 
 there are  also  scalar  bilinears dual to $(2;0,0)$ \ads scalars in type A  theory (see \rf{a5}) 
 and fermion  bilinears  dual to  $(3;0,0)$    \ads  scalar and
  pseudoscalar  in type B  theory (see \rf{a6}).\foot{The  massless $(2;0,0)$ scalars  
 having    $\De-2=0$ will not contribute to the quantities $K=(E_c, \aa, \cc)$  discussed below.} 
 Type A theories  contain   symmetric  tensors    while type B and type C   theories  include also  particular 
 mixed-symmetry representations of massless   higher spin fields in AdS$_5$. 
 %v3
 The second series  of massless  $(2+s; \frac{s}{2},\frac{s}{2})$ fields in $U(N)$  type B and type C 
 theories   are  parity-odd.
 Restriction to real  fields at the boundary   implies projecting out  some  (odd-spin parity even and even-spin parity odd) 
    fields in the bulk  that either vanish or become total derivatives 
 (see \ci{Giombi:2014yra} for discussion of the minimal type B theory case).
 
 Since the content of   type C   theory    dual to   (complex or real) 
 4d  Maxwell fields  
 % does not appear  to be 
 was not explicitly  studied   in the  literature  let us   comment  on  it in some detail.
 It   can be  obtained by taking the product  of two  spin 1 doubletons as in \rf{a7}.\foot{Related  discussions   appeared in 
    \ci{Ferrara:1997dh,Ferrara:1998jm,Anselmi:1999bb}; see also       \ci{Gelfond:2006be} for a
    general construction of higher spin currents as bilinears in higher spin fields in flat space.} 
  In the complex Maxwell field case  the  tower of relevant operators  starts with 
   dimension 4 operators    appearing in the decomposition of 
  $F^*_{\m\n} F_{\k\r} $ into $SO(4)$    irreps:\footnote{Here  *  is  complex conjugation,  tilde  denotes dual tensor 
    and  we suppress  $U(N)$ vector index. }
    
    \noindent 
   (i)  scalar  $F^*_{\m\n} F^{\m\n}$
  and  pseudoscalar $F^*_{\m\n} \tilde F^{\m\n}$  in massive  representation $(4;0,0)$; 
  
  \noindent
  (ii)  antisymmetric tensor $F^*_{\m[\n} F_{\k]\m}$  which is  not conserved  on shell 
   and corresponds to   {massive}  selfdual + anti-selfdual rank 2  tensor, i.e.  representation $(4; 1,0)_c = (4; 1,0)+ (4; 0,1)$;\foot{The corresponding 
   antisymmetric tensor field in \ads  appears, e.g., 
     in $S^5$   compactification of type IIB supergravity 
   and was  discussed in \ci{Kim:1985ez,Arutyunov:1998xt}. 
   Its \ads Lagrangian  has first-derivative  topological kinetic term  plus the standard  mass term.}    
   
   \noindent 
   (iii)   spin 2  conserved  stress tensor $(4; 1, 1)$   and its parity-odd  counterpart   with  one $F_{\m\n}$ replaced by $ \tilde F_{\m\n}$; 
   %%  how come ???? second graviton????  wrong parity ???? 
   
   \noindent 
   (iv)  conserved  current   with symmetries of Weyl tensor, i.e. the massless state $(4; 2,0)_c$
   described  by  the Young tableu with 2 rows and 2 columns. 
   %%%%%%%%%%%%%%%%%%%%%%%%%%%%%
 \iffa  In general, the product of two  doubletons $(j,0) \times (0,j')$  contains  the sum of massless  representations  labelled by $k$ 
 with    Young tableu labels $h_1= j + j' + k, \ h_2 = j-j'$      \ci{Boulanger:2011se} , 
    or  in our   $SU(2) \times SU(2)$  notation 
 $(2+ j  + j' + k ;  \ha  k +  j   ,  \ha k + j')$. 
 In addition,  $(j,0) \times (j',0)$    gives  mixed-symmetry 
 representations with $h_1= j + j' + k, \ h_2 = j+j'$, i.e. 
 $(2+j  + j' + k ;  \ha k + j + j'   ,  \ha k  )$. 
 \fi 
 %%%%%%%%%%%%%%%
 
 In addition,   the product  \rf{a7}  
 of two spin 1 doubletons $\big(  \{1, 0\}  +  \{0, 1\}   \big)  \otimes   \big(  \{1, 0\}  +  \{0, 1\}   \big)$
  (where $\{1, 0\}$ and $ \{0, 1\} $   correspond to  selfdual and antiselfdual 
 parts of $F_{\m\n}$) contains also  higher spin conserved currents dual to  massless  \ads fields. 
 %$ (0,1)  \times (1, 0)$  and $ (1,0) \times (1,0)$ 
 %$(0,1) \times (1,0) $ and $(1,0) \times (1,0) $  give two series  of representations: 
% $(4+k, \frac{k+2}{2},\frac{k+2}{2})$, $k=0, 1, 2, ...$   and 
  % (starting with spin 2  current which is stress tensor)   and 
 % $(4+k, \frac{k+4}{2},\frac{k}{2})$, $k= 0, 1,2,  ... $  %(starting with spin 2 current   with symmetries of Weyl tensor). 
  %There is also a  conjugate pair. 
 % This will be   the complex $U(N)$ vector case.
 The  real  vector case (minimal type C theory)    is found  by  a projection   similar to the one in type B theory case: 
   %Then  the resulting  spectrum given in  Table \ref{T8}. 
   removing  one set  (parity-odd)  of symmetric tensor states and 
   odd-spin  mixed-symmetry states. This  results in the spectrum given in Table \ref{T8}. 
   
   %%%%%%%%%%%%%%%%%%%%%%%%%%%%%%%%%%%%%%%%%%%%%%%%%%%%%
 
The AdS/CFT duality implies  the equality  of the corresponding partition functions. For example,  the singlet-sector partition function  $Z_{\rm CFT}$
of $U(N)$   conformal scalar defined on a curved  space $M^4$  should be equal to the quantum 
partition function  $Z_{\rm  HS}$  of the corresponding  higher spin     theory  with coupling constant  $ N^{-1}$
defined  on an  \ads type  Einstein   space   which is  asymptotic to $M^4$ boundary. 
If  $M^4$   has no non-trivial holonomies  $\log Z_{\rm CFT}$  should  be given  just by the free-theory  one-loop 
contribution.\foot{The singlet   constraint   may be imposed by integrating over an auxiliary  pure-gauge 
 vector   field   gauging the $U(N)$  or $O(N)$  global   symmetry. This constraint  does not change the leading 
 order $N$ term in the partition function, i.e. is not relevant for 
 computing   vacuum energy  and conformal anomaly   coefficients, 
 but  in presence of non-trivial holonomy like in $S^1 \times S^3$ case  it leads to  an additional 
 $O(N^0)$ contribution to the  non-trivial $\b$-dependent part of the 
 partition function (see   \ci{Sundborg:1999ue,Aharony:2003sx,Shenker:2011zf,Giombi:2014yra}   and refs. there).
 Note that  the case of  adjoint-representation  vector fields (cf. \ci{Barabanschikov:2005ri} and also \ci{Lal:2012ax})   is different 
 from the vector-representation  one  we consider here.}
 It  should  match the leading classical term in $\log Z_{\rm HS}$ that should thus scale as $N$. 
 
As the  full non-linear classical    actions for    higher  spin  theories  in \ads  are  presently unknown, 
one is not able  to compare the  leading  large $N$ terms in the corresponding observables like  $(E_c, \aa, \cc)$.
Remarkably, it is still   possible   \cite{Giombi:2013fka}  to perform  %(as in $d=3$ case  \rf{Giombi:2013fka} ) 
non-trivial next-order   checks: as  $O(N^0)$  term  in  $ \log Z_{\rm CFT}$  is   absent in the free theory case, 
  the  
one-loop    contribution to $\ln Z_{\rm  HS}$ should   vanish too. 
%Due to the lack of information about the form of the classical action, in the higher-spin theories there is
% no known way to calculate the leading terms $\sim N$ in the three quantities $E_{c}, \ra, \rc$. Nevertheless, a non trivial test of 
%the above vector dualities is the calculation of the subleading term $\sim N^{0}$. 
This was  explicitly demonstrated  for the $\ra$-coefficient    of type A  theories  in  \cite{Giombi:2014iua}, 
and  for the Casimir energy  of type A and B  theories  in \cite{Giombi:2014yra}.

In the non-minimal type A and B   theories  where one   sums over all spins  one finds the vanishing 
results   for the  one-loop    corrections  to $\aa$-coefficient  (from $Z_{\rm  HS}$ on \ads   with  $S^4$ boundary)   and to 
$E_c$ (from $Z_{\rm  HS}$ on \ads  with  $\mathbb R\times S^{3}$ boundary).
%side gives a vanishing contribution, in agreement with the 
%expectation that there is no $\sim N^{0}$ correction in the free complex scalar  (or free Dirac fermion) theory
%on $S^{4}$ (for $\ra$) or on $\mathbb R\times S^{3}$ (for $E_{c}$). 
In the minimal theories  %, which include  even spins only, 
the one-loop  HS  correction  turns  out to be non-zero  and  equal to that of one  real  4d scalar (in the minimal type A case) 
and one Majorana  fermion (in the minimal type B case).
The proposed interpretation \cite{Giombi:2013fka}   of this  fact 
 is  that the bulk coupling constant in the minimal  HS theory 
is  not $N^{-1}$   but $(N - 1)^{-1}$, 
so that there  is an extra $O(N^0)$  contribution  that   comes from the    corresponding 
 $N-1$ coefficient of the 
tree-level term that  cancels the  non-zero one-loop HS correction.

As for  the $\rc$-coefficient, its    matching   was  not  attempted  so far  (apart from a   remark  in \cite{Giombi:2014iua} that  similar conclusions as for $\aa$-coefficient   may  
apply  in type A theory  if  one    uses  the   expression (\ref{4.4})   with the  special  ``finite'' 
choice   of $r_{b}=-1$   \ci{Tseytlin:2013jya}). 
  Neither $\aa$- nor $\cc$- coefficients  were  discussed  previously 
  in  type B  theories containing    mixed-symmetry 5d fields.

The   expressions for  $\aa$  and $\cc$   coefficients  corresponding to  one-loop corrections 
of general $(\De; j_1,j_2)$  fields in \ads  presented   in section 5.1 
%  (derived  for $\aa$ in section 3 and conjectured for $\cc$  in section 4) 
allow us to  complete the picture and explicitly demonstrate that the   above matching pattern applies  universally 
not only to $E_c$ \cite{Giombi:2014yra}   but also to $\aa$ and $\cc$ in  all type A and type B    cases. 
 The matching  of both  conformal anomaly coefficients
 provides   further   non-trivial 
test of   the consistency of the vectorial AdS/CFT duality.
 Note that here  there is no supersymmetry, so there is no 
a priori reason to expect a  correlation  between the values of $\aa$ and $\cc$  or  $\aa$ and $E_c$   as in \rf{1.13}. 
As we shall see  below, the novel   case of type C theory  appears to
 require  a different matching
  pattern. % as we shall discuss below. 

Since HS  theories    contain   infinite number   of fields, 
   one  needs  a prescription of how  to regularize  the infinite sum  of individual    contributions. 
    In the  computations  of the $\aa$-coefficient    and $E_c$  (from the partition functions in \ads 
with $S^d$  and $\mathbb R\times S^{d-1}$  boundaries  where the heat kernel  is explicitly   known) 
 there is  a  preferred  regularization     equivalent to the use of 
   the spectral $\zeta$-function     \cite{Giombi:2014iua,Giombi:2014yra}. Its  use should be  required by the preservation of 
   symmetries of the theory  at the quantum level. 
  This regularization  amounts to  
   first  doing  the sum over spins  of   individual-field  $\zeta(z)$-functions  for  an arbitrary     $z$ 
  and then  analytically continuing  the result  (or its  derivative)   to the   required  value of  $z$. 
As was   found in  \cite{Giombi:2014iua},  in the case of $d$-dimensional boundary 
  this  regularization  is equivalent 
to introducing  a specific    exponential   cutoff  factor $\exp\left[-\ep\,(s+\textstyle\frac{d-3}{2})\right]$ into the sum  over spins $s$, doing the sum  
  and then dropping all  singular terms  in  the  $\ep \to 0$  limit.  
 
 Below we  shall   apply   the same  prescription 
  also for the summation of  the   contributions  to the $\cc$-coefficient 
 where  a direct  spectral $\zeta$-function regularization is not available.
 In the present $d=4$ case  this  prescription  amounts to 
 %explore both the  $\ra$ and $\rc$ anomaly matching in 
%all cases. To illustrate the detailed results, we have to fix a regularisation procedure for the divergent sums over spin.
%Our choice will be \footnote{These regularisation has been proposed in \cite{Giombi:2014iua} and, in dimension $d$,
%t involves the damping factor $\exp\left[-\ep\,\left(s+\textstyle\frac{d-3}{2}\right)\right]$. 
\be\la{614}
\sum_{s}   K(s)  \equiv % \mbox{finite part of } 
\sum_{s}e^{-\ep\,(s+\frac{1}{2})}\,  K(s) \ \Big|_{\ep\to 0,\ \rm finite\ part} \ , \ \ \ \ \ \ \  \ \ \ \ \ \ \ \    K=(E_c, \aa, \cc) \ . 
\ee
Here  $s= j_1 + j_2$  is the total spin and the sum  includes  summation over all states. 
Let us 
 denote by $K^{+}(\De; j_1,j_2) $ any of the three quantities $E_{c}^{+}, \ra^{+}, \rc^{+}$ corresponding 
 to the one-loop contribution of a   5d field  in   the representation $(\De; j_1,j_2) $.
 Then, as in \rf{662}, \    $K=-2K^{+}$ will give  the  quantities for the associated elementary
  4d  conformal field  with the 
  canonical dimension  equal to  $\De_-=4-\De$. % (shadow field  from the boundary CFT point of view). 
 
%the analogous quantities in 4d. 
%This relation, in the case of $K=\rc$, fixes $r_{b}=-1$ in  (\ref{4.4}).
Starting with the non-minimal type A theory  and 
using   the expressions  in \rf{2.34},\rf{35}  and  \rf{4.4},\rf{4.12} 
together with the regularization  \rf{614}   one finds  that the total one-loop HS contribution
to each of the  three quantities  is indeed   zero % vanishes 
\be
\label{6.14}
\sum_{s=1}^{\infty} K^{+}(2+s;\,\textstyle\frac{s}{2},\textstyle\frac{s}{2}) = 0\  . 
\ee
%As we mentioned, this is consistent with absence of $\sim N^{0}$ renormalisation of the gravitational coupling.
In the minimal type A theory  we  get instead 
%
%
%Free $O(N)$ scalar theory dual to minimal Vasiliev theory in the bulk
%\be
%2\,K^{+}(3;\,0,0)+\sum_{s=2, 4, \dots}^{\infty} K^{+}(2+s;\,\textstyle\frac{s}{2},\textstyle\frac{s}{2}) = 0.
%\ee
%This is interpreted by saying that minimal Vasiliev plus a complex scalar has vanishing $K$. Alternatively,
%writing in the equivalent way
\be\la{616} 
\sum_{s=2, 4, \dots}^{\infty} K^{+}(2+s;\,\textstyle\frac{s}{2},\textstyle\frac{s}{2}) = K(3;\,0,0),
\ee
{ i.e.} the  total \ads  HS theory  one-loop correction   is equal  exactly to the  one-loop contribution of  a single  real massless 
  4d scalar.\foot{Explicitly,  \ $K(3;\,0,0)= ( { 1 \ov 240},  { 1 \ov 360}, { 1 \ov 120})$,  see   Table \ref{T2}.}
As the  contribution  of  $N$ such scalars  should  match the classical  plus one loop minimal type A   higher spin theory 
result, this is consistent with the AdS/CFT duality  provided  the   coefficient in front of the 
classical  minimal HS theory  action is not $N$ but   $N-1$. 

%The interpretation is that in this case the bulk coupling constant is $G_{N}\sim  1/(N-1)$. 

Similarly, in  the   non-minimal  type B  theory  we get   from   \rf{234},\rf{36}   and \rf{4.5},\rf{4.12}
%\foot{Note that here we  have  formally 
%the same mixed representations as in the fermionic CHS theories   but  now for integer instead of  half-integer values of $s$.}  
\be
\label{6.16}
2\,K^{+}(3;\,0,0)+ 2 \sum_{s=1}^{\infty}  K^{+}(2+s;\,\textstyle\frac{s+1}{2},\textstyle\frac{s-1}{2})
 = 0\ .
\ee
Here  the first term  $2\,K^{+}(3;\,0,0)=-K(3;0,0)$   stands  for the contribution of the two 
5d scalars appearing in the  type B spectrum   %we included the contribution from the fields dual to the two singlet scalars in the free CFT 
in  Table \ref{T8}. The  contribution  of  the totally symmetric higher spin  fields  vanishes separately due to   (\ref{6.14}). 
The contributions of $(\De; j_1, j_2)$  and $ (\De; j_2, j_1)$  states are equal so the mixed-symmetry  term doubles. 
For $\cc^+$  this   is equivalent to the  first relation  in \rf{4.14}  (where $\cc= - 2 \cc^+$).

%As before, the interpretation of the vanishing result (\ref{6.16}) is that there is no $\sim N^{0}$ correction 
%to the gravitational coupling here. Eq.~(\ref{6.16}) provides a non-trivial constraint on the expression for $\rc$ and has been one of the requirements
%that we imposed in order to obtain (\ref{4.11}).
In  the  minimal type B  theory   we find 
\be
\label{6.17}
\begin{split}
2\,K^{+}(3;\,0,0) + 2 \sum_{s=2, 4,  \dots}^{\infty} 
 K^{+}(2+s;\,\textstyle\frac{s+1}{2},\textstyle\frac{s-1}{2})
= K(\textstyle\frac{5}{2};\,\textstyle\frac{1}{2},0)_c \ ,
\end{split}
\ee
where the r.h.s.   is  the same as the  contribution of a  single 4d Majorana fermion
(again equivalent to  \rf{4.14} in the  case of $\cc^+$).\foot{Here 
 \ $K(\textstyle\frac{5}{2};\,\textstyle\frac{1}{2},0)_c=2 K(\textstyle\frac{5}{2};\,\textstyle\frac{1}{2},0)=  ( { 17 \ov 960},  { 11 \ov 720}, { 1 \ov 40})$,  \  see  Table \ref{T2}.}

 % in 4d whose interpretation is  as before. Again, (\ref{6.17})
%is a constraint on the function $\rc(\Delta;\,j_{1},j_{2})$ that we used to fix it.
%%%%%%%%%%%%%%%%%%%%%%%%%
%As a final comment, we remark 
%%%%STOP%%%
Repeating the same computations   for the spectrum of the 
non-minimal   type C theory  in Table \ref{T8}  we find  (cf. \rf{a15} and the discussion of  Casimir energy in  Appendix \ref{A:FF})
\ba\no
2\,K^{+}(4;\,0,0)&+  K^{+}(4;\,1,0)_{c}+ 2\,\sum_{s=2}^{\infty}K^{+}(2+s; {\ts\frac{s}{2},\frac{s}{2}}) 
+    \sum_{s=2}^{\infty}K^{+}(2+s; {\ts\frac{s+2}{2},\frac{s-2}{2}})_{c} \\
& = 2\,  K(3; {\ts\frac{1}{2}, \frac{1}{2}}) =  - 4\,K^{+}(3; {\ts\frac{1}{2}, \frac{1}{2}}) \ . \label{6.18}
\end{align}
 Here  the sum of all \ads    one-loop   contributions is no longer  zero but is   twice 
$K(3; {\ts\frac{1}{2}, \frac{1}{2}}) =  ( { 11 \ov 120},  { 31 \ov 180}, { 1 \ov 10})$, i.e.  is  the same as 
   the contribution of  one complex 4d Maxwell field. 
This   suggests that already  in the non-minimal   type C theory  case 
   one  needs  to assume  that the coefficient  in front of the corresponding HS  classical  action  in  \ads 
  is not $N$    but  $N-1$.\foot{An alternative  possibility 
   may be  to add 4 real  massless 5d vectors   to the  bulk theory,  i.e. to put the r.h.s.  term  in  \rf{6.18}   to the l.h.s.
   as in \rf{a15}, but  it is  unclear  why  that would lead to a consistent  HS theory
   (and also which should be the corresponding conserved spin 1 currents in the boundary theory).}

In the minimal type C theory  %with truncated spectrum   
we  get a  relation similar  to \rf{6.18}
%In particular, the analogue of (\ref{X1}) turns out to be 
\ba
2\,K^{+}(4;\,0,0) &+\sum_{s=2}^{\infty}K^{+}(2+s; {\ts\frac{s}{2},\frac{s}{2}}) +\sum_{s=2,4,\dots}^{\infty}K^{+}(2+s; {\ts\frac{s+2}{2},\frac{s-2}{2}})_{c} \no 
\\ &=  2\,  K(3; {\ts\frac{1}{2}, \frac{1}{2}})  =  - 4\,K^{+}(3; {\ts\frac{1}{2}, \frac{1}{2}})  \ . \label{6.19}
\end{align}
Since here  the boundary vector field is real,  this  non-vanishing result    could be accommodated   by the  shift 
   $N\to N-2$  in  the coefficient of  the classical HS action.
    This is   analogous  to  what happened  in the  
type A and B theories where one required   an extra   -1   shift of the coefficient of the HS action 
when  going from non-minimal to minimal case.   
The reason for the $N\to N-1$  shift  required already in the non-minimal type C
 case remains to be  understood.

Let us  mention  also that as discussed in Appendix \ref{A:FF},   the one-particle partition functions on $S^1 \times S^3$ 
 in the non-minimal  and minimal type C theories satisfy  the relations \rf{a20} and \rf{a21} 
which are the direct analogs of the relations  \rf{a166},\rf{a167} and  \rf{a16},\rf{a17}    in  the type A and type B theories 
\ci{Giombi:2014yra}. % (see Appendix \ref{A:FF} for details).
It is  straightforward  to derive these relations from the large $N$ limit  of  the 
 singlet-sector   partition function for the boundary  spin 1 theory just  like 
 that was done  in the spin 0 and spin $\ha$  cases in
   \ci{Shenker:2011zf,Giombi:2014yra}.\foot{We shall present  details of this derivation elsewhere.}

%\subsubsection{Supersymmetric cases} 

%%%%%%%%%%%%%%%
Finally, let us   note that   while 
   supersymmetry is not a   necessary ingredient in vectorial AdS/CFT duality,  it  is  possible to consider  also 
supersymmetric    AdS$_{5}$/CFT$_{4}$   dual pairs.\foot{Supersymmetric 
AdS$_{4}$/CFT$_{3}$ cases  were discussed, e.g., in \ci{Sezgin:2003pt,Leigh:2003gk,Chang:2012kt}. }
An  example  of  $\N=1$   supersymmetric higher spin  theory in \ads was constructed  in \ci{Alkalaev:2002rq}. 
The 4d  boundary theory   should be represented by $N$ free    spin $(0,\ha)$   $\N=1$ supermultiplets
having   bosonic integer  spin and fermionic   half integer spin conserved  currents.  Equivalently, 
in addition  to the  bosonic HS  5d fields there will be the 
fermionic  ones coming from the product of  spin 0 and spin $\ha$ doubleton
representations (cf. \rf{a24} for $n_1=0, \ n_0=n_{1/2}$). 
The  analog of \rf{6.14} for the non-minimal theory  should  then  be 
given by the   sum of  the  bosonic and fermionic  5d  field contributions. 
The bosonic   part    vanishes  separately due to \rf{6.14}  while the fermionic  part   
can be   verified to satisfy  the required identity
(here we use  $\s= s- \ha$  as in \rf{216} which takes integer values for the fermions)
%Let us note that although supersymmetry is not a crucial ingredient in vector AdS/CFT dualities, 
%it is certainly  possible to include it in the discussion. A simple example is  the supersymmetric Vasiliev theory 
%discussed in Section 2.4 of \cite{Chang:2012kt}.
%A further relation, possibly related to the fermionic bulk fermion and fermionic currents 
%in the supersymmetric Vasiliev theory discussed in Section 2.4 of \cite{Chang:2012kt} is 
\be\la{619}
\begin{split}
2\,K^{+}(\textstyle\frac{5}{2};\,\textstyle\frac{1}{2},0) &+\sum_{\s=1}^{\infty} %\Big[
K^{+}(2+\s+\textstyle\frac{1}{2};\,\textstyle\frac{\s}{2},\textstyle\frac{\s+1}{2})_c
%+  K^{+}(2+\s+\textstyle\frac{1}{2};\,\textstyle\frac{\s+1}{2},\textstyle\frac{\s}{2}) \Big] 
= 0\ .
\end{split}
\ee
There  is  also a minimal-theory analog of this  relation 
%In this case, a minimal-version adds nothing new because the sum of the currents over odd $\s$ is zero
\be\la{620}
\sum_{\s=1, 3, 5, \dots}^{\infty} %\Big[
K^{+}(2+\s+\textstyle\frac{1}{2};\,\textstyle\frac{\s}{2},\textstyle\frac{\s+1}{2})_c
%+K^{+}(2+\s+\textstyle\frac{1}{2};\,\textstyle\frac{\s+1}{2},\textstyle\frac{\s}{2})\Big]
 = 0\ .
\ee
It   should be possible also  to consider  the  case of supersymmetric  boundary theory  
containing spin 1 fields. This will generalize  the type A, B and C theory examples considered above. %  (see Appendix \ref{A:FF}). 

The most  supersymmetric  case of the free unitary  boundary  CFT 
 will be  a collection of $N$ free $\N=4$  Maxwell supermultiplets. 
The spectrum of the   dual \ads   HS theory will  then be given by the product of  two $\N=4$ superdoubletons 
\ci{Gunaydin:1998jc,Gunaydin:1998sw,Sezgin:2001yf,Sezgin:2001zs,Sezgin:2002rt} with the low-spin $\le 2$ part  \ci{Gunaydin:1984vz} being
the same as 
the set of fields of type IIB supergravity compactified on $S^5$    given  in  Table \ref{T5}.
 This  HS  theory with \ads vacuum 
should  correspond  to the ``leading Regge  trajectory''  part of the zero tension  limit of    AdS$_{5} \times $S$^5$ superstring 
(cf.  \ci{Bianchi:2003wx,Beisert:2003te}). This may 
suggest a way  to consider  a particular maximally supersymmetric case   of the 
 vectorial AdS/CFT duality as a  truncation of  zero gauge coupling    limit of  the adjoint AdS/CFT. 
As  we have seen   in sections 5.1 and  6.1, when 5d fields   are combined into supermultiplets  many cancellations happen, 
and this   should  especially  be  true   in the maximally supersymmetric case. 

We  postpone    detailed    discussion  of the supersymmetric   case  for the future,  presenting here only the result of 
the computation of   $K^{+} = (E_{c}^{+}, {\rm a}^{+}, {\rm c}^{+})$  corresponding  to the
infinite set of higher spin  5d fields 
appearing in the product of two superdoubletons  $\{{\mc N}\}$   representing $\N$-supersymmetric Maxwell theory (see  
  Appendix \ref{A:FF}). In general,  if 
 $\{\mc N\}$  contains $n_1$   vector, $n_{1\ov 2}$ fermion and $n_0$ scalar doubletons  \rf{nn}  then 
 we find from \rf{a25}\footnote{Here we    again use the regularization \rf{614}
   with $s= j_1 +j_2$. It turns out that 
 in  $\N=4$ supersymmetric case the total result  has no poles in $\epsilon \to 0$.
 This is due to supersymmetry and can be understood  as follows. 
 Here   we are summing the contributions  of bosonic  and fermionic fields, 
 $\sum_s   K_b(s)   + \sum_{\rm s}   K_f(\rm s)$, where in the fermionic case ${\rm s}= s - { 1 \ov 2} $ is  an  integer. 
 Ignoring regularization and separating  finite number of  low-spin terms, the remaining   sum  can be rewritten  as
 $\sum_s  \big[ K_b(s)   +   K_f(s-{ 1 \ov 2} )\big]$   and  happens to vanish, implying  finiteness of the total result.
 } 
% The implicit regularisation amounts to taking the finite part of the spin sums weighted with the factor
 %$e^{-\epsilon\,\left(j_{1}+j_{2}+\frac{1}{2}\right)}$. 
 %We remark that the divergent part, i.e. powers of $1/\epsilon$, vanishes. 
\ba
%\begin{split}
 K^{+}(\{\mc N\}\otimes \{\mc N\}) &= \te  n_{1}\,\Big(
\frac{4\,n_{0}+17\,n_{1\ov 2}+88\,n_{1}}{480},
\frac{2\,n_{0}+11\,n_{1\ov 2}+124\,n_{1}}{360},
\frac{n_{0}+3\,n_{1\ov 2}+12\,n_{1}}{60}
\Big) \no \\
&=2\,n_{0}\,n_{1}\,K(3;\,0,0)+2\,n_{1\ov 2}\,n_{1}\, K({\ts\frac{5}{2}};\,\ha,0)_c  %+K({\ts\frac{5}{2}};\,0,\ha)\right]
+2\,n_{1}^{2}\,K(3;\,\ha, \ha)  \ . \la{6.23}
\end{align}
This   generalizes the above results   \rf{6.14} ($n_0=1,\, n_{1\ov 2}=n_1=0$), \rf{6.16} ($n_{1\ov 2}=1, n_0=n_1=0$)
and \rf{6.18} ($n_1=1,\, n_{1\ov 2}=n_0=0$) in 
 non-minimal type A, B, and C theories:  the r.h.s   of  \rf{6.23} contains no 
 $n_{0}^{2}$ or $n_{1\ov 2}^{2}$ terms, but there is  $n_{1}^{2}$ term. 
For the  particular choices of $n_i$ corresponding to   $\N\le 4$ supersymmetric Maxwell theory, i.e.  
$(n_1, n_{1\ov 2},n_0) = (1,1,0), (1,2,2), (1,4,6)$ we thus get  a remarkable relation 
%For a single  Maxwell  supermultiplet, we have 
%
%Replacing the values of $n_{S}, n_{1\ov 2}, n_{1}$, we obtain in the supersymmetric cases
%\be
%K^{+}(\{\mc N\}\otimes \{\mc N\}) = 
%\left(\frac{7}{32}, \frac{3}{8},\frac{1}{4}\right), \  
%\left(\frac{13}{48}, \frac{5}{12},\frac{1}{3}\right), \  
%\left(\frac{3}{8}, \frac{1}{2},\frac{1}{2}\right),\ \mc N = 1,2,4.
%\ee
%So, in all cases we have 
\be
K^{+}(\{{\mc N}\}\otimes \{{\mc N}\}) = 2 \, K(\{\N\}) = 2\,K(\mbox{$\mc N$-Maxwell}) \ . \la{6.24}
\ee
Here the r.h.s. %$K(\mbox{$\mc N$-Maxwell})$
   is  twice   the contribution of the  $\N$-supersymmetric Maxwell theory,  or, 
   % i.e. 
%$K(\mbox{$4$-Maxwell}) =  6 K(3;\,0,0)+4 K({\ts\frac{5}{2}};\,\ha,0)_c 
%+K(3;\,\ha, \ha) $, etc.,  or, 
which is the same, 
 the contribution of  the $\N$-superdoubleton  (cf.  \rf{a101}, see \rf{5177} for $\N=4$).
This is the direct  super-generalization   of the relation \rf{6.19}  in type C theory.

Eq. \rf{6.24}  (i.e.  ``anomaly of a product is twice  anomaly of a factor'')    
may be   viewed  as  the  analog of the  relation  for the characters or  partition functions 
 $ \Z(\{{\mc N}\}\otimes \{{\mc N}\})= [ \Z(\{\N\}) ]^2 $  and 
 also admits the following   interpretation. 
 As was  observed  above in \rf{520}, the  one-loop contribution of  the states of $\N=8$ 5d supergravity  is 
 already equal to the contribution of two $\N=4$ Maxwell multiplets. Thus all  other states appearing in the product 
 $\{{\mc N}\}\otimes \{{\mc N}\}$  (i.e. in  \rf{a25}  with $n_1=1, n_{1\ov 2} =4, n_0=6$)   should give 
 zero contribution  to \rf{6.24}. As they should form massless supermultiplets of $PSU(2,2|4)$, this is indeed  consistent with 
 what was found in \rf{519}. 
 
 KK states   of type  IIB supergravity on $S^5$   are contained  in tensor products of more than two $\N=4$
  superdoubletons \ci{Gunaydin:1998jc}. 
  Their contribution   \rf{66} was  computed  in section 6.1  above.  
   We leave  the discussion of the   contributions of  their partner  higher spin states for the future. % and their  string theory for the future. 

%The   proper interpretation of  \rf{6.24} remains to be understood. 

\iffa 
n view of \rf{518} this   also  implies that one-loop  contribution of $\N=8$  5d supergravity  is the same  as of  two  
 $\N=4$ Maxwell  multiplets, 
 \be 
  K^+ (\N=8\ {\rm 5d\ SG})  = 2\, K({\N=4 \ \rm Maxwell}) \ . \la{520} \ee 
  Remarkably, this  non-trivial  relation may be interpreted  as  expressing the fact that 
   the states of $\N=8$  5d supergravity  appear in the product of two $\N=4$ superdoubletons  \ci{Gunaydin:1998jc}.
  We shall return to this  observation  in section 6 below. 
 
higher  KK states in products of more superdoubletons 

first,  N=8, d=5 sugra gives=    2  N=4 YM contributions -- we know that. 
But  the product of 2 superdoubletons  contains its states; taht means you result that product of 2 superdoubletons is equal  to 2 YM   is interpreted by saying that all contribution comes just from sugra states,   while all other states presumably fall into long massless  supermultiplets and thus give  zero contribution. 

 I thought that KK tower of states on type
IIB on S5  should also be  contained in product of  two
superdoubletons ?!
But your decomposition in app A  in A.24   certainly does not contain
 massive states labelled by index p.  Why?    -- \ci{Gunaydin:1998jc}  ,
\fi

\iffa 
In superdoubleton case we have SU(4) symmetry/reps 6 and 4 etc  and in
their product other SU(4) reps and these are KK states.
So bottom line is that  analogs of KK states would be those coming
from extra flavour indices in vectorial duality.  such theories were
mentioned  in GKT and we discussed them
but did not find anything new as i recall from their identities.
is same is true for type C theory with extra flavour ? probably...
tensionless limit of superstring.
full tensionless string has many more states than  product of two N=4
superdoubletons --  the point is that SYM  should be adjoint and that
should lead to extra reps.  that will be extra regge trajectories
becoming massless.
we need to look at Bianchi and Beisert et al  to understand that.
> Your answer suggests that in tensionless limit, all states (massive string ones, KK etc) are all captured by VÕs theory
-- no, I meant just leading Regge trajectory states.
as for usual KK states in 10d supergravity they are simply at massless level.
> or perhaps some supersymmetric extension of it. Is this correct ? Does this mean that vector duality may be regarded as tensionless limit of adjoint ads/cft ?
-- no, only as a subsector probably.    but that remains to be understood.
\fi

%then combining  these 0, 1/2, 1  theories we can build in superdoubleton product theory that should be zero tension limit of string theory  -- and spectrum of it  is  Gunaydin et al -- but there are extra subtlety of SU(4) product representations ignored in  naive product. 
%in the case of higher spin  theory  corresponding to product of two superdoubletons we expect similar cancellations. 
%we postpone detailed discussion of this case  for the future.in view of sect 5 for supermultiplets we expect 0. 

%%%%%%%%%%%%%%%%%%%%%%%%%%%%%%%%%%
\subsection{Conformal higher spin theories} 
%%%%%%%%%%%%%%%%%%%%%%%%%%%%%%%%%%%%%

The relation \rf{6.14}   written in terms of $K= - 2  K^+= (E_c, \aa,\cc)$   has also another interpretation: 
it expresses 
the vanishing of the total    Casimir energy and  the total   conformal  anomaly  coefficients 
in the  4d conformal higher spin (CHS)  theory  of  all   symmetric bosonic gauge  fields. 
The vanishing of the total $\aa$-coefficient   was first observed   in the  5d context    \ci{Giombi:2013yva}
and then understood also directly  from the 4d  perspective \ci{Tseytlin:2013jya}.  
The  cancellation  of  $E_c$ was  demonstrated  in \ci{Beccaria:2014jxa}.
The vanishing of the total $\cc$-coefficient  requires the use  of our proposed
 expression \rf{4.11}   leading to \rf{4.4} 
with the specific  choice of the  parameter $r_b=-1$    \ci{Tseytlin:2013jya}     in \rf{4.12}.
%\foot{The ``finite''   value for $r_b$ originally suggested in 
%was   $r_b=-1$   as the assumption about  the preferred regularization was  different from \rf{614}.} 
Similar   conclusion  applies  also  to  the  fermionic   CHS 
theory with the individual field contributions given in \rf{234},\rf{36},\rf{4.5},\rf{4.12}, 
generalizing   earlier  demonstration of  the vanishing of  its   total $\aa$-coefficient   \ci{Tseytlin:2013jya}. 

The consistency of the vectorial AdS/CFT
  is thus tightly related with  the consistency (cancellation of  anomalies) 
of the associated CHS theories. 
 This is  not  completely  surprising  in view   of  the direct 
 connection %(see Introduction)
 of   the CHS theory   (viewed as  induced  by the boundary CFT
   \cite{Tseytlin:2002gz,Segal:2002gd,Bekaert:2010ky,Giombi:2013yva}) 
   to     the  CFT conserved currents     (CHS fields are shadow  fields 
 for the CFT currents)  and then,  via AdS/CFT,      to  the 
  5d  higher spins (CHS fields   are  effectively boundary  values for  the 5d fields). 

While it still  remains to   prove  our   conjecture  for   the $\cc$  coefficient in 
\rf{4.11}  (implying the values in \rf{4.4},\rf{4.5},\rf{4.12}) 
 this is a  strong indication that  
 in addition 
to the $\N=4$ supersymmetric  theory of conformal 
 supergravity coupled to 4 Maxwell multiplets containing finite number of fields,  
 the 
  theory of an infinite collection  of  conformal higher spins  is  also a  consistent  quantum conformal 
   theory  with no  Weyl    anomalies  (both theories are of course  perturbatively  non-unitary). 
The same should be true also for the SU$(2,2|\N)$  supersymmetric  conformal higher spin theories  like the one 
constructed in \ci{Fradkin:1989md,Fradkin:1990ps}  and its truncations \ci{Alkalaev:2002rq}. 

%question: in HS type B theories mixed reps are actually bosonic, while 
%in  discussing CHS fields  we had exactly same reps  but for 1/2 integer spins 

%%%%%%%%%%%%%%%%%%%%%%%%%%%%%%%
\section{Concluding remarks}
%%%%%%%%%%%%%%%%
There  are many open questions. 
One   interesting  question is to understand better  the  vectorial AdS/CFT duality  in the spin 1  boundary theory case, 
% computing the corresponding singlet-sector  partition  function,  
  clarifying 
  the structure of  the dual   type C theory in Table \ref{T8}   and providing the  
   interpretation  for  the equation  \rf{6.18}.   

It remains to explore further  the  relation  between  vectorial AdS/CFT duality  setup for 
$\N=4$ superdoubleton as boundary theory and a tensionless  limit of  the AdS$_5 \times $ S$^5$   string theory,  
computing, in particular,  the quantities $(E_c, \aa,\cc)$   
and also  the  twisted and thermodynamic one-particle partition functions
for the string  spectrum of  5d fields. 

Another direction  is to  attempt to build an example of vectorial  AdS/CFT duality 
by starting   with spin $ > 1$   conformal   fields  at the boundary and considering the set of  (in general, non-unitary) 
5d  higher spin fields  corresponding to their conserved currents.

\iffa 
open questions: 
(i)  product of two superdoubletons -- relation to   zero tension limit and first Regge trajectory; 
compute  all quantities including  twisted but also  thermodynamic partition function 
(ii) same as application to string theory 
\ci{Bianchi:2003wx,Beisert:2003te}
(iii)  use  higher spin CHS as  fundamental and build     massless  HS fields out of them; non-unitary, will involve  
partially massless  fields.\foot{We are grateful to  E. Skvortsov   for discussion of this case.} 
\fi
%singlet constraing and Z  for type C theory. 

%\ci{Metsaev:1994ys,Metsaev:2003cu,Metsaev:2002vr,Alkalaev:2003qv,Alkalaev:2012rg,Alkalaev:2010af,Alkalaev:2006rw,Alkalaev:2006hq}

%%%%%%%%%%%%%%%%%%%%%%%%%%%%%%%%%%%%
\section*{Acknowledgments}

We  thank 
 S. Giombi, M. G\"unaydin, I. Klebanov,  S. Lal, M. Vasiliev and  D. Vassilevich  for   useful   discussions and comments. 
We  are particularly grateful  to K. Alkalaev, R. Metsaev   and  E. Skvortsov  for important explanations of related   questions.
The  work of A.A.T was supported by the ERC Advanced grant No.290456.
%``Gauge theory -- string theory duality'' %and also by the STFC grant ST/J000353/1.
%v2
The work of  M.B.  and  A.A.T.   was  supported 
 by  the  Russian Science Foundation grant 14-42-00047 in association  with Lebedev Physical Institute.

\newpage

%%%%%%%%%%%%%%%%%%%%%%%%%%%%%%%%%%%%%%%%%%%%
\appendix 
%%%%%%%%%%%%%%%%%%%%%%%%%%%%%%%%%%%%%%%%%%

\section{$SO(2,4)$ representations,  characters and generalised\\ Flato-Fronsdal relations}\label{A:FF}

%\newcommand{\ha}{{\ts\frac{1}{2}}}
%%%%%%%%%%%%%%%%%

Below  we shall summarize  some relations  for relevant representations  of the $d=4$  conformal group and their characters
 using  some results of  \cite{Dolan:2005wy}. We shall  then   consider
 the relations  between characters that have the  interpretation in terms  of one-particle partition functions in the context of vectorial AdS/CFT discussed in section 6.2. We shall also  discuss 
  the case of supersymmetric combination of representations. 
 
%The unitary irreducible representations of $SO(2,4)$ are massive, massless, and doubleton \cite{Dolan:2005wy}. 
We  shall adopt the following short-hand  notation for the   unitary irreducible representations of $SO(2,4)$ 
%$SO(2,4)$  representations
\be\la{a1} 
\begin{split}
{\rm ``massive"}: & \quad(\Delta;\,j_{1}, j_{2})     \hskip 3.26cm\ \Delta >   2+j_{1}+j_{2}\ \ \  \ \ {\rm or} \\
 & \hskip 5.45cm % \mbox{or}\
  \Delta>1+j_{1}+j_{2}\ \mbox{\underline{and}}\  j_{1}j_{2}=0\\
{\rm ``massless"}:  & \quad %(j_{1}, j_{2})\equiv
 (2 + j_1 + j_2;\,j_{1}, j_{2})  \hskip 1.94cm   \Delta = 2+j_{1}+j_{2}\ \mbox{\underline{and}}\  j_{1}j_{2}>0 \\
{\rm ``doubleton"}:  & \quad \{j, 0\},\  \{0,j\} \hskip 2.95cm\Delta = 1+j  \ , 
\end{split}
\ee
where  $j$ can take   integer or half-integer values  and the names refer to \ads  interpretation of the corresponding
%v2
 fields.\foot{As  we  consider the 
 \ads  case  we   use  name doubleton  \ci{Gunaydin:1998km}  instead of singleton.
 The massive case with $j_{1}j_{2}=0$ was  called {\em massive self-dual} in \cite{Metsaev:2008ba} where it is
shown that, contrary to the doubleton case,  this representation  admits a realisation in terms of local fields in AdS$_5$.
Examples of such fields are $(3; 1,0)$ in Table \ref{T2}   and $(4;1,0)$ in non-minimal type C theory in Table \ref{T8}.}  
We shall   also use    $ (\De; j_1,j_2)_{c}\equiv  (\De; j_1,j_2)  + (\De; j_2,j_1)   $.

Products of two doubleton representations decompose as follows   \ci{Vasiliev:2004cm,Dolan:2005wy}
\begin{align}
\{j, 0\}\otimes \,\{j', 0\} &= \bigoplus_{k=|j-j'|}^{j+j'} (2+j+j';\,k ,  0)+\bigoplus_{k=1}^{\infty}(2 + j + j' + k;  j+j'+\ts\frac{k}{2}, \frac{k}{2}) \ ,\la{a2} \\
\{0,j\}\otimes \,\{0,j'\} &= \bigoplus_{k=|j-j'|}^{j+j'}  (2+j+j';\, 0,k)+\bigoplus_{k=1}^{\infty}(2 + j + j' + k; \ts \frac{k}{2}, j+j'+\ts\frac{k}{2}) \ ,\la{a3} \\
\{j, 0\}\otimes \{0, j'\} &= \bigoplus_{k=0}^{\infty}(2+ j + j' + k;  j+\ts\frac{k}{2}, j'+\frac{k}{2})\ , \la{a4}
\end{align}
where  the first term in \rf{a2},\rf{a3}  is   the finite sum   over  representations 
 corresponding to states  appearing 
in the product    $j\otimes j' = (j+j')\oplus(j+j'-1)\oplus\cdots\oplus |j-j'|$. 
For example,   the product of two spin 0 doubletons  gives the  Flato-Fronsdal type  relation \ci{Flato:1978qz,Vasiliev:2004cm}
\be
\{0,0)\otimes \,\{0,0\} = \mc (2;\,0,0)+\bigoplus_{s=1}^{\infty}\mc (2+s; \ts\frac{s}{2},\frac{s}{2}) \ . \la{a5}
\ee
For   the product of  two spin $\ha$ doubletons we get 
\be\la{a6}
\begin{split}
&\big(  \{\ha, 0\}  +  \{0, \ha\}   \big)  \otimes   \big(  \{\ha, 0\}  +  \{0, \ha\}   \big) \\
&\qquad = 2\,(3;\,0,0) + (3;\,1,0)_{c}+
{2\,\bigoplus_{k=0}^{\infty}(3+k;  \ts\frac{k+1}{2},\frac{k+1}{2})}  + \bigoplus_{k=1}^{\infty}(3+k; 1+{\ts \frac{k}{2}},{\ts\frac{k}{2}})_{c}\\
&\qquad = 2\,(3;\,0,0) +
{2\,\bigoplus_{s=1}^{\infty}(2+s;  \ts\frac{s}{2},\frac{s}{2})}+ \bigoplus_{s=1}^{\infty}(2+s; {\ts \frac{s+1}{2},\frac{s-1}{2}})_{c} \ .
 \end{split}
\ee
%where $ (j_1,j_2)_{c} \equiv  (j_1,j_2) + (j_2,j_1)$ and similarly for $ (\De; j_1,j_2)_{c}$. 
For two spin 1  doubletons one finds 
\begin{align}
&
\big(  \{1, 0\}  +  \{0, 1\}   \big)  \otimes   \big(  \{1, 0\}  +  \{0, 1\}   \big)  \no \\
&\qquad =2\,(4;\,0,0)+(4;\,1,0)_{c}+  (4;\,2,0)_{c}  + {
2\,\bigoplus_{k=0}^{\infty}(4+k; \ts\frac{k+2}{2},\frac{k+2}{2}) }+
\bigoplus_{k=1}^{\infty}(4+k; {\ts 2+\frac{k}{2},\frac{k}{2}})_{c}\no  \\
&\qquad = 2\,(4;\,0,0)+(4;\,1,0)_{c}+
2\,\bigoplus_{s=2}^{\infty}(2+s;  {\ts\frac{s}{2},\frac{s}{2})} +\bigoplus_{s=2}^{\infty}(2+s; {\ts\frac{s+2}{2},\frac{s-2}{2}})_{c}\ . \la{a7}
 \end{align}
 
 \def \wZ {{ \mZ}}
 \subsection{Characters of products of doubletons} 
 %%%%%%%%%%%%%%%%%%%%%%%%%%%%%
 
Above  relations have  immediate counterparts in terms of (``blind'') characters 
  for the basic   representations in \rf{a1}\footnote{%v2
  Note  that 
  %We remind that in the main text we have noticed that 
  the  expression for  the character of the massless representation \rf{a9} 
  formally applies   also for $j_1 j_2=0$ when it gives the character of the corresponding  massive self-dual   representation, cf. \rf{a1}.}
  % is   also valid for the case $\Delta=2+j_{1}+j_{2}$ with $j_{1}j_{2}=0$ as is clear from the explicit expression in   (\ref{a9}).}
% (we use the same notation as for 
%representations adding simply a $q$ label in the arguments)
\begin{align}
{\rm  ``massive"}:&  \ \ \  \mZ(\Delta; j_{1}, j_{2}) =\frac{q^{\Delta}}{(1-q)^{4}} (2j_{1}+1)\,(2j_{2}+1)\ ,\la{a8} \\
{\rm ``massless"}:& \ \ \    \mZ(2 + j_1 + j_2; j_{1}, j_{2}) = \frac{q^{2+j_{1}+j_{2}}}{(1-q)^{4}}\,\big[(2j_{1}+1)(2j_{2}+1)-4\,q\,j_{1}\,j_{2}\big]\ ,\la{a9} \\
{\rm ``doubleton"}:& \ \ \    \mZ ( \{j, 0\}) = \mZ(\{0, j\}) = \frac{q^{1+j}}{(1-q)^{3}}\big[2j+1 -q\,(2j-1)\big]\ . \la{a10}
\end{align}
The character  \rf{a8}   has the interpretation of   one-particle partition function 
$\widehat \Z^+ $ in \rf{2.10}   corresponding to a massive 5d field while 
the one in \rf{a9}  is the one-particle partition function 
$ \Z^+$ in \rf{2.10}   corresponding to a massless  5d field \rf{2.9} with $\De_0= 2 + j_1 + j_2$.
 For the doubleton  partition function we shall also use the notation  $ \Z  ( \{j, 0\})$, i.e. 
\be \la{a00}
\mZ(\Delta; j_{1}, j_{2}) = \widehat \Z^+(\Delta; j_{1}, j_{2}) , \ \ \
 \mZ(\De_0 ; j_{1}, j_{2}) = \Z^+ (\Delta_0; j_{1}, j_{2})  , \ \ \ 
  \mZ ( \{j, 0\}) \equiv \Z  ( \{j, 0\})  \ee
%\foot{In   this Appendix we  shall mostly omit  ${}^+$    labels on $\Z$.} 
Note that  the massless   and  doubleton   characters satisfy the following identity
\ba\la{a1011}
&\mZ(2 + 2j; j, j) = \big[ \mZ ( \{j, 0\}) \big]^2 -   \big[ \mZ(  \{j+ \ha , 0\})  \big]^2 \ , \ \ \ \ \\
& \ \ {\rm i.e.} \ \ \  \   \mZ(3; \ha , \ha ) =  \big[ \mZ ( \{\ha, 0\})  \big]^2 -   \big[ \mZ  (\{1 , 0\})  \big]^2 \ , ... \la{a102} \ . 
\end{align}
There are also the following relations  %between  the low-spin doubleton  and massless  
%partition functions
 implying that  doubletons  can be 
identified with the corresponding  boundary conformal  fields  (cf. \rf{1.9},\rf{2.3},\rf{2.6},\rf{227}):
\ba 
 & \mZ ( \{0, 0\}) = \Z(3;0,0) \ , \ \ \ \  \te \mZ ( \{\ha, 0\}) = \Z ({5\ov 2} ;{1\ov 2},0) \ , \ \ \ \ 
  \mZ  (\{1, 0\}_c) =  \Z (3;\ha ,\ha) \ , \la{a101} \\
 & \qquad \qquad  \Z(\De; j_1, j_2 ) \equiv   \Z^-(\De; j_1, j_2 ) - \Z^+(\De; j_1, j_2 )  \ . \la{a103}
 \end{align} 
  %  \Z^- (3;0,0) &-  \mZ^+ (3;0,0)\ , \ \ \    \ \ \ 
%   \mZ ( \{\ha, 0\}) =   \te \mZ^- ({5\ov 2} ;{1\ov 2},0)  - \te \mZ^- ({5\ov 2} ;{1\ov 2},0) \ ,\no  \\
%&     \mZ  (\{1, 0\}_c) = \mZ^- (3;\ha ,\ha) -  \mZ^+ (3;\ha,\ha)\ . \la{a101}  \end{align}
The relations for one-particle partition functions  of 
non-minimal type A and type B theories in Table \ref{T8}  are direct character counterparts of  \rf{a5} and \rf{a6}:
 \ba
&  \big[\mZ  ( \{0, 0\})\big] ^{2}= \mZ(2; 0, 0)+
\sum_{s=1}^{\infty} \mZ(2+s; \ts\frac{s}{2}, \frac{s}{2})\  ,
\la{a166}\\
&     \big[2\mZ  ( \{\ha, 0\})\big] ^{2} = 2\,\wZ(3; 0, 0)
+2\sum_{s=1}^{\infty} \mZ(2+s;  {\ts\frac{s}{2}, \frac{s}{2}})
+\sum_{s=1}^{\infty} \mZ(2+s;  \ts\frac{s+1}{2}, \frac{s-1}{2})_{c} \ . \la{a167} 
\end{align}
We also get the   following  character identities that 
express the relations between   one-particle partition functions   in 
 minimal type A and type B theories    \cite{Giombi:2014yra}\foot{Here 
the notation $\big[ \mZ  ( \{0, 0\})\big]_{q\to  q^2} $ stands for $  \frac{q^{2}}{(1-q^2)^{3}}\big(1 +q^2\big)$, etc.}
%can be written  as 
\ba
&\frac{1}{2}   \big[\mZ  ( \{0, 0\})\big] ^{2}+ \frac{1}{2}\, \big[ \mZ  ( \{0, 0\})\big]_{q\to  q^2} = \mZ(2; 0, 0)+
\sum_{s=2, 4, \dots}^{\infty} \mZ(2+s; \ts\frac{s}{2}, \frac{s}{2})\  ,
\la{a16}\\
&    \frac{1}{2}   \big[2\mZ  ( \{\ha, 0\})\big] ^{2}-   \frac{1}{2}\, \big[ 2  \mZ  ( \{\ha, 0\})\big]_{q\to  q^2}     
% \frac{1}{2}\,\left[2\,\SS(\ha; q)\right]^{2}-\frac{1}{2}\left[2\,\SS(\ha; q^{2})\right] 
 \no \\ 
 & \qquad \qquad = 2\,\wZ(3; 0, 0)
+\sum_{s=1}^{\infty} \mZ(2+s;  {\ts\frac{s}{2}, \frac{s}{2}})
+\sum_{s=2, 4, 6, \dots}^{\infty} \mZ(2+s;  \ts\frac{s+1}{2}, \frac{s-1}{2})_{c} \ . \la{a17} 
\end{align}
For spin 1  doubleton characters  we find the following identities 
%For spin 1, we find empirically, 
\ba   
&
%and also the following ones whose relation with tensor product decompositions is unclear
 \big[\mZ  ( \{1, 0\})\big] ^{2} = \sum_{s=2}^{\infty}\mZ (2+s; {\ts\frac{s}{2}, \frac{s}{2}})  = 4\, \wZ(4; 0, 0)+ {1 \ov 2} \sum_{s=2}^{\infty} 
\mZ(2+s; {\ts\frac{s+2}{2}, \frac{s-2}{2}})_{c} \ ,    \la{a18} \\
&  \big[\mZ  ( \{1, 0\})\big] ^{2}+  \big[ \mZ  ( \{1, 0\})\big]_{q\to  q^2}
    = 2\, \wZ(4; 0, 0)+\sum_{s=2,4,\dots}^{\infty} 
\mZ(2+s; {\ts\frac{s+2}{2}, \frac{s-2}{2}})_{c}  \la{a19} \ . 
\end{align}
Since \rf{a8} implies that $ \wZ(4; 1, 0)=  \wZ(4; 0, 1) = 3 \wZ(4; 0,0)$   we get the relation  which is the counterpart  of  
   \rf{a7} at the character level:
\ba
\no 
 \big[2 \mZ  ( \{1, 0\})\big] ^{2} =\ \ & 2\, \wZ(4; 0, 0)+   \wZ(4; 1,0)_{c} \\ 
  &+    2  \sum_{s=2}^{\infty}\mZ (2+s; {\ts\frac{s}{2}, \frac{s}{2}})  + 
  \sum_{s=2}^{\infty} 
\mZ(2+s; {\ts\frac{s+2}{2}, \frac{s-2}{2}})_{c} \ . \la{a20}   \end{align}
It has the  direct interpretation as the relation of one-particle partition functions in 
 non-minimal  type C theory in Table \ref{T8}. 
Similarly, from \rf{a18} and \rf{a19}   we get  the  minimal type C theory  counterpart 
  of 
 the relations \rf{a16} and \rf{a17} in the minimal type A and type B theories 
\ba
&{1 \ov 2 }   \big[2\mZ  ( \{1, 0\})\big] ^{2}+ {1 \ov 2 } \big[2 \mZ  ( \{1, 0\})\big]_{q\to  q^2} \no \\
  &  =  \  2\, \wZ(4; 0, 0)+\sum_{s=2}^{\infty}\mZ (2+s; {\ts\frac{s}{2}, \frac{s}{2}})   
   +  \sum_{s=2,4,\dots}^{\infty} \mZ(2+s; {\ts\frac{s+2}{2}, \frac{s-2}{2}})_{c}  \ . \la{a21} \end{align}
It would be interesting to  know  a  group theoretic interpretation  of this relation.
It is possible  to show, just like this was done in the scalar case in 
\ci{Giombi:2014yra}, 
 that the   l.h.s.  of \rf{a21}   corresponds to the 
leading large $N$  term  in the singlet-sector  partition function of  $N$ real  4d   Maxwell vectors.

Let us now  comment on the  corresponding \rf{2.29},\rf{2281}  Casimir energy. 
Note that the expressions 
\begin{align}
&\big[\mZ ( \{0, 0\})\big]^2   %= \big[\mZ(0; q)\big]^{2}
 = \frac{q^{2}(1 + q)^{2}}{(1-q)^{6}}, \la{a11} \   \ \ \ \ \qquad 
\big[2\mZ ( \{\ha, 0\}) \big]^{2}%= 4\big[ \mZ(\ha; q) \big]^{2} 
= \frac{16\,q^{3}}{(1-q)^{6}}\ ,
\end{align}
are  invariant under $q\to   q^{-1} $. This implies  that the total  Casimir  energy   of the 
5d fields  appearing in  the r.h.s. of \rf{a5},\rf{a6} or \rf{a166},\rf{a167}. 
 % to the  sum of representations  in the r.h.s of \rf{a5} and \rf{a6}  should  vanish. 
%computed in  right hand sides must vanish.
The presence of the additional $\mZ_{q\to q^2}$ terms  in the r.h.s. 
of \rf{a16},\rf{a17}   which  change sign under $q\to  q^{-1}$  implies   that the Casimir energy for the representations  in the 
 r.h.s. is  no  longer  vanishing  in   minimal  type A and type B   theories  suggesting the 
  $N\to N-1$   shift  in the 5d classical action of the dual HS   theory 
  for a consistent AdS/CFT  interpretation\cite{Giombi:2014yra} (see  section 6.2). 

In  contrast, for spin 1 doubleton  product \rf{a7}  we  get  $q\to q^{-1} $   non-invariant 
expression
\be
\big[2\mZ  ( \{1, 0\})\big]^{2} %= 4\,\big[\mZ(1; q)\big]^{2}
 = \frac{4 q^4\,(3-q)^{2}}{(1-q)^{6}}\ \la{a13} \ 
\ee
 already in the r.h.s  \rf{a20}.  This implies  that the Casimir energy   in type C theory 
does not vanish even in the  non-minimal case.  Observing that one can form  a $q\to q^{-1} $
 invariant    combination as 
\be
\big[2\mZ  ( \{1, 0\})\big]^{2} +  4 \mZ(3; \ha, \ha)  %= 
   % 4 \big[\mZ(1; q)\big]^{2}+ 4 \mZ(\ha, \ha;q)
    = \frac{16q^{3}}{(1-q)^{6}}\ ,   \la{a14}
\ee
we  conclude  that one can  formally make the  Casimir energy vanish   by 
adding  four  $(3; {\ts\frac{1}{2}, \frac{1}{2}})$   to the 
representations in   \rf{a7},  getting   a  theory with field content 
%  that the following    sum of representations    %   this implies 
\be
4\,(3; {\ts\frac{1}{2}, \frac{1}{2}})+2\,(4;\,0,0)+ (4;\,1,0)_{c} +
2\,\bigoplus_{s=2}^{\infty}(2+s; {\ts\frac{s}{2},\frac{s}{2}}) 
+ \bigoplus_{s=2}^{\infty}(2+s; {\ts\frac{s+2}{2},\frac{s-2}{2}})_{c}\ . 
\la{a15} 
\ee
%has    total Casimir energy equal   to zero. 
%v2
In the case of the minimal  type C theory the l.h.s. of \rf{a21} contains  half of the  same term  plus  an extra    $q\to q^{-1} $  
non-invariant term    $  \mZ  ( \{1, 0\})_{q\to  q^2}$, and the two combined together 
 give  the same     Casimir energy as in the non-minimal theory  (see section 6.2). 

%%%%%%%%%%%%%%%%%%%%%%%%%%%%%%%%%%%%%
\subsection{Product of two $\mc N\le 4$   superdoubletons}

A natural extension of the  above  discussion  is to consider a supersymmetric combination of the 
 0, $\ha$, 1 doubletons  forming a superdoubleton $\{\mc N\}$ representing 
$\mc N$-supersymmetric Maxwell theory  
\cite{Gunaydin:1998jc,Gunaydin:1998sw,Sezgin:2001yf,Sezgin:2002rt}. 
One can then  study the $SO(2,4)$ representation content of 
the tensor product of two superdoubletons $\{\mc N\}$.
More generally, let us  define
\be\la{nn}
\{\mc N\} = n_{0}\,\{0,0\}  +     n_{1\ov 2}\,\left[\{\ha, 0\}+\{0,\ha\}\right]+ n_{1}\,\left[\{1,0\}+\{0,1\}\right] \ ,
\ee
where  $(n_1, n_{1\ov 2},n_0) = (1,1,0), (1,2,2), (1,4,6)$ for one vector  multiplet  with $\mc N=1, 2, 4$ supersymmetries.
The representations appearing in the tensor product
 $\{\mc N\}\otimes \{\mc N\}$ are easily found by using  the above  expressions for  the 
 doubletons\foot{Here we ignore 
 details of $SU(\N)$  index structure, i.e. just count different representations of $SO(2,4)$. 
 We shall also 
  not discuss  in detail  the organization  into representations of the 
  superconformal group $SU(2,2|\N)$,  
   see \ci{Gunaydin:1998jc}.}
\ba
%\begin{split}
&\{\mc N\}\otimes \{\mc N\} = n_{0}^{2}\,(2\; 0,0)+2\,n_{1\ov 2}^{2}\,(3;\,0,0)+2\,n_{1}^{2}\,(4;\,0,0) 
+ 2\,n_{0}\,n_{1\ov 2}\,({\ts \frac{5}{2}};\,0,\ha)_{c} \no \\ &
+ 2\,n_{1\ov 2}\,n_{1}\,({\ts \frac{7}{2}};\,0,\ha)_{c}+(2\,n_{0}\,n_{1}+n_{1\ov 2}^{2})\,(3;\,0,1)_{c}+n_{1}^{2}\,(4;\,0,1)_{c}
+2\,n_{1\ov 2}\,n_{1}\,({\ts \frac{7}{2}};\,0,{\ts\frac{3}{2}})_{c}+n_{1}^{2}\,(4;\,0,2)_{c}
\no \\ &+2\,n_{1\ov 2}^{2}\,\sum_{k=0}^{\infty}({\ts 3+k;\,\frac{k+1}{2},\frac{k+1}{2}})
+n_{0}^{2}\,\sum_{k=1}^{\infty}({\ts 2+k;\,\frac{k}{2},\frac{k}{2}})
+2\,n_{1}^{2}\,\sum_{k=0}^{\infty}({\ts 4+k;\,\frac{k+2}{2},\frac{k+2}{2}})\no \\
&+2\,n_{1\ov 2}\,n_{1}\,\sum_{k=0}^{\infty}({\ts \frac{7}{2}+k;\,\frac{k+1}{2},\frac{k+2}{2}})_{c}
+2\,n_{0}\,n_{1\ov 2}\,\sum_{k=1}^{\infty}({\ts \frac{5}{2}+k;\,\frac{k}{2},\frac{k+1}{2}})_{c}\la{a24}  \\
&+(2\,n_{0}\,n_{1}+n_{1\ov 2}^{2})\,\sum_{k=1}^{\infty} ({\ts 3+k;\,\frac{k}{2},\frac{k+2}{2}})_{c}
+2\,n_{1\ov 2}\,n_{1}\,\sum_{k=1}^{\infty} ({\ts \frac{7}{2}+k;\,\frac{k}{2},\frac{k+3}{2}})_{c}
+n_{1}^{2}\,\sum_{k=1}^{\infty} ({\ts 4+k;\,\frac{k}{2},\frac{k+4}{2}})_{c}\no 
\end{align}
%\ee
Grouping terms  together, this can be also  written  as\footnote{The term $-2\,n_{1}^{2}\,(3;\,\ha, \ha)$
appears  as a consequence of $\sum_{s=1}^{\infty}(2+s;\,\frac{s}{2},\frac{s}{2})+\sum_{s=1}^{\infty}(3+s;\,\frac{s+1}{2},\frac{s+1}{2}) = 
-(3;\,\ha, \ha)+2\,\sum_{s=1}^{\infty}(2+s;\,\frac{s}{2},\frac{s}{2})$. Also, terms   labelled  by $s$ are 
bosonic while those labeled by $\s=s- \ha $ are fermionic.}
\ba
%\begin{split}
\{\mc N\}&\otimes \{\mc N\} = n_{0}^{2}\,(2\; 0,0)+2\,n_{1\ov 2}^{2}\,(3;\,0,0)+2\,n_{1}^{2}\,(4;\,0,0)\no \\
&+ 2\,n_{0}\,n_{1\ov 2}\,({\ts \frac{5}{2}};\,0,\ha)_{c}
+ 2\,n_{1\ov 2}\,n_{1}\,({\ts \frac{7}{2}};\,0,\ha)_{c}+(2\,n_{0}\,n_{1}+n_{1\ov 2}^{2})\,(3;\,0,1)_{c}+n_{1}^{2}\,(4;\,0,1)_{c}\no\\
&+2\,n_{1\ov 2}\,n_{1}\,({\ts \frac{7}{2}};\,0,{\ts\frac{3}{2}})_{c}+n_{1}^{2}\,(4;\,0,2)_{c}-2\,n_{1}^{2}\,(3;\,\ha, \ha)\no \\
&+(n_{0}^{2}+2\,n_{1\ov 2}^{2}+2\,n_{1}^{2})\,\sum_{s=1}^{\infty}({\ts 2+s;\,\frac{s}{2},\frac{s}{2}})
+2\,n_{1\ov 2}\,(n_{0}+n_{1})\,\sum_{\s=1}^{\infty}({\ts \frac{5}{2}+\s;\,\frac{\s}{2},\frac{\s+1}{2}})_{c}\no \\
&
+(2\,n_{0}\,n_{1}+n_{1\ov 2}^{2})\,\sum_{s=2}^{\infty}({\ts 2+s;\,\frac{s-1}{2},\frac{s+1}{2}})_{c}
+2\,n_{1\ov 2}\,n_{1}\,\sum_{\s=2}^{\infty}({\ts \frac{5}{2}+\s;\,\frac{\s-1}{2},\frac{\s+2}{2}})_{c}\la{a25}\\
&
+n_{1}^{2}\sum_{s=3}^{\infty}({\ts 2+s;\,\frac{s-2}{2},\frac{s+2}{2}})_{c}.\no 
\end{align}
%\ee
The previous expressions  \rf{a5},\rf{a6},\rf{a7} for products of doubletons   with the same spin 
 are obtained as special cases -- as the coefficients of $n_{0}^{2}$, $n_{1\ov 2}^{2}$  and $n_{1}^{2}$ terms. 
 %We  use the relation in section 6.2.
 
The r.h.s. of (\ref{a25}) could be reorganised in order to  make manifest the   supersymmetry, i.e. rewritten in terms of  multiplets of the superconformal group  $SU(2,2|\N)$. Doing so, for example, for $\N=4$ 
one would get an infinite sum of  massless finite-dimensional $PSU(2,2|4)$ 
multiplets. Each of them is fully characterised by its lowest weight state as discussed  in  
 \cite{Gunaydin:1998sw,Gunaydin:1998jc}; further details are illustrated in Appendix  A of
\cite{Sezgin:2001yf} (for superconformal characters see   \ci{Bianchi:2006ti,Dobrev:2012me}).

% and the $n_{0}=n_{1\ov 2}=0$ limit (spin 1).
%\be
%\begin{split}
%\mc S\otimes \mc S &= 36\,(2\; 0,0)+32\,(3;\,0,0)+2\,(4;\,0,0) \\
%&+ 48\,({\ts \frac{5}{2}};\,0,\ha)_{c}
%+ 8\,({\ts \frac{7}{2}};\,0,\ha)_{c}+28\,(3;\,0,1)_{c}+(4;\,0,1)_{c}\\
%&+8\,({\ts \frac{7}{2}};\,0,{\ts\frac{3}{2}})_{c}+(4;\,0,2)_{c}-2\,(3;\,\ha, \ha)\\
%&+70\,\sum_{s=1}^{\infty}({\ts 2+s;\,\frac{s}{2},\frac{s}{2}})
%+56\,\sum_{\s=1}^{\infty}({\ts \frac{5}{2}+\s;\,\frac{\s}{2},\frac{\s+1}{2}})_{c}
%+28\,\sum_{s=2}^{\infty}({\ts 2+s;\,\frac{s-1}{2},\frac{s+1}{2}})_{c}\\
%&+8\,\sum_{\s=2}^{\infty}({\ts \frac{5}{2}+\s;\,\frac{\s-1}{2},\frac{\s+2}{2}})_{c}
%+\sum_{s=3}^{\infty}({\ts 2+s;\,\frac{s-2}{2},\frac{s+2}{2}})_{c}.
%\end{split}
%\ee

Let us    note  that,  as follows from \rf{a101}, the partition functions of  superdoubletons  are the same as 
of the corresponding super Maxwell theories. For example, for the $\N=4$  case 
with $\{\N=4\} = \{ 1,0\}_c + 4 \{ \ha, 0\}_c + 6  \{0,0\}$ we get  (see  \rf{a101})
\ba    \la{a26} 
 \Z(\{\N=4\} ) = \Z( \N=4 \ {\rm Maxwell}) = \Z ( 3; \ha, \ha ) + 4\,  \Z (\te  {5\ov 2} ; \ha, 0 )_c  + 6\, \Z ( 3; 0,0 ) \ . 
 \end{align}

%%%%%%%%%%%%%%%%%%%%%%%%%%%%%%%%%%%%%%%%%%%%%%%%%%%%%%%%%
\section{Partition functions of  free conformal supergravity fields on $S^{1}\times S^{3}$}
\label{A:CSG}

Here we  shall  explicitly compute  the one-loop partition functions for low-spin fields that appear in $\N \le 4$ 
conformal supergravities (see Tables  \ref{T2}   and \ref{T3}). 
The resulting expressions   for the one-particle partition functions 
 will   be the same  that  follow  from the  operator counting method. 
The cases of  the  standard scalar, vector   and  Weyl graviton   were already discussed in   \cite{Beccaria:2014jxa}.
For example, for the Maxwell vector (cf. \rf{2.4},\rf{2.6},\rf{2.8},\rf{2.15},\rf{a13})
\be \la{B1} 
\Z_1=\Z( \{1,0\}_c) =  \Z(3; \ha, \ha) =  \Z^-(3; \ha, \ha) -  \Z^+(3; \ha, \ha) = { 2 (3-q) q^2 \ov (1-q)^3} \ . \ee
Let us  start with  the  familiar  case of  spin $\frac{1}{2}$ Majorana  fermion, i.e.  
 $ \mathscr L_{1\ov 2} = \overline{\psi}\,e^{\mu}_{a}\,\gamma^{a}\,\DD_{\mu}\psi, $  
\ \  $\DD_{\mu} = \partial_{\mu}+\frac{1}{2}\,\sigma_{ab}\,\omega_{\mu}^{ab}(e),
\ \  \sigma_{ab} = \frac{1}{2}\gamma_{[a}\gamma_{b]}$. 
The corresponding partition function is 
%\subsection{Majorana spin $\frac{1}{2}$ fermion}
%As a warm-up, we begin with a Majorana spin $\frac{1}{2}$ fermion. This will illustrate how a direct computation 
%reproduces the known one-particle partition function obtained in \cite{Kutasov:2000td} by the operator counting
%method. 
\iffa The conformally invariant Lagrangian is 
\begin{align}
\label{A.1}
\mathscr L_{1\ov 2} &= e \overline{\psi}\,e^{\mu}_{a}\,\gamma^{a}\,\DD_{\mu}\psi, 
\qquad \DD_{\mu} = \partial_{\mu}+\frac{1}{2}\,\sigma_{ab}\,\omega_{\mu}^{ab}(e),
\qquad \sigma_{ab} = \frac{1}{2}\gamma_{[a}\gamma_{b]},
\end{align} \fi 
%and the associated partition function is then 
\be
Z_{1\ov 2}  = (\det \slashed{\DD})^{1/2} = (\det \slashed{\DD}^{2})^{1/4} \ , \qquad 
\slashed{\DD}^{2} = \DD^{2}-\frac{1}{4}R = \del_{0}^{2}+\pmb{\DD}^{2}-\frac{1}{4}R\ ,
\ee
\iffa By standard manipulations, we can write
\be
\label{A.3}
\slashed{\DD}^{2} = \DD^{2}-\frac{1}{4}\,R = \del_{0}^{2}+\pmb{\DD}^{2}-\frac{R}{4},
\ee \fi 
where $R=6$ is the scalar curvature  of  the unit-radius $S^{3}$ and $  \del_0$ is  
derivative along the Euclidean time with period $\beta=-\ln q $.
% In (\ref{A.3}), we have split 
%the operator  $\DD^{2}$ in a part acting on the flat $S^{1}$ and a part on the sphere $S^{3}$.
%On $S^{1}$, parametrised by the the 0-th coordinate, we have clearly $\DD_{0} \equiv \partial_{0}$. 
In general,  the spectrum of  the square of the Dirac operator on  unit-radius $S^{d-1}$ with odd  $d-1$
%the spectrum of the spatial part (including the curvature) 
is \cite{Camporesi:1995fb}
\be
-\pmb{\DD}^{2}+\frac{1}{4}R  \ \to \  {\te  \left(n+\frac{d-1}{2}\right)^{2}},\qquad 
{\rm d}_{n} = 2^{d/2}\,\frac{(n+d-2)!}{n!\,(d-2)!}\ , \ \  \ \ \ \ \ \ \ \  n = 0, 1, 2, \dots\,  \ . 
\ee
Then  by the standard arguments the  corresponding one-particle partition function  in \rf{22}  is  given by (see, e.g., 
 \cite{Beccaria:2014jxa})
%the one particle partition function is found as 
\be
\Z_{1\ov 2}  = \sum_{n=0}^{\infty}{\rm d}_{n}\,q^{n+\frac{d-1}{2}} = 2^{d/2}\,\frac{q^{\frac{d-1}{2}}}{(1-q)^{d-1}}
 = 2^{d/2}\,\frac{q^{\frac{d-1}{2}}-q^{\frac{d+1}{2}}}{(1-q)^{d}} \ . \la{A3} 
\ee
 This has direct operator-counting interpretation in $\mathbb R^d$:   counting components of $\psi$ (and their derivative  descendants) 
  minus equations of motion $\slashed{\del}\psi=0$. For $d=4$  this    gives as in  \cite{Kutasov:2000td} 
%\iffa As anticipated, the above expression matches the 4d result (2.10) in \cite{Kutasov:2000td}
\be
\Z_{1\ov 2}  =  \Z({\te{5\ov 2}}; \ha, 0)_c   %  \sum_{n=0}^{\infty} 2\,(n+1)(n+2)\,q^{n+\frac{3}{2}} = 
=\frac{4\,q^{3/2}}{(1-q)^{3}} \ . \la{B2}
\ee
%\fi 
%\subsection{Conformal gravitino}
Next, let us consider the  conformal gravitino  \ci{Kaku:1978nz}  with  the following quadratic  Lagrangian    in curved 
background (we omit  $\DD R\overline\psi\psi$ terms) 
\ba
\label{A.9}
& \mathscr L_{3\ov 2} = -4\,e^{-1}\,\epsilon^{\mu\nu\rho\sigma}\,\overline\phi_{\rho}\,\gamma_{5}\,\gamma_{\sigma}
\,\DD_{\mu}\phi_{\nu}\no \\
& \qquad\  \  -R^{\mu\nu}\left[
2\,\overline\psi^{\lambda}\sigma_{\lambda\nu}\phi_{\mu}-2\,\overline\psi_{\mu}\sigma_{\lambda\nu}\phi^{\lambda}
+2\,\overline\psi^{\lambda}\gamma_{\nu}\left(\DD_{[\mu}\psi_{\lambda]}-\gamma_{[\mu}\phi_{\lambda]}\right)
\right]  +\frac{4}{3}\,R\,\overline\psi^{\lambda}\sigma_{\lambda\nu}\phi_{\nu},\\
& \phi_{\mu} \equiv  \frac{1}{3}\gamma^{\nu}\big(\DD_{\nu}\psi_{\mu}-\DD_{\mu}\psi_{\nu}
+\frac{1}{2}\gamma_{5}\epsilon_{\nu\mu\alpha\beta}\,\DD^{\alpha}\psi^{\beta}
\big), \qquad 
\DD_{\mu}\psi_{\nu} = \big(\partial_{\mu}+\frac{1}{2}\sigma_{ab}\omega^{ab}_{\mu}(e)\big)\,\psi_{\nu}.
\end{align}
Considering a Bach  (e.g., an  Einstein) space background  
we may  fix the gauge symmetries by 
%We want to give the form of the quadratic part of the Lagrangian expanded around a general conformally flat space.
%This will lead to a conformal gravitino third order operator.
%To this aim, we consider a $\gamma$-transverse traceless (TT) field obeying the gauge conditions 
$\gamma^{\mu}\psi_{\mu}=0$ and $\DD_{\mu}\psi^{\mu}=0$, i.e. restrict to transverse  $\gamma$-traceless  field. 
Then  we get 
%Later, we shall explain how the full partition function
%can be obtained from the TT one. After a long but straightforward calculation, we arrive at the following compact form 
%for the quadratic part of the Lagrangian (\ref{A.9})
\be
\begin{split}
\mathscr L_{\frac{3}{2}} = \overline\psi^{\lambda} \mc O_{\frac{3}{2}}\,\psi_{\lambda},\qquad \qquad 
\mc O_{\frac{3}{2}} = -\slashed{\DD}^{3}-R^{\mu\nu}\gamma_{\nu}\DD_{\mu}+\frac{1}{6}R\slashed{\DD}.
\end{split}
\ee
This operator factorizes   \cite{Tseytlin:1984wj,Fradkin:1983zz,Deser:1983tm} 
 on an Einstein space  background ($R_{\mu\nu}=\frac{1}{4} Rg_{\mu\nu}$) as
%An important check of this expression is its form on an Einstein space with $R_{\mu\nu}=\frac{R}{4} g_{\mu\nu}$.
%In this case, we find 
\ba
\mc O_{\frac{3}{2}} = -\slashed{\DD}\,\big(\slashed{\DD}^{2}+\frac{1}{12}R\big) %= 
%-\big(\slashed{\DD}^{2}\big)^{1/2}\,\big(\slashed{\DD}^{2}+\frac{R}{12}\big)
%=  \Big( \DD^{2}- \frac{R}{4}-\frac{R}{12} \Big)^{1/2}    \Big(\DD^{2}- \frac{R}{4}\Big) \ , 
\label{A.14}\  , \qquad \quad
\slashed{\DD}^{2} = \DD^{2}+\frac{1}{2}[\DD_{\mu},\DD_{\nu}]\,\gamma^{\mu\nu}=  \DD^{2}- \frac{1}{4}R-\frac{1}{12}R \ . 
\end{align}
%%%%%%%%%%%%%%%%%%%%%%%%%%%%%
Specializing to   the $S^{1}\times S^{3}$ case, we have 
$R_{\mu\nu}\to R_{ij} = \frac{R}{3}\,g_{ij}= 2 g_{ij}$  i.e.
\be
\label{A.16}
\mc O_{\frac{3}{2}} = -\slashed{\DD}^{3}-2 \vD+\slashed{\DD}=
 -(\gamma^{0}\partial_{0}+\vD)^{3}+\gamma^{0}\partial_{0}- \vD \ , \qquad  \vD \equiv  \gamma_{i}\,\DD^{i} \ . 
\ee
Taking into account that $\{\vD, \gamma^{0}\}=0$ the determinant of this operator can  written as 
\be
\det\mc O_{\frac{3}{2}} = \Big( \det
(\del_0^{2}+\vD^{2})\det \big[(\del_0+1)^{2}+\vD^{2}\big] \det \big[(\del_0-1)^{2}+\vD^{2}\big]
\Big)^{1/2}\ . 
\ee
From (\ref{A.14})  we get  
%Finally, the relation between $\pmb{\DD}^{2}$ and $\vD^{2}$, is (from (\ref{A.14}))
$
\vD^{2}\psi_{i} = \big(\pmb{\DD}^{2}-\frac{R}{4}\big)\psi_{i}+\frac{1}{2}(\gamma_{ij}R^{j}_{k}
-\gamma_{kj}R^{j}_{i}-\frac{R}{3}\gamma_{ik})\psi^{k}  $ 
so that for   $\gamma_{i}\psi^{i}=0$ 
%Using  $R_{ij} = \frac{R}{3}g_{ij}$ and $\gamma_{i}\psi^{i}=0$, this gives
\be
\vD^{2} = \pmb{\DD}^{2}-\frac{R}{4}-\frac{R}{6} = \pmb{\DD}^{2}-\frac{5}{2} \ . \ee 
The spectrum of $\pmb{\DD}^{2} $  for a general spin $s$ field on $S^{3}$  is  (see, e.g., 
\cite{David:2009xg})
\be \la{A11} 
-\pmb{\DD}^{2}\   \to \ (n+s)(n+s+2)-s, \ \qquad \qquad 
{\rm d}_{n} = 2(n+1)(n+2s+1)\ ,
\ee
so that  for $s={3\ov 2}$ we get 
\be
\vD^{2}  \to   -\te \big(n+\frac{5}{2}\big)^{2} \ , \ \ \ \ \ \ \qquad \ 
{\rm d}_{n} = 2(n+1)(n+4) \ . 
\ee
Thus  the   contribution of the spatially transverse   and traceless   gravitino  $\psi_i$ 
 to  the one-particle partition function is 
\be\la{A13} 
\Z_{\frac{3}{2}}^{\rm TT}(q) = \sum_{n=0}^{\infty}2\,(n+1)(n+4)(q^{n+\frac{3}{2}}+q^{n+\frac{5}{2}}+q^{n+\frac{7}{2}}) = \frac{4\,q^{3\ov 2}\,(2+q+q^{2}-q^{3})}{(1-q)^{3}}.
\ee
To get the full partition function we  still  need  to add  the contribution of one Majorana spinor degree 
of freedom.\foot{To recall, in covariant  gauge  the  conformal gravitino 
partition function may be  written as  $ Z= \Big[  (\det  \mc{O}_{\frac{1}{2}})^{2}/\det \mc{O}_{\frac{3}{2}}   \Big]^{-1/4}$, 
where $\mc{O}_{\frac{3}{2}}$ is defined  on transverse $\gamma_\mu$-traceless   field $\psi_\mu$  (see, e.g., \cite{Tseytlin:2013jya}).
 This correctly  accounts for  $-8$ degrees  dynamical degrees of  freedom:  transverse traceless (TT) 
 field $\psi_\mu$ contributes $2\times 4$ (with 
extra factor of  3  being   due to the degree  of the kinetic operator) and  the  fermion contributes $2\times 4$.} 
%. So we have $3\times 2\times 4-2\times 4=16$
%and then there is extra $-1/2$  overall power.}
On  $\mathbb R \times S^3$,  we  may further split TT  $\psi_\mu$ into  TT  $\psi_i$ and a 
spinor. This gives 
\be
Z= \Big[  \det \mc{O}^{\rm TT}_{\frac{3}{2}}\ {\rm \det}' \mc{O}_{\frac{1}{2}}  \Big]^{1/4}\ , 
\ee
where  $\mc O^{\rm TT}_{\frac{3}{2}}$ is now defined on  transverse $\gamma_i$-traceless  $\psi_i$ field. 
Adding together  \rf{A13}   and the contribution of a Majorana fermion \rf{A3} without  the $n=0$ 
zero mode term\footnote{This mode must be dropped for the same reason as discussed 
in Appendix D of  \cite{Beccaria:2014jxa}.}
we  arrive  at  the  following  conformal gravitino one-particle partition function
\be
\begin{split}
\Z_{\frac{3}{2}}(q) =   \frac{4\,q^{3\ov 2}\,(2+2\,q-6q^{2}+2q^{3})}{(1-q)^{4}} \ .\la{A15} 
\end{split}
\ee
This expression admits the following operator counting interpretation in flat space. 
The natural gravitino 
analog of the covariant  Weyl tensor field strength  for the  conformal graviton is  its superpartner 
 \cite{Ferrara:1977mv}  (tilde denotes the dual field)
%predicts that the super partner of the Weyl tensor is the gravitino field strength that reads in flat space 
%(here, notation is with 4d signature $(-1,1,1,1)$, and $\star$ denotes the dual.)
\be
\Phi_{\mu\nu}=\frac{1}{3}\big(\psi_{\mu\nu}-\gamma_{5}\td \psi_{\mu\nu}%^{\star}
+2\,\gamma_{[\nu}^{\ \ \lambda}
\psi_{\lambda\mu]}\big),\qquad  \quad\psi_{\mu\nu}=\partial_{\mu}\psi_{\nu}-\partial_{\nu}\psi_{\mu} \ , 
\ee
obeying $
\Phi_{\mu\nu} = \gamma_{5}\, \td \Phi_{\mu\nu}, %^{\star},
\ \  \gamma^{\mu}\,\Phi_{\mu\nu}=0.
$
These  conditions imply that $\Phi_{\mu\nu}$ has $4\times 2$ components (the $\gamma_{5}$ self-duality
reduces  6 to 3 and the $\gamma$-tracelessness  adds one additional constraint). 
This  explains the first  $ 4 \times 2  q^{3\ov 2}$ term  in the numerator of \rf{A15}. 
The  next term $4\times 2q^{5\ov 2}$
is associated with $\slashed{\partial}\Phi_{\mu\nu}$. The equations of motion  and the Bianchi 
identities remove the term $4 \times (3+3) q^{7\ov 2}$;  the term  $4 \times 2  q^{9\ov 2}$   compensates for overcounting in this subtraction
(cf. \cite{Beccaria:2014jxa}). 

The Lagrangian of the   conformal  fermion $\Psi$ with $\slashed{\partial}^{3}$ kinetic term  is   \cite{Bergshoeff:1980is,Fradkin:1981jc}
\be
\mathscr L_{\Psi} =  \overline\Psi\,\mc O_{\Psi}\,\Psi, \qquad \qquad 
\mc O_{\Psi} = \slashed{\DD}^{3}+ \big(R_{\mu\nu}-\frac{1}{6}Rg_{\mu\nu}\big)\gamma^{\mu}\DD^{\nu}.
\ee
On $S^{1}\times S^{3}$  the kinetic operator takes the form 
$\mc O_{\Psi} = \slashed{\DD}^{3}-\slashed{\DD}+2\,\vD
$, i.e.  is the same  as the one in \rf{A.16} but now defined on a Majorana spinor.  Using \rf{A11} for $s=\ha $ we get
%This is the same operator encountered in the gravitino case, see (\ref{A.16}). Using the spectrum of $\vD^{2}$ on spinors, we obtain 
\be
\Z_{\Psi}(q) = \sum_{n=0}^{\infty}2(n+1)(n+2)(q^{n+\frac{1}{2}}+
q^{n+\frac{3}{2}}+q^{n+\frac{5}{2}}) = \frac{4\,q^{1\ov 2}\,(1-q^{3})}{(1-q)^{4}} \ , \la{A18}
\ee
which admits the same counting interpretation as in the case of the $\slashed \del $ spinor \rf{A3}.

The Lagrangian for  the  conformal scalar $\Phi$ with $\partial^{4}$ kinetic term  is \cite{Fradkin:1981jc}
\be
\mathscr L _{\Phi}=  D^{2}\Phi D^{2}\Phi-2\big(R_{\mu\nu}-\frac{1}{3}Rg_{\mu\nu}\big)D^{\mu}\Phi D^{\nu}\Phi.
\ee
On $S^1 \times S^3$ the  kinetic operator  becomes 
\be
\mc O_{\Phi} = D^{4}-4 D^{2}+4 \mathbf{D}^{2} \ \to \  (\partial_{0}^{2}-n^{2}) \big[\partial_{0}^{2}-(n+2)^{2}\big] \ , 
\ee
where  $D^{2} = \partial_{0}^{2}+\mathbf{D}^{2}$, and  we used that $\mathbf{D}^{2}$ has the spectrum 
$-n(n+2)$  with multiplicity $(n+1)^{2}$. As a result, 
\be
\Z_{\Phi}(q) = \sum_{n=0}^{\infty}(n+1)^{2}(q^{n}+q^{n+2}) = \frac{1-q^{4}}{(1-q)^{4}} \ . 
\ee
Similar computation can be done in the   case of the non-gauge  conformal antisymmetric 
tensor field  $T_{\mu\nu}$    \ci{Bergshoeff:1980is,Fradkin:1981jc,Fradkin:1985am} with the   Lagrangian 
(corresponding to the Weyl-invariant action)
\be
\mathscr L_{T} = (D^{\mu}T_{\mu\nu})^{2}-\frac{1}{4}(D_{\mu}T_{\rho\sigma})^{2}-
R_{\mu\nu}T^{\mu\lambda}T^{\nu}_{\ \ \lambda}+\frac{1}{8}R T_{\mu\nu}^{2}+\frac{1}{2}R_{\mu\alpha\nu\beta}
T^{\mu\nu}T^{\alpha\beta} \ .
\ee
Here we shall just quote  the result   for   the corresponding partition function    which  
  is  much easier to  find  by the counting method in flat space. 
$T_{\mu\nu}$ has 6 components with dimension 1. The  equations of motion 
\be
E_{\mu\nu} \equiv  \partial_{\mu}\partial_{\lambda}T^{\lambda}_{\ \ \nu}-
\partial_{\nu}\partial_{\lambda}T^{\lambda}_{\ \ \mu}-\frac{1}{2}\partial^{2}T_{\mu\nu}=0
\ee
reprsent  6 conditions   with dimension 3. Thus 
%We have 6 components of $T_{\mu\nu}$, with dimension 1 and 6 equations of motion $E_{\mu\nu}=0$,  with dimension 3.  So
\be
\Z_{T}(q) = \frac{6q-6q^{3}}{(1-q)^{4}} \ .\la{A19}
\ee

\def \rmR {{\cal R}}  \def \rmH {{\cal H}}

%%%%%%%%%%%%%%%%%%%%%%%%%%%%%%%%%%%%%%%%%%%%%%%%

\section{Spectral $\zeta$-function for 2nd-order    operator on  $(\Delta;\,j_{1},j_{2})$ fields   in AdS$_{5}$}
\label{A:heatkernel}
%%%%%%%%%%%%%%%%%%%%%%%%%

The computation of $\aa$-coefficient requires  consideration of (in general, massive)   higher  spin  field partition function 
in Euclidean \ads   with boundary $S^4$. The relevant  kinetic operator $\OO$  given  in \rf{1.16}
is defined on   transverse   fields.

In general, for the operator $\OO$ on a space $\M$  one  can express  the corresponding 
$\zeta$-function in terms of  heat kernel as 
\be
\zeta(z) =  \frac{1}{\Gamma(z)}\int_{0}^{\infty}dt\, t^{z-1}\,{\rm Tr }\, K \ , \ \ \ \ \qquad 
K(x, y; t) = \langle x|e^{-t\,\mc O}|y\rangle  \ .  \la{B111} 
\ee
%where $\Tr$
\iffa  %%%%%%%%%%%%%%%%%%%%%%%
The heat kernel of a differential operator $\mc O$, defined on a certain manifold $\M$, is  the function 
\be
K(x, y; t) = \langle x|e^{t\,\mc O}|y\rangle, 
\ee
where we do not show possible Lorentz indices. The associated spectral zeta function is the Mellin transform of 
its trace. Formally,
\be
\zeta(z) =  \frac{1}{\Gamma(z)}\int_{0}^{\infty}dt\, t^{z-1}\,{\rm Tr }\, K.
\ee\fi %%%%%%%%%%%%%%%%%%%%%%%%%
For  a homogeneous manifold $\M$ the trace  over the  position $x$  gives a factor of (regularized) volume, i.e. 
%, it is convenient to split the trace in the Lorentz and coordinate part and write
\be
\label{B.3}
\zeta(z) = \mbox{Vol}(\M)\,\zeta(z; x) \ , \ \ \ \ \ \ 
 \zeta(z; x) \equiv   \frac{1}{\Gamma(z)}\int_{0}^{\infty}dt\, t^{z-1}\,{\rm tr }\, K(x,x; t) \ .
\ee
Here  $\tr$ is the  trace over the  Lorentz indices of the operator  and $\zeta(z; x)$ does not actually depend on $x$.

%In our application, $\mc O$ will be the wave operator for $(\Delta;\,j_{1},j_{2})$ fields on AdS$_{5}$
%whose form has been discussed in the Introduction.
To determine $\zeta(z)$  in our case we shall use  the results  for  the heat kernel 
of the Laplacian  in AdS$_{2n +1}$   with  even  $n$   in 
\cite{Camporesi:1994ga,Gopakumar:2011qs} 
(see also \cite{Lal:2012ax}) specialising   to the case  of  $n=2$. 
Following 
\cite{Camporesi:1994ga,Gopakumar:2011qs}, 
we shall start  with   heat-kernel for  the sphere S$^{5}$ and then  analytically continue to  AdS$_{5}$. 
Let us consider a field on S$^{5}$ transforming under the tangent space rotations  in a  representation $\rm H$ of $SO(5)$.
Since the  sphere is a homogeneous space S$^{5}=SO(6)/SO(5)$ 
 the heat kernel receives contributions from each representation 
$\rmR$ of $SO(6)$ that contains $\rmH$ when restricted to $SO(5)$.
 Let us  denote  $\rmR$ and $\rmH$  by  the corresponding weights as
%in more details as 
\be
\begin{split}
\rmR = (r_{1}, r_{2}, r_{3}), \qquad r_{1}\ge r_{2}\ge |r_{3}|, \qquad \qquad 
\rmH = (h_{1}, h_{2}),  \qquad \ \ h_{1}\ge h_{2} \ge 0,
\end{split}
\ee
were all labels are integer or half integer. The  branching condition on the representation 
$\rmR$ is  \be \la{B.6} 
r_{1}\ge h_{1} \ge r_{2}\ge h_{2} \ge |r_{3}|
\ee 
with the additional requirement that $r_{i}-h_{i}\in \mathbb Z$. 
The heat kernel at the coincident points, traced over
representation indices, can be written  as %(it is indeed independent on $x$)
\be
{\rm tr}\,  K(x,x; t) = \frac{1}{\pi^{3}}\,\sum_{r_{i}}\, d_{\rm R}\, e^{-t\,E_{\rmR}^{(\rmH)}},  \la{B7} 
\ee
where  $E_{\rmR}^{(\rmH)}$  are  the eigenvalues of the Laplacian $ -D^2$  on  S$^5$
expressed in terms of the second  Casimir   values for the two representations  and  $d_{\rmR}$ is the dimension of $\rmR$
\ba
%\begin{split}
& \left. -  D^{2}\right|_{\rm S^{5}}\  \to \    E_{\rm R}^{(\rmH)} =  C_{2}({\rmR})  -  C_{2}({\rmH}) \  , \\
& C_{2}({\rmR}) = r_{1}\,(r_{1}+4)+r_{2}\,(r_{2}+2)+r_{3}^{2}, \qquad\quad
C_{2}({\rmH}) =h_{1}\,(h_{1}+3)+h_{2}(h_{2}+1), \\
%C_{2}(k_{1},\dots, k_{n+1}) &= \vec{k}\cdot\vec{k}+2\,\vec{r}\cdot\vec{k}, \qquad
%r_{i} = \left\{ 
%\begin{array}{ll}
%n-i+1, & \qquad SO(2n+2), \\
%n+\frac{1}{2}-i, & \qquad SO(2n+1),
%\end{array}
%\right.
&d_{\rmR} = \frac{1}{12} \left[\left(r_1+2\right){}^2-\left(r_2+1\right){}^2\right]
   \left[\left(r_1+2\right){}^2-r_3^2\right] \left[\left(r_2+1\right){}^2-r_3^2\right]. \la{B19}
%d_{\rm R} = \prod_{i<j}^{3} \frac{\ell_{i}^{2}-\ell_{j}^{2}}{\mu_{i}^{2}-\mu_{j}^{2}},\qquad
%\ell_{i}=r_{i}+3-i, \ \ \mu_{i}=3-i.
\end{align}
The analytic continuation from S$^{5}$ to AdS$_{5}$  amounts to the replacement  \cite{Camporesi:1994ga,Gopakumar:2011qs} \be
\label{B.10}
r_{1} \ \ \to \ \  i\,\lambda-2\ ,
\ee 
with the sum over $r_{1}$ becoming an integral over the positive real $\lambda$.
For the $(\Delta;\,j_{1},j_{2})$ representation of $SO(2,4)$ 
 we have $h_{1}=j_{1}+j_{2}$ and $h_{2}=j_{1}-j_{2}$.
The analytically continued $E_{\rmR}^{({\rmH})}$ is then
%(there is a minus sign discussed in 
\cite{Gopakumar:2011qs}
\be
\label{B.11}
\left. -D^{2}\right|_{\rm AdS_{5}} \ \to \   E_{\rmR}^{(\rmH)} = \lambda^{2}-r_{2}(r_{2}+2)-r_{3}^{2}+2\,j_{1}(j_{1}+2)+2j_{2}(j_{2}+1)+4.
\ee
For the general mixed-symmetry fields  which are   traceless and transverse  (on which our operator $\OO$ is defined) 
 the branching condition (\ref{B.6}) imposes the following restriction\footnote{For the special case of totally symmetric fields
  see the discussion after (2.17)  in  \cite{Gopakumar:2011qs}.}  
\be
r_{2} = h_{1} = j_{1}+j_{2},\qquad |r_{3}|=h_{2} = j_{1}-j_{2} \ . 
\ee
Then   (\ref{B.11}) becomes
\be
\left. -D^{2}\right|_{\rm AdS_{5}}  \ \ \to \ \  \lambda^{2}+2\,j_{1}+4 \ . 
\ee
 Thus  finally  for the full     operator  $\OO$   in   (\ref{1.16})  with $X= \De (\De-4) - 2 j_1  $
  we get the following eigenvalue
\be
\left. (-D^{2}+X)\right|_{\rm AdS_{5}}  \ \to \ \  \lambda^{2}+(\Delta-2)^{2} \ .  \la{B14}
\ee
%
%
%It is instructive to look at (\ref{B.12}) for totally symmetric bosonic tensors that have $(s_{1},s_{2}) = (s,0)$, or $j_{1}=j_{2}=\frac{s}{2}$, and the branching condition (\ref{B.7})
%is further refined for transverse traceless fields~\footnote{See comment around (2.17) of \cite{Gopakumar:2011qs}.}
%and reads
%\be
%m_{2} = s, \quad m_{3}=0.
%\ee
%Hence,  (\ref{B.12}) can be written 
%\be
%-D^{2} = \lambda^{2}+s+4.
%\ee
%Adding the known mass term  $M^{2}=m^2-2+(s-2)(s+1)$~\footnote{
%The wave equation is in this case $(-D^2+M^{2})\,\Phi_{s}=0$ with $M^{2} = m^2-2+(s-2)(s+1)$. The relation between $m$ and
%$\Delta$ is $m^{2} = (\Delta-2)^{2}-s^{2}$. This is obtained replacing $-D^{2}$ by the Casimir of $SO(2,4)/SO(5)$
%for the two representations $(\Delta\;j_{1},j_{2})$ and $(s_{1}=j_{1}+j_{2},s_{2}=j_{1}-j_{2})$. They are 
%$\Delta(\Delta-4)+2j_{1}(j_{1}+1)+2j_{2}(j_{2}+1)$ and $s_{1}(s_{1}+3)+s_{2}(s_{2}+1) = 2j_{1}^{2}+2j_{2}^{2}+4j_{1}+2j_{2}$ respectively. The difference gives $-D^{2}=-(\Delta-2)^{2}+2j_{1}+4$. Thus, for $(j_{1},j_{2})=(\frac{s}{2},\frac{s}{2})$, we find $-D^{2}+M^{2} = -(\Delta-2)^{2}+s^{2}+m^{2}=0$.
%} we obtain 
%\be
%\begin{split}
%-D^{2}+M^{2} &= \lambda^{2}+s+4+m^{2}-2+(s-2)(s+1) = \lambda^{2}+m^{2}+s^{2} \\
%&= \lambda^{2}+(\Delta-2)^{2}.
%\end{split}
%\ee
%This result is universal and holds for all $(\Delta;\,j_{1},j_{2})$. Its follows by consistency with the thermal partition 
%function and, in principle, could be proved from the explicit form of the AdS mass operator $M^{2}$ for the mixed-symmetric case.
The  
the regularised volume  of the Euclidean  \ads  or hyperboloid $\mathbb H^{5}$ 
 may be written as 
$\mbox{Vol}(\mathbb H^{5}) = \pi^{2}\, \log \RR +...$ where  $\RR$ is an IR cutoff
(the radius of $S^4$ measured in 5d metric  $d\rho^{2} + \sinh^{2}\rho\, d\Omega^{2}_{\rm S^{4}}$ at  large $\rho$).
Doing the analytic continuation (\ref{B.10}) in  the dimension $d_{\rmR}$ in \rf{B19}
we then finally obtain from (\ref{B.3}),\rf{B7} and \rf{B14} 
\ba
\label{B.16}
&\zeta(z) =   \mbox{Vol}(\mathbb H^{5})\,  \zeta(z; x) \no \\
&\qquad \to  \   - \log \RR\, \frac{(2j_{1}+1)(2j_{2}+1)}{12\pi^{2}} 
\,\int_{0}^{\infty}d\lambda\, 
\frac{\big[ \lambda^{2}+(j_{1}-j_{2})^{2}\big] \big[ \lambda^{2}+(j_{1}+j_{2}+1)^{2}\big] }{\big[\lambda^{2}+(\Delta-2)^{2}\big]^{z}}.
\end{align}

%%%%%%%%%%%%%%%%%%%%%%%%
\section{One-parameter ansatz for $\rc$-coefficient}  \label{A:cc}
%%%%%%%%%%%%%%%%%%%%%%%%

Here we  present a  generalization of  our  proposal  for the $\cc$-coefficient \rf{4.11}  
that preserves   correspondence with all  known   results in special cases.  It turns out that 
this leaves  just one-parameter freedom.  The remaining free parameter is   fixed  once we 
assume in addition  the   consistency conditions required  for vectorial AdS/CFT. 

Let us  start with the  following ansatz
\ba
%\begin{split}
\widehat{\rc}(\Delta;\,j_{1},j_{2}) = &\frac{1}{720} (-1)^{2 (j_1+j_2)} (2 j_1+1)
   (2 j_2+1) (\Delta-2) \Big[
   k_{1}\,(\Delta-2)^4\no \\
   &+\big[ k_{2}\,(j_{1}^{2}+j_{2}^{2})+k_{3}\,j_{1}\,j_{2}+k_{4}\,(j_{1}+j_{2})+k_{5}\big]\,(\Delta-2)^2\no \\
   &+ k_{6}\,(j_{1}^{4}+j_{2}^{4})+k_{7}\,(j_{1}^{3}j_{2}+j_{1}j_{2}^{3})+k_{8}\,j_{1}^{2}j_{2}^{2}
   +k_{9}\,(j_{1}^{3}+j_{2}^{3})+k_{10}\,(j_{1}^{2}j_{2}+j_{1}j_{2}^{2})\no \\
   &+
   k_{11}\,(j_{1}^{2}+j_{2}^{2})+k_{12}\,j_{1}j_{2}+k_{13}\,(j_{1}+j_{2})+k_{14}\Big], \la{d1} 
   \end{align}
   where $k_n$ are   some constants to be determined. We shall then 
 require that   this expression    should   reproduce  (i)  the  values of 
 $\rc$ for the   conformal  supergravity fields in Table \ref{T2}; 
 (ii) the representation (\ref{4.4}) and (\ref{4.5}) for $\cc$  of totally symmetric fields (with any $r_b$, $r_f$); 
 (iii) the value of $\rc-\ra$ for all long and short  $SU(2,2|1)$  supermultiplets  as obtained in section  \ref{sec:multiplets}.
 %%%%%%%%%%%%%%%%%%%%%%%
\iffa \begin{enumerate}
\item[a.] Value of $\rc$ for the extended supergravity fields,
\item[b.] parametrisation (\ref{4.4}) and (\ref{4.5}) for totally symmetric fields,
\item[c.] value of $\rc-\ra$ for all long and short $\N=1$ multiplets of $SU(2,2|1)$ according to the 
results obtained from Sec. (\ref{sec:multiplets}).
\end{enumerate}
\fi 
%%%%%%%%%%%%%%%%%%%%%
Remarkably, these conditions fix   all constants in \rf{d1}   apart from   one constant that can be identified with 
the  parameter $r_{b}$ in \rf{4.4}, i.e. we get   %The explicit solution is 
\ba
\widehat{\rc}(\Delta;\,j_{1},j_{2}) = &\frac{1}{720} (-1)^{2 (j_1+j_2)} (2 j_1+1)
   (2 j_2+1) (\Delta-2) \Big[
   2\,(r_{b}-2)\,(\Delta-2)^4\no  \\
   &+\big[ \frac{20}{3}\,(r_{b}+1)\,(j_{1}^{2}+j_{2}^{2})+\frac{20}{3}\,(r_{b}+1)\,(j_{1}+j_{2})
   -10\,(r_{b}-1)\big]\,(\Delta-2)^2\no   \\
   & +2\,(r_{b}+4)\,(j_{1}^{4}+j_{2}^{4})+\frac{20}{3}\,(r_{b}+4)\,j_{1}^{2}j_{2}^{2}
   +4\,(r_{b}+4)\,(j_{1}^{3}+j_{2}^{3})\no   \\
   &+\frac{20}{3}\,(r_{b}+4)\,(j_{1}^{2}j_{2}+j_{1}j_{2}^{2})+\frac{20}{3}\,(r_{b}+4)\,j_{1}j_{2}
  \no  \\
   & -\frac{2}{3}(13r_{b}+22)\,(j_{1}^{2}+j_{2}^{2})
   -\frac{4}{3}\,(8r_{b}+17)\,(j_{1}+j_{2})+8\,r_{b}\Big] \ .  \la{d2}
   \end{align}
The expression   (\ref{4.5}) for the totally symmetric fermionic fields 
then has  %holds with 
\be
r_{f} = \frac{16}{3}\,r_{b}+\frac{169}{3} \ ,   \la{d3}
\ee
which is a generalization of both \rf{46} and \rf{4.12}.  
Our  proposal \rf{4.11}  corresponds to the choice of $r_b$ in \rf{4.12}, i.e.
\be r_b=-1  \ ,  \la{d4} \ee 
while \rf{4.2}  is reproduced  if $r_b= \ha$ as in \rf{46}. 
Our choice \rf{d4}   ensures, in particular,  that the consistency  conditions for vectorial AdS/CFT   discussed in section 6.2 that 
hold for $\aa$-coefficient  and $E_c$   are  valid also   for the  $\cc$-coefficient.

%%%%%%%%%%%%%%%%%%%%%%%%%%%%%
\section{\ads field content  of type IIB  10d supergravity compactified on S$^{5}$}

In Table \ref{T5}  we  summarize the  field 
   content  of  S$^{5}$ compactification of IIB supergravity \cite{Gunaydin:1984fk,Kim:1985ez}. 
   For each  KK  level $p$  we list  the   corresponding 
   $SO(2,4)$  and      $SU(4)$ representations. 
   
   The   dimension of $SU(4)$  representation $[a,b,c]$ (where $a,b,c$ are Dynkin labels)\foot{Dynkin labels  are related 
   to the 
   Young tableu labels or highest weights  $h_i$ by  $[a,b,c]=  (h_{2}-h_{3}, h_{1}-h_{2}, h_{2}+h_{3})$.}
   is %given by 
\be
\label{C.1}
\dd(a,b,c)=\frac{1}{12} %(1+a-b)(1+b-c)(1+b+c)(2+a-c) (2+a+c)(3+a+b) \ . 
(a+1)(b+1)(c+1)(a+b+2)(b+c+2)(a+b+c+3) \ .
\ee
We recall that the level $p=1$ states (doubleton multiplet) 
 are  decoupled from the  physical  spectrum. 
  The level $p=2$ is the massless multiplet  of  gauged  $\N=8$ 5d supergravity; 
  it is isomorphic to  the  multiplet of states  of   $\N=4$ conformal supergravity in Tables \ref{T2} and  \ref{T3}.\foot{There we 
    ignored  the  auxiliary scalar in the $SU(4)$ representation $(0,2,0)$ of dimension 20.} 
  The states with $p\ge 3$  form  shortened  massive multiplets with spin $\le 2$.

   %%%%%%%%%%%%%%%%%%%%%%%%%%%%%%%%%%%%%%%
\label{A:KK}
 \begin{table}[H]
\be
\begin{array}{|c|c|c}
\hline
 & (\Delta;\, j_{1},j_{2}) & SU(4)  \\
 \hline
  && \\
 & (p;\,0,0) & [0,p,0]  \\
 & (p+\frac{1}{2};\,\frac{1}{2},0) & [0,p-1,1]_{c} \\
 & (p+1;\,1,0) & [0,p-1,0]_{c}  \\
p\ge 2 & (p+1;\,0,0) & [0,p-2,2]_{c}  \\
 & (p+2;\,0,0) & [0,p-2,0]_{c}  \\
 & (p+\frac{3}{2};\,\frac{1}{2},0) & [0,p-2,1]_{c}  \\
 & (p+1;\,\frac{1}{2},\frac{1}{2}) & [1,p-2,1]  \\
 & (p+\frac{3}{2};\,1,\frac{1}{2}) & [1,p-2,0]_{c}  \\
 & (p+2;\,1,1) & [0,p-2,0]  \\ 
 && \\
 && \\
\hline
\end{array}
\begin{array}{||c|c|c|}
\hline
 & (\Delta;\,j_{1},j_{2}) & SU(4)  \\
 \hline
   & (p+\frac{3}{2};\,\frac{1}{2},0) & [2,p-3,1]_{c}  \\
  & (p+\frac{5}{2};\,\frac{1}{2},0) & [0,p-3,1]_{c}  \\
p\ge 3  & (p+2;\,\frac{1}{2},\frac{1}{2}) & [1,p-3,1]_{c}  \\
  & (p+2;\,1,0) & [2,p-3,0]_{c} \\
  & (p+3;\,1,0) & [0,p-3,0]_{c}  \\
  & (p+\frac{5}{2};\,1,\frac{1}{2}) & [1,p-3,0]_{c}  \\
  \hline
  & (p+2;\,0,0) & [2,p-4,2]  \\
  & (p+3;\,0,0) & [0,p-4,2]_{c}  \\
p\ge 4  & (p+4;\,0,0) & [0,p-4,0]  \\
  & (p+\frac{5}{2};\,\frac{1}{2},0) & [2,p-4,1]_{c}  \\
  & (p+\frac{7}{2};\,\frac{1}{2},0) & [0,p-4,1]_{c}  \\
  & (p+3;\,\frac{1}{2},\frac{1}{2}) & [1,p-4,1] \\
  \hline
  \end{array}
\nonumber
\ee
\caption{Field  content of  compactification of type  IIB supergravity on $S^5$.}
\label{T5}
\end{table}
%%%%%%%%%%%%%%%%%%%%%

\bibliography{CHS-Biblio}

\providecommand{\href}[2]{#2}\begingroup\raggedright\begin{thebibliography}{100}

\bibitem{Giombi:2013yva}
S.~Giombi, I.~R. Klebanov, S.~S. Pufu, B.~R. Safdi, and G.~Tarnopolsky, {\it
  {AdS Description of Induced Higher-Spin Gauge Theory}},  {\em JHEP} {\bf
  1310} (2013) 016, [\href{http://xxx.lanl.gov/abs/1306.5242}{{\tt
  arXiv:1306.5242}}].

\bibitem{Giombi:2013fka}
S.~Giombi and I.~R. Klebanov, {\it {One Loop Tests of Higher Spin AdS/CFT}},
  {\em JHEP} {\bf 1312} (2013) 068,
  [\href{http://xxx.lanl.gov/abs/1308.2337}{{\tt arXiv:1308.2337}}].

\bibitem{Tseytlin:2013jya}
A.~A. Tseytlin, {\it {On partition function and Weyl anomaly of conformal
  higher spin fields}},  {\em Nucl.Phys.} {\bf B877} (2013) 598--631,
  [\href{http://xxx.lanl.gov/abs/1309.0785}{{\tt arXiv:1309.0785}}].

\bibitem{Tseytlin:2013fca}
A.~A. Tseytlin, {\it {Weyl anomaly of conformal higher spins on six-sphere}},
  {\em Nucl.Phys.} {\bf B877} (2013) 632--646,
  [\href{http://xxx.lanl.gov/abs/1310.1795}{{\tt arXiv:1310.1795}}].

\bibitem{Giombi:2014iua}
S.~Giombi, I.~R. Klebanov, and B.~R. Safdi, {\it {Higher Spin
  AdS$_{d+1}$/CFT$_d$ at One Loop}},  {\em Phys.Rev.} {\bf D89} (2014) 084004,
  [\href{http://xxx.lanl.gov/abs/1401.0825}{{\tt arXiv:1401.0825}}].

\bibitem{Giombi:2014yra}
S.~Giombi, I.~R. Klebanov, and A.~A. Tseytlin, {\it {Partition Functions and
  Casimir Energies in Higher Spin $AdS_{d+1}/CFT_d$}},
  \href{http://xxx.lanl.gov/abs/1402.5396}{{\tt arXiv:1402.5396}}.

\bibitem{Beccaria:2014jxa}
M.~Beccaria, X.~Bekaert, and A.~A. Tseytlin, {\it {Partition function of free
  conformal higher spin theory}},  {\em JHEP} {\bf 1408} (2014) 113,
  [\href{http://xxx.lanl.gov/abs/1406.3542}{{\tt arXiv:1406.3542}}].

\bibitem{Barvinsky:2005ms}
A.~Barvinsky and D.~Nesterov, {\it {Quantum effective action in spacetimes with
  branes and boundaries}},  {\em Phys.Rev.} {\bf D73} (2006) 066012,
  [\href{http://xxx.lanl.gov/abs/hep-th/0512291}{{\tt hep-th/0512291}}].

\bibitem{Diaz:2007an}
D.~E. Diaz and H.~Dorn, {\it {Partition functions and double-trace deformations
  in AdS/CFT}},  {\em JHEP} {\bf 0705} (2007) 046,
  [\href{http://xxx.lanl.gov/abs/hep-th/0702163}{{\tt hep-th/0702163}}].

\bibitem{Diaz:2008hy}
D.~E. Diaz, {\it {Polyakov formulas for GJMS operators from AdS/CFT}},  {\em
  JHEP} {\bf 0807} (2008) 103, [\href{http://xxx.lanl.gov/abs/0803.0571}{{\tt
  arXiv:0803.0571}}].

\bibitem{Witten:2001ua}
E.~Witten, {\it {Multitrace operators, boundary conditions, and AdS / CFT
  correspondence}},  \href{http://xxx.lanl.gov/abs/hep-th/0112258}{{\tt
  hep-th/0112258}}.

\bibitem{Gubser:2002zh}
S.~S. Gubser and I.~Mitra, {\it {Double trace operators and one loop vacuum
  energy in AdS / CFT}},  {\em Phys.Rev.} {\bf D67} (2003) 064018,
  [\href{http://xxx.lanl.gov/abs/hep-th/0210093}{{\tt hep-th/0210093}}].

\bibitem{Gubser:2002vv}
S.~S. Gubser and I.~R. Klebanov, {\it {A Universal result on central charges in
  the presence of double trace deformations}},  {\em Nucl.Phys.} {\bf B656}
  (2003) 23--36, [\href{http://xxx.lanl.gov/abs/hep-th/0212138}{{\tt
  hep-th/0212138}}].

\bibitem{Hartman:2006dy}
T.~Hartman and L.~Rastelli, {\it {Double-trace deformations, mixed boundary
  conditions and functional determinants in AdS/CFT}},  {\em JHEP} {\bf 0801}
  (2008) 019, [\href{http://xxx.lanl.gov/abs/hep-th/0602106}{{\tt
  hep-th/0602106}}].

\bibitem{Duff:1977ay}
M.~Duff, {\it {Observations on Conformal Anomalies}},  {\em Nucl.Phys.} {\bf
  B125} (1977) 334.

\bibitem{Christensen:1978md}
S.~Christensen and M.~Duff, {\it {New Gravitational Index Theorems and
  Supertheorems}},  {\em Nucl.Phys.} {\bf B154} (1979) 301.

\bibitem{Cappelli:1988vw}
A.~Cappelli and A.~Coste, {\it {On the Stress Tensor of Conformal Field
  Theories in Higher Dimensions}},  {\em Nucl.Phys.} {\bf B314} (1989) 707.

\bibitem{Fradkin:1983tg}
E.~S. Fradkin and A.~A. Tseytlin, {\it {Conformal Anomaly in Weyl Theory and
  Anomaly Free Superconformal Theories}},  {\em Phys.Lett.} {\bf B134} (1984)
  187.

\bibitem{Herzog:2013ed}
C.~P. Herzog and K.-W. Huang, {\it {Stress Tensors from Trace Anomalies in
  Conformal Field Theories}},  {\em Phys.Rev.} {\bf D87} (2013) 081901,
  [\href{http://xxx.lanl.gov/abs/1301.5002}{{\tt arXiv:1301.5002}}].

\bibitem{Huang:2013lhw}
K.-W. Huang, {\it {Weyl Anomaly Induced Stress Tensors in General Manifolds}},
  {\em Nucl.Phys.} {\bf B879} (2014) 370--381,
  [\href{http://xxx.lanl.gov/abs/1308.2355}{{\tt arXiv:1308.2355}}].

\bibitem{Henningson:1998gx}
M.~Henningson and K.~Skenderis, {\it {The Holographic Weyl anomaly}},  {\em
  JHEP} {\bf 9807} (1998) 023,
  [\href{http://xxx.lanl.gov/abs/hep-th/9806087}{{\tt hep-th/9806087}}].

\bibitem{Mansfield:2000zw}
P.~Mansfield and D.~Nolland, {\it {Order $1/ N^{2}$ test of the Maldacena
  conjecture: Cancellation of the one loop Weyl anomaly}},  {\em Phys.Lett.}
  {\bf B495} (2000) 435--439,
  [\href{http://xxx.lanl.gov/abs/hep-th/0005224}{{\tt hep-th/0005224}}].

\bibitem{Mansfield:2003gs}
P.~Mansfield, D.~Nolland, and T.~Ueno, {\it {The Boundary Weyl anomaly in the
  $\mathcal N=4$ SYM / type IIB supergravity correspondence}},  {\em JHEP} {\bf
  0401} (2004) 013, [\href{http://xxx.lanl.gov/abs/hep-th/0311021}{{\tt
  hep-th/0311021}}].

\bibitem{Ardehali:2013xya}
A.~A. Ardehali, J.~T. Liu, and P.~Szepietowski, {\it {$1/N^2$ corrections to
  the holographic Weyl anomaly}},  {\em JHEP} {\bf 1401} (2014) 002,
  [\href{http://xxx.lanl.gov/abs/1310.2611}{{\tt arXiv:1310.2611}}].

\bibitem{Ardehali:2013gra}
A.~A. Ardehali, J.~T. Liu, and P.~Szepietowski, {\it {The spectrum of IIB
  supergravity on $AdS_5\times S^5/\mathbb Z_3$ and a $1/N^2$ test of
  AdS/CFT}},  {\em JHEP} {\bf 1306} (2013) 024,
  [\href{http://xxx.lanl.gov/abs/1304.1540}{{\tt arXiv:1304.1540}}].

\bibitem{Ardehali:2014zba}
A.~A. Ardehali, J.~T. Liu, and P.~Szepietowski, {\it {c-a from the $\mathcal
  N=1$ superconformal index}},  \href{http://xxx.lanl.gov/abs/1407.6024}{{\tt
  arXiv:1407.6024}}.

\bibitem{Fradkin:1981jc}
E.~S. Fradkin and A.~A. Tseytlin, {\it {One Loop Beta Function in Conformal
  Supergravities}},  {\em Nucl.Phys.} {\bf B203} (1982) 157.

\bibitem{Paneitz:1983}
S.~Paneitz, {\it {A Quartic Conformally Covariant Differential Operator for
  Arbitrary Pseudo-Riemannian Manifolds (Summary)}},
  \href{http://xxx.lanl.gov/abs/0803.4331}{{\tt arXiv:0803.4331}}.

\bibitem{Deser:1983tm}
S.~Deser and R.~I. Nepomechie, {\it {Anomalous Propagation of Gauge Fields in
  Conformally Flat Spaces}},  {\em Phys.Lett.} {\bf B132} (1983) 321.

\bibitem{Erdmenger:1997wy}
J.~Erdmenger and H.~Osborn, {\it {Conformally covariant differential operators:
  Symmetric tensor fields}},  {\em Class.Quant.Grav.} {\bf 15} (1998) 273--280,
  [\href{http://xxx.lanl.gov/abs/gr-qc/9708040}{{\tt gr-qc/9708040}}].

\bibitem{Achour:2013afa}
J.~B. Achour, E.~Huguet, and J.~Renaud, {\it {Conformally invariant wave
  equation for a symmetric second rank tensor ("spin-2") in d-dimensional
  curved background}},  {\em Phys.Rev.} {\bf D89} (2014) 064041,
  [\href{http://xxx.lanl.gov/abs/1311.3124}{{\tt arXiv:1311.3124}}].

\bibitem{Erdmenger:1997gy}
J.~Erdmenger, {\it {Conformally covariant differential operators: Properties
  and applications}},  {\em Class.Quant.Grav.} {\bf 14} (1997) 2061--2084,
  [\href{http://xxx.lanl.gov/abs/hep-th/9704108}{{\tt hep-th/9704108}}].

\bibitem{Liu:1998bu}
H.~Liu and A.~A. Tseytlin, {\it {D = 4 superYang-Mills, D = 5 gauged
  supergravity, and D = 4 conformal supergravity}},  {\em Nucl.Phys.} {\bf
  B533} (1998) 88--108, [\href{http://xxx.lanl.gov/abs/hep-th/9804083}{{\tt
  hep-th/9804083}}].

\bibitem{Branson:1999jz}
T.~P. Branson, P.~B. Gilkey, K.~Kirsten, and D.~V. Vassilevich, {\it {Heat
  kernel asymptotics with mixed boundary conditions}},  {\em Nucl.Phys.} {\bf
  B563} (1999) 603--626, [\href{http://xxx.lanl.gov/abs/hep-th/9906144}{{\tt
  hep-th/9906144}}].

\bibitem{Metsaev:1994ys}
R.~R. Metsaev, {\it {Lowest eigenvalues of the energy operator for totally
  (anti)symmetric massless fields of the n-dimensional anti-de Sitter group}},
  {\em Class.Quant.Grav.} {\bf 11} (1994) L141--L145.

\bibitem{Metsaev:1995re}
R.~R. Metsaev, {\it {Massless mixed symmetry bosonic free fields in
  d-dimensional anti-de Sitter space-time}},  {\em Phys.Lett.} {\bf B354}
  (1995) 78--84.

\bibitem{Metsaev:2003cu}
R.~R. Metsaev, {\it {Massive totally symmetric fields in AdS$_{d}$}},  {\em
  Phys.Lett.} {\bf B590} (2004) 95--104,
  [\href{http://xxx.lanl.gov/abs/hep-th/0312297}{{\tt hep-th/0312297}}].

\bibitem{Metsaev:1998xg}
R.~Metsaev, {\it {Fermionic fields in the d-dimensional anti-de Sitter
  space-time}},  {\em Phys.Lett.} {\bf B419} (1998) 49--56,
  [\href{http://xxx.lanl.gov/abs/hep-th/9802097}{{\tt hep-th/9802097}}].

\bibitem{Metsaev:2013wza}
R.~R. Metsaev, {\it {CFT adapted approach to massless fermionic fields,
  AdS/CFT, and fermionic conformal fields}},
  \href{http://xxx.lanl.gov/abs/1311.7350}{{\tt arXiv:1311.7350}}.

\bibitem{David:2009xg}
J.~R. David, M.~R. Gaberdiel, and R.~Gopakumar, {\it {The Heat Kernel on
  AdS$_{3}$ and its Applications}},  {\em JHEP} {\bf 1004} (2010) 125,
  [\href{http://xxx.lanl.gov/abs/0911.5085}{{\tt arXiv:0911.5085}}].

\bibitem{Gupta:2012he}
R.~K. Gupta and S.~Lal, {\it {Partition Functions for Higher-Spin theories in
  AdS}},  {\em JHEP} {\bf 1207} (2012) 071,
  [\href{http://xxx.lanl.gov/abs/1205.1130}{{\tt arXiv:1205.1130}}].

\bibitem{Zinoviev:2008ck}
Y.~Zinoviev, {\it {On spin 3 interacting with gravity}},  {\em
  Class.Quant.Grav.} {\bf 26} (2009) 035022,
  [\href{http://xxx.lanl.gov/abs/0805.2226}{{\tt arXiv:0805.2226}}].

\bibitem{Boulanger:2008tg}
N.~Boulanger, S.~Leclercq, and P.~Sundell, {\it {On The Uniqueness of Minimal
  Coupling in Higher-Spin Gauge Theory}},  {\em JHEP} {\bf 0808} (2008) 056,
  [\href{http://xxx.lanl.gov/abs/0805.2764}{{\tt arXiv:0805.2764}}].

\bibitem{Fradkin:1985am}
E.~S. Fradkin and A.~A. Tseytlin, {\it {Conformal supergravity}},  {\em
  Phys.Rept.} {\bf 119} (1985) 233--362.

\bibitem{Gunaydin:1998jc}
M.~Gunaydin, D.~Minic, and M.~Zagermann, {\it {Novel supermultiplets of
  $SU(2,2|4)$ and the $AdS_{5}/ CFT_{4}$ duality}},  {\em Nucl.Phys.} {\bf
  B544} (1999) 737--758, [\href{http://xxx.lanl.gov/abs/hep-th/9810226}{{\tt
  hep-th/9810226}}].

\bibitem{Cardy:1991kr}
J.~L. Cardy, {\it {Operator content and modular properties of higher
  dimensional conformal field theories}},  {\em Nucl.Phys.} {\bf B366} (1991)
  403--419.

\bibitem{Kutasov:2000td}
D.~Kutasov and F.~Larsen, {\it {Partition sums and entropy bounds in weakly
  coupled CFT}},  {\em JHEP} {\bf 0101} (2001) 001,
  [\href{http://xxx.lanl.gov/abs/hep-th/0009244}{{\tt hep-th/0009244}}].

\bibitem{Gopakumar:2011qs}
R.~Gopakumar, R.~K. Gupta, and S.~Lal, {\it {The Heat Kernel on $AdS$}},  {\em
  JHEP} {\bf 1111} (2011) 010, [\href{http://xxx.lanl.gov/abs/1103.3627}{{\tt
  arXiv:1103.3627}}].

\bibitem{Dolan:2005wy}
F.~Dolan, {\it {Character formulae and partition functions in higher
  dimensional conformal field theory}},  {\em J.Math.Phys.} {\bf 47} (2006)
  062303, [\href{http://xxx.lanl.gov/abs/hep-th/0508031}{{\tt
  hep-th/0508031}}].

\bibitem{Gibbons:2006ij}
G.~Gibbons, M.~Perry, and C.~Pope, {\it {Partition functions, the Bekenstein
  bound and temperature inversion in anti-de Sitter space and its conformal
  boundary}},  {\em Phys.Rev.} {\bf D74} (2006) 084009,
  [\href{http://xxx.lanl.gov/abs/hep-th/0606186}{{\tt hep-th/0606186}}].

\bibitem{Bergshoeff:1980is}
E.~Bergshoeff, M.~de~Roo, and B.~de~Wit, {\it {Extended Conformal
  Supergravity}},  {\em Nucl.Phys.} {\bf B182} (1981) 173.

\bibitem{Basar:2014hda}
G.~Basar, A.~Cherman, D.~A. McGady, and M.~Yamazaki, {\it {Casimir energy of
  confining large $N$ gauge theories}},
  \href{http://xxx.lanl.gov/abs/1408.3120}{{\tt arXiv:1408.3120}}.

\bibitem{Camporesi:1994ga}
R.~Camporesi and A.~Higuchi, {\it {Spectral functions and zeta functions in
  hyperbolic spaces}},  {\em J.Math.Phys.} {\bf 35} (1994) 4217--4246.

\bibitem{Costa:2014kfa}
M.~S. Costa, V.~Gonçalves, and J.~Penedones, {\it {Spinning AdS Propagators}},
   {\em JHEP} {\bf 1409} (2014) 064,
  [\href{http://xxx.lanl.gov/abs/1404.5625}{{\tt arXiv:1404.5625}}].

\bibitem{Nutma:2014pua}
T.~Nutma and M.~Taronna, {\it {On conformal higher spin wave operators}},  {\em
  JHEP} {\bf 1406} (2014) 066, [\href{http://xxx.lanl.gov/abs/1404.7452}{{\tt
  arXiv:1404.7452}}].

\bibitem{Dobrev:1985qv}
V.~Dobrev and V.~Petkova, {\it {All Positive Energy Unitary Irreducible
  Representations of Extended Conformal Supersymmetry}},  {\em Phys.Lett.} {\bf
  B162} (1985) 127--132.

\bibitem{Gunaydin:1984fk}
M.~Gunaydin and N.~Marcus, {\it {The Spectrum of the $S^{5}$ Compactification
  of the Chiral $\mathcal N=2$, D=10 Supergravity and the Unitary
  Supermultiplets of $U(2, 2|4)$}},  {\em Class.Quant.Grav.} {\bf 2} (1985)
  L11.

\bibitem{Howe:1981qj}
P.~S. Howe, K.~Stelle, and P.~Townsend, {\it {Supercurrents}},  {\em
  Nucl.Phys.} {\bf B192} (1981) 332.

\bibitem{Gunaydin:1984vz}
M.~Gunaydin and N.~Marcus, {\it {The Unitary Supermultiplet of $\mathcal N=8$
  Conformal Superalgebra Involving Fields of Spin $\le 2$}},  {\em
  Class.Quant.Grav.} {\bf 2} (1985) L19.

\bibitem{Ferrara:1998ej}
S.~Ferrara, C.~Fronsdal, and A.~Zaffaroni, {\it {On N=8 supergravity on AdS(5)
  and N=4 superconformal Yang-Mills theory}},  {\em Nucl.Phys.} {\bf B532}
  (1998) 153--162, [\href{http://xxx.lanl.gov/abs/hep-th/9802203}{{\tt
  hep-th/9802203}}].

\bibitem{Gunaydin:1998sw}
M.~Gunaydin, D.~Minic, and M.~Zagermann, {\it {4-D doubleton conformal
  theories, CPT and IIB string on $AdS_{5}\times S^{5}$}},  {\em Nucl.Phys.}
  {\bf B534} (1998) 96--120,
  [\href{http://xxx.lanl.gov/abs/hep-th/9806042}{{\tt hep-th/9806042}}].

\bibitem{Maldacena:1997re}
J.~M. Maldacena, {\it {The Large N limit of superconformal field theories and
  supergravity}},  {\em Int.J.Theor.Phys.} {\bf 38} (1999) 1113--1133,
  [\href{http://xxx.lanl.gov/abs/hep-th/9711200}{{\tt hep-th/9711200}}].

\bibitem{Gubser:1998bc}
S.~Gubser, I.~R. Klebanov, and A.~M. Polyakov, {\it {Gauge theory correlators
  from noncritical string theory}},  {\em Phys.Lett.} {\bf B428} (1998)
  105--114, [\href{http://xxx.lanl.gov/abs/hep-th/9802109}{{\tt
  hep-th/9802109}}].

\bibitem{Witten:1998qj}
E.~Witten, {\it {Anti-de Sitter space and holography}},  {\em
  Adv.Theor.Math.Phys.} {\bf 2} (1998) 253--291,
  [\href{http://xxx.lanl.gov/abs/hep-th/9802150}{{\tt hep-th/9802150}}].

\bibitem{Balasubramanian:1999re}
V.~Balasubramanian and P.~Kraus, {\it {A Stress tensor for Anti-de Sitter
  gravity}},  {\em Commun.Math.Phys.} {\bf 208} (1999) 413--428,
  [\href{http://xxx.lanl.gov/abs/hep-th/9902121}{{\tt hep-th/9902121}}].

\bibitem{Bianchi:2003wx}
M.~Bianchi, J.~F. Morales, and H.~Samtleben, {\it {On stringy $AdS_{5}\times
  S^{5}$ and higher spin holography}},  {\em JHEP} {\bf 0307} (2003) 062,
  [\href{http://xxx.lanl.gov/abs/hep-th/0305052}{{\tt hep-th/0305052}}].

\bibitem{Kim:1985ez}
H.~Kim, L.~Romans, and P.~van Nieuwenhuizen, {\it {The Mass Spectrum of Chiral
  $\mathcal N=2$ D=10 Supergravity on $S^{5}$}},  {\em Phys.Rev.} {\bf D32}
  (1985) 389.

\bibitem{Bilal:1999ph}
A.~Bilal and C.-S. Chu, {\it {A Note on the chiral anomaly in the AdS / CFT
  correspondence and $1/N^2 $ correction}},  {\em Nucl.Phys.} {\bf B562} (1999)
  181--190, [\href{http://xxx.lanl.gov/abs/hep-th/9907106}{{\tt
  hep-th/9907106}}].

\bibitem{Allen:1983an}
B.~Allen and S.~Davis, {\it {Vacuum Energy in Gauged Extended Supergravity}},
  {\em Phys.Lett.} {\bf B124} (1983) 353.

\bibitem{Duff:1982ev}
M.~Duff, {\it {Supergravity, the Seven Sphere, and Spontaneous Symmetry
  Breaking}},  {\em Nucl.Phys.} {\bf B219} (1983) 389.

\bibitem{Gibbons:1984dg}
G.~Gibbons and H.~Nicolai, {\it {One Loop Effects on the Round Seven Sphere}},
  {\em Phys.Lett.} {\bf B143} (1984) 108--114.

\bibitem{Inami:1984vp}
T.~Inami and K.~Yamagishi, {\it {Vanishing Quantum Vacuum Energy in
  Eleven-dimensional Supergravity on the Round Seven Sphere}},  {\em
  Phys.Lett.} {\bf B143} (1984) 115--120.

\bibitem{Vasiliev:2001zy}
M.~Vasiliev, {\it {Conformal higher spin symmetries of 4-d massless
  supermultiplets and osp(L,2M) invariant equations in generalized
  (super)space}},  {\em Phys.Rev.} {\bf D66} (2002) 066006,
  [\href{http://xxx.lanl.gov/abs/hep-th/0106149}{{\tt hep-th/0106149}}].

\bibitem{Vasiliev:2003ev}
M.~Vasiliev, {\it {Nonlinear equations for symmetric massless higher spin
  fields in (A)dS(d)}},  {\em Phys.Lett.} {\bf B567} (2003) 139--151,
  [\href{http://xxx.lanl.gov/abs/hep-th/0304049}{{\tt hep-th/0304049}}].

\bibitem{Bekaert:2005vh}
X.~Bekaert, S.~Cnockaert, C.~Iazeolla, and M.~Vasiliev, {\it {Nonlinear higher
  spin theories in various dimensions}},
  \href{http://xxx.lanl.gov/abs/hep-th/0503128}{{\tt hep-th/0503128}}.

\bibitem{Didenko:2014dwa}
V.~Didenko and E.~Skvortsov, {\it {Elements of Vasiliev theory}},
  \href{http://xxx.lanl.gov/abs/1401.2975}{{\tt arXiv:1401.2975}}.

\bibitem{Klebanov:2002ja}
I.~Klebanov and A.~Polyakov, {\it {AdS dual of the critical O(N) vector
  model}},  {\em Phys.Lett.} {\bf B550} (2002) 213--219,
  [\href{http://xxx.lanl.gov/abs/hep-th/0210114}{{\tt hep-th/0210114}}].

\bibitem{Sezgin:2003pt}
E.~Sezgin and P.~Sundell, {\it {Holography in 4D (super) higher spin theories
  and a test via cubic scalar couplings}},  {\em JHEP} {\bf 0507} (2005) 044,
  [\href{http://xxx.lanl.gov/abs/hep-th/0305040}{{\tt hep-th/0305040}}].

\bibitem{Leigh:2003gk}
R.~G. Leigh and A.~C. Petkou, {\it {Holography of the N=1 higher spin theory on
  AdS(4)}},  {\em JHEP} {\bf 0306} (2003) 011,
  [\href{http://xxx.lanl.gov/abs/hep-th/0304217}{{\tt hep-th/0304217}}].

\bibitem{Giombi:2009wh}
S.~Giombi and X.~Yin, {\it {Higher Spin Gauge Theory and Holography: The
  Three-Point Functions}},  {\em JHEP} {\bf 1009} (2010) 115,
  [\href{http://xxx.lanl.gov/abs/0912.3462}{{\tt arXiv:0912.3462}}].

\bibitem{Didenko:2012vh}
V.~Didenko and E.~Skvortsov, {\it {Towards higher-spin holography in ambient
  space of any dimension}},  {\em J.Phys.} {\bf A46} (2013) 214010,
  [\href{http://xxx.lanl.gov/abs/1207.6786}{{\tt arXiv:1207.6786}}].

\bibitem{Bekaert:2013zya}
X.~Bekaert and M.~Grigoriev, {\it {Higher order singletons, partially massless
  fields and their boundary values in the ambient approach}},  {\em Nucl.Phys.}
  {\bf B876} (2013) 667--714, [\href{http://xxx.lanl.gov/abs/1305.0162}{{\tt
  arXiv:1305.0162}}].

\bibitem{Konstein:2000bi}
S.~Konstein, M.~Vasiliev, and V.~Zaikin, {\it {Conformal higher spin currents
  in any dimension and AdS / CFT correspondence}},  {\em JHEP} {\bf 0012}
  (2000) 018, [\href{http://xxx.lanl.gov/abs/hep-th/0010239}{{\tt
  hep-th/0010239}}].

\bibitem{Mikhailov:2002bp}
A.~Mikhailov, {\it {Notes on higher spin symmetries}},
  \href{http://xxx.lanl.gov/abs/hep-th/0201019}{{\tt hep-th/0201019}}.

\bibitem{Schnitzer:2003zr}
H.~J. Schnitzer, {\it {Gauged vector models and higher spin representations in
  AdS(5)}},  {\em Nucl.Phys.} {\bf B695} (2004) 283--300,
  [\href{http://xxx.lanl.gov/abs/hep-th/0310210}{{\tt hep-th/0310210}}].

\bibitem{Bekaert:2012vt}
X.~Bekaert and M.~Grigoriev, {\it {Notes on the ambient approach to boundary
  values of AdS gauge fields}},  {\em J.Phys.} {\bf A46} (2013) 214008,
  [\href{http://xxx.lanl.gov/abs/1207.3439}{{\tt arXiv:1207.3439}}].

\bibitem{Maldacena:2011jn}
J.~Maldacena and A.~Zhiboedov, {\it {Constraining Conformal Field Theories with
  A Higher Spin Symmetry}},  {\em J.Phys.} {\bf A46} (2013) 214011,
  [\href{http://xxx.lanl.gov/abs/1112.1016}{{\tt arXiv:1112.1016}}].

\bibitem{Vasiliev:2004cm}
M.~Vasiliev, {\it {Higher spin superalgebras in any dimension and their
  representations}},  {\em JHEP} {\bf 0412} (2004) 046,
  [\href{http://xxx.lanl.gov/abs/hep-th/0404124}{{\tt hep-th/0404124}}].

\bibitem{Boulanger:2013zza}
N.~Boulanger, D.~Ponomarev, E.~Skvortsov, and M.~Taronna, {\it {On the
  uniqueness of higher-spin symmetries in AdS and CFT}},  {\em Int.J.Mod.Phys.}
  {\bf A28} (2013) 1350162, [\href{http://xxx.lanl.gov/abs/1305.5180}{{\tt
  arXiv:1305.5180}}].

\bibitem{Stanev:2012nq}
Y.~S. Stanev, {\it {Correlation Functions of Conserved Currents in Four
  Dimensional Conformal Field Theory}},  {\em Nucl.Phys.} {\bf B865} (2012)
  200--215, [\href{http://xxx.lanl.gov/abs/1206.5639}{{\tt arXiv:1206.5639}}].

\bibitem{Stanev:2013qra}
Y.~S. Stanev, {\it {Constraining conformal field theory with higher spin
  symmetry in four dimensions}},  {\em Nucl.Phys.} {\bf B876} (2013) 651--666,
  [\href{http://xxx.lanl.gov/abs/1307.5209}{{\tt arXiv:1307.5209}}].

\bibitem{Alba:2013yda}
V.~Alba and K.~Diab, {\it {Constraining conformal field theories with a higher
  spin symmetry in d=4}},  \href{http://xxx.lanl.gov/abs/1307.8092}{{\tt
  arXiv:1307.8092}}.

\bibitem{Bekaert:2009fg}
X.~Bekaert and M.~Grigoriev, {\it {Manifestly conformal descriptions and higher
  symmetries of bosonic singletons}},  {\em SIGMA} {\bf 6} (2010) 038,
  [\href{http://xxx.lanl.gov/abs/0907.3195}{{\tt arXiv:0907.3195}}].

\bibitem{Gunaydin:1998km}
M.~Gunaydin and D.~Minic, {\it {Singletons, doubletons and M theory}},  {\em
  Nucl.Phys.} {\bf B523} (1998) 145--157,
  [\href{http://xxx.lanl.gov/abs/hep-th/9802047}{{\tt hep-th/9802047}}].

\bibitem{Ferrara:1998jm}
S.~Ferrara and C.~Fronsdal, {\it {Gauge fields as composite boundary
  excitations}},  {\em Phys.Lett.} {\bf B433} (1998) 19--28,
  [\href{http://xxx.lanl.gov/abs/hep-th/9802126}{{\tt hep-th/9802126}}].

\bibitem{Sundborg:2000wp}
B.~Sundborg, {\it {Stringy gravity, interacting tensionless strings and
  massless higher spins}},  {\em Nucl.Phys.Proc.Suppl.} {\bf 102} (2001)
  113--119, [\href{http://xxx.lanl.gov/abs/hep-th/0103247}{{\tt
  hep-th/0103247}}].

\bibitem{Sezgin:2001zs}
E.~Sezgin and P.~Sundell, {\it {Doubletons and 5-D higher spin gauge theory}},
  {\em JHEP} {\bf 0109} (2001) 036,
  [\href{http://xxx.lanl.gov/abs/hep-th/0105001}{{\tt hep-th/0105001}}].

\bibitem{Boulanger:2011se}
N.~Boulanger and E.~Skvortsov, {\it {Higher-spin algebras and cubic
  interactions for simple mixed-symmetry fields in AdS spacetime}},  {\em JHEP}
  {\bf 1109} (2011) 063, [\href{http://xxx.lanl.gov/abs/1107.5028}{{\tt
  arXiv:1107.5028}}].

\bibitem{Vasiliev:2001wa}
M.~Vasiliev, {\it {Cubic interactions of bosonic higher spin gauge fields in
  AdS(5)}},  {\em Nucl.Phys.} {\bf B616} (2001) 106--162,
  [\href{http://xxx.lanl.gov/abs/hep-th/0106200}{{\tt hep-th/0106200}}].

\bibitem{Alkalaev:2002rq}
K.~Alkalaev and M.~Vasiliev, {\it {N=1 supersymmetric theory of higher spin
  gauge fields in AdS(5) at the cubic level}},  {\em Nucl.Phys.} {\bf B655}
  (2003) 57--92, [\href{http://xxx.lanl.gov/abs/hep-th/0206068}{{\tt
  hep-th/0206068}}].

\bibitem{Alkalaev:2003qv}
K.~Alkalaev, O.~Shaynkman, and M.~Vasiliev, {\it {On the frame - like
  formulation of mixed symmetry massless fields in (A)dS(d)}},  {\em
  Nucl.Phys.} {\bf B692} (2004) 363--393,
  [\href{http://xxx.lanl.gov/abs/hep-th/0311164}{{\tt hep-th/0311164}}].

\bibitem{Alkalaev:2006rw}
K.~Alkalaev, O.~Shaynkman, and M.~Vasiliev, {\it {Frame-like formulation for
  free mixed-symmetry bosonic massless higher-spin fields in AdS(d)}},
  \href{http://xxx.lanl.gov/abs/hep-th/0601225}{{\tt hep-th/0601225}}.

\bibitem{Alkalaev:2012ic}
K.~Alkalaev, {\it {Massless hook field in AdS(d+1) from the holographic
  perspective}},  {\em JHEP} {\bf 1301} (2013) 018,
  [\href{http://xxx.lanl.gov/abs/1210.0217}{{\tt arXiv:1210.0217}}].

\bibitem{Alkalaev:2012rg}
K.~Alkalaev, {\it {Mixed-symmetry tensor conserved currents and AdS/CFT
  correspondence}},  {\em J.Phys.} {\bf A46} (2013) 214007,
  [\href{http://xxx.lanl.gov/abs/1207.1079}{{\tt arXiv:1207.1079}}].

\bibitem{Metsaev:2005ar}
R.~R. Metsaev, {\it {Cubic interaction vertices of massive and massless higher
  spin fields}},  {\em Nucl.Phys.} {\bf B759} (2006) 147--201,
  [\href{http://xxx.lanl.gov/abs/hep-th/0512342}{{\tt hep-th/0512342}}].

\bibitem{Alkalaev:2010af}
K.~Alkalaev, {\it {FV-type action for $AdS_5$ mixed-symmetry fields}},  {\em
  JHEP} {\bf 1103} (2011) 031, [\href{http://xxx.lanl.gov/abs/1011.6109}{{\tt
  arXiv:1011.6109}}].

\bibitem{Boulanger:2011qt}
N.~Boulanger, E.~Skvortsov, and Y.~Zinoviev, {\it {Gravitational cubic
  interactions for a simple mixed-symmetry gauge field in AdS and flat
  backgrounds}},  {\em J.Phys.} {\bf A44} (2011) 415403,
  [\href{http://xxx.lanl.gov/abs/1107.1872}{{\tt arXiv:1107.1872}}].

\bibitem{Lopez:2012pr}
L.~Lopez, {\it {On cubic AdS interactions of mixed-symmetry higher spins}},
  \href{http://xxx.lanl.gov/abs/1210.0554}{{\tt arXiv:1210.0554}}.

\bibitem{Joung:2012hz}
E.~Joung, L.~Lopez, and M.~Taronna, {\it {Generating functions of
  (partially-)massless higher-spin cubic interactions}},  {\em JHEP} {\bf 1301}
  (2013) 168, [\href{http://xxx.lanl.gov/abs/1211.5912}{{\tt
  arXiv:1211.5912}}].

\bibitem{Ferrara:1997dh}
S.~Ferrara and C.~Fronsdal, {\it {Conformal Maxwell theory as a singleton field
  theory on adS(5), IIB three-branes and duality}},  {\em Class.Quant.Grav.}
  {\bf 15} (1998) 2153--2164,
  [\href{http://xxx.lanl.gov/abs/hep-th/9712239}{{\tt hep-th/9712239}}].

\bibitem{Anselmi:1999bb}
D.~Anselmi, {\it {Higher spin current multiplets in operator product
  expansions}},  {\em Class.Quant.Grav.} {\bf 17} (2000) 1383--1400,
  [\href{http://xxx.lanl.gov/abs/hep-th/9906167}{{\tt hep-th/9906167}}].

\bibitem{Gelfond:2006be}
O.~Gelfond, E.~Skvortsov, and M.~Vasiliev, {\it {Higher spin conformal currents
  in Minkowski space}},  {\em Theor.Math.Phys.} {\bf 154} (2008) 294--302,
  [\href{http://xxx.lanl.gov/abs/hep-th/0601106}{{\tt hep-th/0601106}}].

\bibitem{Arutyunov:1998xt}
G.~Arutyunov and S.~Frolov, {\it {Antisymmetric tensor field on AdS(5)}},  {\em
  Phys.Lett.} {\bf B441} (1998) 173--177,
  [\href{http://xxx.lanl.gov/abs/hep-th/9807046}{{\tt hep-th/9807046}}].

\bibitem{Sundborg:1999ue}
B.~Sundborg, {\it {The Hagedorn transition, deconfinement and N=4 SYM theory}},
   {\em Nucl.Phys.} {\bf B573} (2000) 349--363,
  [\href{http://xxx.lanl.gov/abs/hep-th/9908001}{{\tt hep-th/9908001}}].

\bibitem{Aharony:2003sx}
O.~Aharony, J.~Marsano, S.~Minwalla, K.~Papadodimas, and M.~Van~Raamsdonk, {\it
  {The Hagedorn - deconfinement phase transition in weakly coupled large N
  gauge theories}},  {\em Adv.Theor.Math.Phys.} {\bf 8} (2004) 603--696,
  [\href{http://xxx.lanl.gov/abs/hep-th/0310285}{{\tt hep-th/0310285}}].

\bibitem{Shenker:2011zf}
S.~H. Shenker and X.~Yin, {\it {Vector Models in the Singlet Sector at Finite
  Temperature}},  \href{http://xxx.lanl.gov/abs/1109.3519}{{\tt
  arXiv:1109.3519}}.

\bibitem{Barabanschikov:2005ri}
A.~Barabanschikov, L.~Grant, L.~L. Huang, and S.~Raju, {\it {The Spectrum of
  Yang Mills on a sphere}},  {\em JHEP} {\bf 0601} (2006) 160,
  [\href{http://xxx.lanl.gov/abs/hep-th/0501063}{{\tt hep-th/0501063}}].

\bibitem{Lal:2012ax}
S.~Lal, {\it {CFT$_{4}$ Partition Functions and the Heat Kernel on AdS$_{5}$}},
   {\em Phys.Lett.} {\bf B727} (2013) 325--329,
  [\href{http://xxx.lanl.gov/abs/1212.1050}{{\tt arXiv:1212.1050}}].

\bibitem{Chang:2012kt}
C.-M. Chang, S.~Minwalla, T.~Sharma, and X.~Yin, {\it {ABJ Triality: from
  Higher Spin Fields to Strings}},  {\em J.Phys.} {\bf A46} (2013) 214009,
  [\href{http://xxx.lanl.gov/abs/1207.4485}{{\tt arXiv:1207.4485}}].

\bibitem{Sezgin:2001yf}
E.~Sezgin and P.~Sundell, {\it {Towards massless higher spin extension of D=5,
  N=8 gauged supergravity}},  {\em JHEP} {\bf 0109} (2001) 025,
  [\href{http://xxx.lanl.gov/abs/hep-th/0107186}{{\tt hep-th/0107186}}].

\bibitem{Sezgin:2002rt}
E.~Sezgin and P.~Sundell, {\it {Massless higher spins and holography}},  {\em
  Nucl.Phys.} {\bf B644} (2002) 303--370,
  [\href{http://xxx.lanl.gov/abs/hep-th/0205131}{{\tt hep-th/0205131}}].

\bibitem{Beisert:2003te}
N.~Beisert, M.~Bianchi, J.~Morales, and H.~Samtleben, {\it {On the spectrum of
  AdS / CFT beyond supergravity}},  {\em JHEP} {\bf 0402} (2004) 001,
  [\href{http://xxx.lanl.gov/abs/hep-th/0310292}{{\tt hep-th/0310292}}].

\bibitem{Tseytlin:2002gz}
A.~A. Tseytlin, {\it {On limits of superstring in $AdS_{5}\times S^{5}$}},
  {\em Theor.Math.Phys.} {\bf 133} (2002) 1376--1389,
  [\href{http://xxx.lanl.gov/abs/hep-th/0201112}{{\tt hep-th/0201112}}].

\bibitem{Segal:2002gd}
A.~Y. Segal, {\it {Conformal higher spin theory}},  {\em Nucl.Phys.} {\bf B664}
  (2003) 59--130, [\href{http://xxx.lanl.gov/abs/hep-th/0207212}{{\tt
  hep-th/0207212}}].

\bibitem{Bekaert:2010ky}
X.~Bekaert, E.~Joung, and J.~Mourad, {\it {Effective action in a higher-spin
  background}},  {\em JHEP} {\bf 1102} (2011) 048,
  [\href{http://xxx.lanl.gov/abs/1012.2103}{{\tt arXiv:1012.2103}}].

\bibitem{Fradkin:1989md}
E.~S. Fradkin and V.~Y. Linetsky, {\it {Cubic Interaction in Conformal Theory
  of Integer Higher Spin Fields in Four-dimensional Space-time}},  {\em
  Phys.Lett.} {\bf B231} (1989) 97.

\bibitem{Fradkin:1990ps}
E.~S. Fradkin and V.~Y. Linetsky, {\it {Superconformal Higher Spin Theory in
  the Cubic Approximation}},  {\em Nucl.Phys.} {\bf B350} (1991) 274--324.

\bibitem{Metsaev:2008ba}
R.~Metsaev, {\it {Conformal self-dual fields}},  {\em J.Phys.} {\bf A43} (2010)
  115401, [\href{http://xxx.lanl.gov/abs/0812.2861}{{\tt arXiv:0812.2861}}].

\bibitem{Flato:1978qz}
M.~Flato and C.~Fronsdal, {\it {One Massless Particle Equals Two Dirac
  Singletons: Elementary Particles in a Curved Space. 6.}},  {\em
  Lett.Math.Phys.} {\bf 2} (1978) 421--426.

\bibitem{Bianchi:2006ti}
M.~Bianchi, F.~Dolan, P.~Heslop, and H.~Osborn, {\it {N=4 superconformal
  characters and partition functions}},  {\em Nucl.Phys.} {\bf B767} (2007)
  163--226, [\href{http://xxx.lanl.gov/abs/hep-th/0609179}{{\tt
  hep-th/0609179}}].

\bibitem{Dobrev:2012me}
V.~Dobrev, {\it {Explicit character formulae for positive energy unitary
  irreducible representations of D = 4 conformal supersymmetry}},  {\em
  J.Phys.} {\bf A46} (2013) 405202,
  [\href{http://xxx.lanl.gov/abs/1208.6250}{{\tt arXiv:1208.6250}}].

\bibitem{Camporesi:1995fb}
R.~Camporesi and A.~Higuchi, {\it {On the Eigen functions of the Dirac operator
  on spheres and real hyperbolic spaces}},  {\em J.Geom.Phys.} {\bf 20} (1996)
  1--18, [\href{http://xxx.lanl.gov/abs/gr-qc/9505009}{{\tt gr-qc/9505009}}].

\bibitem{Kaku:1978nz}
M.~Kaku, P.~Townsend, and P.~van Nieuwenhuizen, {\it {Properties of Conformal
  Supergravity}},  {\em Phys.Rev.} {\bf D17} (1978) 3179.

\bibitem{Tseytlin:1984wj}
A.~A. Tseytlin, {\it {Effective action in de Sitter space and conformal
  supergravity}},  {\em Yad.Fiz.} {\bf 39} (1984), no.~6 1606--1615.

\bibitem{Fradkin:1983zz}
E.~S. Fradkin and A.~A. Tseytlin, {\it {Instanton zero modes and beta functions
  in supergravities. 2. Conformal supergravity}},  {\em Phys.Lett.} {\bf B134}
  (1984) 307.

\bibitem{Ferrara:1977mv}
S.~Ferrara and B.~Zumino, {\it {Structure of Conformal Supergravity}},  {\em
  Nucl.Phys.} {\bf B134} (1978) 301.

\end{thebibliography}\endgroup

\bibliographystyle{JHEP}

\end{document}